\documentclass[fleqn,usenatbib]{mnras}

\usepackage{newtxtext,newtxmath}

\usepackage[T1]{fontenc}

\DeclareRobustCommand{\VAN}[3]{#2}
\let\VANthebibliography\thebibliography
\def\thebibliography{\DeclareRobustCommand{\VAN}[3]{##3}\VANthebibliography}
\renewcommand{\emph}[1]{\textit{#1}}
\newcommand{\lre}{\ensuremath{\lambda_{R_{\rm{e}}}}}
\newcommand{\re}{\ensuremath{\rm{R}_{\rm{e}}}}
\newcommand{\hot}{\ensuremath{f_{\rm{Hot, e}}}}
\newcommand{\warm}{\ensuremath{f_{\rm{Warm, e}}}}
\newcommand{\cold}{\ensuremath{f_{\rm{Cold, e}}}}
\newcommand{\counter}{\ensuremath{f_{\rm{Counter, e}}}}
\newcommand{\hotcr}{\ensuremath{f_{\rm{Hot+CR, e}}}}
\newcommand{\boxorbit}{\ensuremath{f_{\rm{Box, e}}}}
\newcommand{\age}{\ensuremath{\rm{Age}_{\rm{LW}}}}
\newcommand{\env}{\ensuremath{\log(\Sigma_5/\rm{Mpc}^{-2})}}
\newcommand{\lr}{\ifmmode{\lambda_{R_e}}\else{$\lambda_{R_e}$}\fi}
\newcommand{\kms}{\ifmmode{\,\rm{km}\, \rm{s}^{-1}}\else{$\,$km$\,$s$^{-1}$}\fi}
\newcommand{\logm}{\ensuremath{\log_{10}\!\left(\rm{M}_*/\rm{M}_\odot\right)}}
\newcommand{\update}[1]{#1}

\usepackage{graphicx}	
\usepackage{amsmath}	

\title[SAMI Schwarzschild Modelling]{The SAMI Galaxy Survey: Linking Tidal Features and Orbit Populations Using Schwarzschild Modelling}

\author[T. H. Rutherford et al.]{
T. H. Rutherford,$^{1,2}$\thanks{E-mail: trut2989@uni.sydney.edu.au}
J. van de Sande,$^{3,1,2}$
S. M. Croom,$^{1,2}$
A. Fraser-McKelvie,$^{2,4}$
G. Santucci,$^{5}$\newauthor
$\text{ }$G. van de Ven,$^{6}$
S. M. Sweet,$^{7,2}$
S. Brough,$^{3,2}$
Y. Mai,$^{8,2,9}$
J. Bryant,$^{1,2,10}$
J. Bland-Hawthorn,$^{1,2}$\newauthor
$\text{ }$M. Goodwin,$^{8}$
J. Lawrence,$^{8}$
N. P. F. Lorente$^{8}$
\\
$^{1}$Sydney Institute for Astronomy, School of Physics, A28, The University of Sydney, NSW, 2006, Australia\\
$^{2}$ARC Centre of Excellence for All Sky Astrophysics in 3 Dimensions (ASTRO 3D), Australia\\
$^{3}$School of Physics, University of New South Wales, NSW, 2052, Australia\\
$^{4}$European Southern Observatory, Karl-Schwarzschild-Stra\ss e 2, Garching, 85748, Germany\\
$^{5}$CSIRO Space \& Astronomy, PO Box 1130, Bentley, Western Australia 6102, Australia\\
$^{6}$Department of Astrophysics, University of Vienna, T\"urkenschanzstra\ss e 17, 1180 Vienna, Austria\\
$^{7}$School of Mathematics and Physics, University of Queensland, St Lucia, Queensland 4072, Australia\\
$^{8}$Australian Astronomical Optics, Macquarie University, Sydney, NSW 2109, Australia\\
$^{9}$Astrophysics and Space Technologies Research Centre, Macquarie University, Sydney, NSW 2109, Australia\\
$^{10}$Astralis-USyd, Sydney Institute for Astronomy, School of Physics, The University of Sydney, Sydney, NSW 2006, Australia\\
}

\date{Accepted XXX. Received YYY; in original form ZZZ}

\pubyear{\the\year{}}

\begin{document}
\label{firstpage}
\pagerange{\pageref{firstpage}--\pageref{lastpage}}
\maketitle

\begin{abstract}
The evolution of angular momentum in galaxies is shaped by a combination of internal secular processes and external mechanisms such as mergers. Orbit-superposition based dynamical modelling provides a powerful means of linking the intrinsic orbital structures of galaxies to their global properties and merger histories. We construct Schwarzschild orbit-superposition models of massive (\logm>10) SAMI galaxies using the DYNAMITE code, utilising deep KiDS photometry to accurately reproduce each galaxy's luminosity distribution. We find that the fractions of hot, cold, warm, and counter-rotating orbits all show significant correlations with the spin parameter proxy \lre, with the strongest correlation arising from the combined hot plus counter-rotating fraction. When controlling for stellar mass and environment, we find that the fraction of hot and cold orbits show significant correlations with stellar age, whereas warm orbits do not. We further find that the lower values of \lre\ for young galaxies with shell merger features as compared to the full sample is driven by an excess of hot orbits and a deficit of cold orbits, with no dependence on warm orbits. We suggest that the kinematic transformation in this SAMI sample proceeds through stars transitioning directly from cold to hot orbits. As warm orbits are expected to arise from secular heating processes, these findings indicate that merger-driven heating is the dominant mechanism governing the redistribution of angular momentum and the reduction of rotational support in massive galaxies.
\end{abstract}




\section{Introduction}

The mass assembly of galaxies in the Universe is the result of hierarchical processes \citep{1978MNRAS.183..341W}, where many systems merge together over cosmic time. The merger history of galaxies can impact both the internal stellar dynamics \citep[e.g.][]{1979MNRAS.189..831W,1980MNRAS.193..189F,2014MNRAS.444.3357N}, and the external morphology, where tidal features formed from stellar debris serve as long-lived evidence of interactions \citep[e.g.][]{1972ApJ...178..623T,2005AJ....130.2647V,2018ApJ...857..144H,2019A&A...632A.122M,2022ApJS..262...39H}. These features are often classified into two broad categories: shells and streams. Streams form when material is tidally stripped either from the disc of a massive, gas-rich system or from a low-mass satellite, giving rise to extended features \citep{1992AJ....103.1089B,2008ApJ...683...94O, 2018ApJ...857..144H}. Further, shells form from accretion events along nearly radial paths, in contrast to streams, which require a much more circular infall \citep{2019MNRAS.487..318K}. These classifications of tidal debris provide a method for interpreting the kinematic signatures (e.g. spin-down) observed in merger remnants, linking visible features to the intrinsic stellar dynamics.

Simulations have shown that mergers can reduce the rotational support of galaxies, as quantified by the spin parameter proxy \lre\ \citep[e.g.,][]{2009A&A...501L...9D, 2009MNRAS.397.1202J, 2011MNRAS.416.1654B, 2014MNRAS.444.3357N, 2017MNRAS.464.3850L, 2018MNRAS.473.4956L, 2017MNRAS.468.3883P,2020IAUFM..30A.208L}\update{, where \lre\ is the luminosity-weighted rotational velocity $V$ normalised by the second moment $V^2+\sigma^2$, defined as $\lre = {\langle R|V|\rangle} / {\langle R \sqrt{V^2 + \sigma^2}\rangle}$}. However, the evolution of stellar dynamics is not solely governed by major mergers. Secular evolution, harassment \citep[e.g.][]{1999MNRAS.304..465M}, fly-by encounters, dynamical friction \citep{2017ApJ...837...68C}, minor mergers \citep{2020MNRAS.493.3778S}, and cold accretion \citep{2017MNRAS.465.2895L} have also been shown in simulations to drive spin evolution, as well as spurious disc heating effects \citep{2023MNRAS.525.5614L}. Simulations also display a trend for slow rotator galaxies to be preferentially created in dry rather than wet mergers \citep[e.g.][]{2018MNRAS.473.4956L}. Observational studies have confirmed a link between mergers and kinematic spin-down \citep[e.g.][]{2023A&A...672A..27B,2024ApJ...965..158Y}, but the relation shows a large amount of scatter \citep[e.g.][]{2016ApJ...832...69O} and depends on parameters such as stellar age and environmental density \citep[e.g.][]{2024MNRAS.529..810R,2024MNRAS.529.3446C}. 

While simulations and observations both point to merger-driven spin evolution, the large variability in the relation suggests that the integrated dynamical support measured by \lre\ cannot fully capture the variety of dynamical responses to mergers. Indeed, although \lre\ has become the standard for distinguishing between fast and slow rotators and has enabled many important advances in our understanding of galaxy kinematics and evolution \citep[e.g.][]{2011MNRAS.414..888E, 2016ARA&A..54..597C, 2021MNRAS.505.3078V,2021MNRAS.508.2307V,2024MNRAS.529.3446C}, it is sensitive to effects such as inclination, kinematic twists, decoupled components, and beam-smearing \citep[e.g.][]{2011MNRAS.414..888E,2017MNRAS.472.1272V,2020MNRAS.497.2018H}. Furthermore, galaxies with different orbital structures can have the same value of \lre\ \citep[e.g.][]{2009MNRAS.397.1202J}, and while merger remnants change the orbital distribution of galaxies, \lre\ is not always strongly affected \citep[][]{2011MNRAS.416.1654B,2014MNRAS.444.3357N}. Therefore, a full understanding of how merger processes impact stellar dynamics requires an understanding of the detailed orbital structures within a galaxy, beyond what is captured by an average value of $\lambda_R$ within one effective radius. 

A commonly used method of classifying orbits is by using their circularity parameter $\lambda_z=J_z/J_c(E)$, which is the ratio of the specific angular momentum to that of a circular orbit with the same energy \citep{2003ApJ...597...21A}. \cite{2018MNRAS.473.3000Z} classified orbits with $\lambda_z>0.8$ as cold (near circular orbits), orbits with $-0.25<\lambda_z<0.25$ as hot (mostly radial orbits), orbits with $0.25<\lambda_z<0.8$ as warm (a mixture) and orbits with $\lambda_z<-0.25$ as counter-rotating. Galaxies with a large proportion of cold and warm orbits over hot orbits will display \lre$\gtrapprox 0.5$, whereas galaxies with a large proportion of hot orbits will show \lre$\lessapprox 0.3$ \citep{2022ApJ...930..153S}. However, such orbital classifications cannot be derived from velocity moment maps. Measuring orbital distributions from observations requires dynamical modelling techniques that recover the intrinsic dynamical structure \citep[e.g.][]{2018MNRAS.473.3000Z,2019MNRAS.486.4753J}.

The Schwarzschild orbit-superposition method \citep{1979ApJ...232..236S,1982ApJ...263..599S} can be used to create dynamical models of galaxies that gives insight into their intrinsic orbital structures. By populating a triaxial gravitational potential with a comprehensive library of stellar orbits and fitting to the observed velocity maps, a representation of the galaxy's internal dynamics can be obtained. 
There have been many implementations of Schwarzschild models over time, both axisymmetric and triaxial \citep[e.g.][]{1999ApJS..124..383C,2003ApJ...583...92G,2004ApJ...602...66V,2008MNRAS.385..647V,2015MNRAS.450.2842V,2020ApJ...889...39V,2021MNRAS.500.1437N}. In this work, we employ DYNAMITE\footnote{DYnamics, Age and Metallicity Indicators Tracing Evolution} \citep{2020ascl.soft11007J,2022A&A...667A..51T} due to its ability to model triaxial potentials and its flexible, documented\footnote{\url{https://dynamics.univie.ac.at/dynamite_docs/index.html}} python wrapper. By reconstructing the three-dimensional orbital distribution, these Schwarzschild models provide the means to quantify the variety of internal dynamics within a galaxy.

In order to understand the complex relationships between a galaxy's internal dynamics and its assembly history \citep[e.g.][]{2014MNRAS.444.3357N}, dynamical models of a statistically significant sample of galaxies is required. Therefore, we take advantage of the recent wealth of deep imaging from the Hyper-Suprime-Cam (HSC) \citep{2018PASJ...70S...8A} and Kilo-Degree Survey (KiDS) \citep{2019A&A...625A...2K} optical galaxy surveys, complemented by kinematic data from the SAMI Galaxy Survey \citep{2012MNRAS.421..872C,2021MNRAS.505.3078V,2021MNRAS.505..991C} to construct orbit-superposition Schwarzschild models of a sample of massive, early-type galaxies. We base our sample on the sample in \cite{2024MNRAS.529..810R}, who use deep HSC imaging to investigate merger features around ETGs with $\rm{M}_*>10^{10}\rm{M}_\odot$. We re-create and expand the sample of models of SAMI galaxies created by \cite{2022ApJ...930..153S}, by modelling our luminosity density with KiDS imaging rather than Sloan Digital Sky Survey (SDSS) imaging \citep{2009ApJS..182..543A}. As KiDS goes to deeper surface brightness limits than SDSS with better seeing, we expect the outer isophotes of each galaxy's luminosity density, and thus stellar mass distribution, to be more accurately modelled. We aim to investigate the impact of mergers on a galaxy's internal dynamics by combining the merger feature classifications with orbital distributions derived from these dynamical models.

This paper is structured as follows. Section \ref{sec:data_4} describes the data used in this paper and sample selection. Section \ref{sec:method_4} discusses our method. Section \ref{sec:results_4} presents the results. Section \ref{sec:discussion_4} discusses the results in the context of previous work. Section \ref{sec:conclusion_4} presents a conclusion to this work. Throughout this paper, we adopt a flat $\Lambda$CDM cosmology, with $H_0=70$km s$^{-1}$ Mpc$^{-1}$, $\Omega_m=0.3$, $\Omega_\Lambda=0.7$. We further assume a Chabrier \citep{2003PASP..115..763C} initial mass function (IMF), and the AB magnitude system \citep{1983ApJ...266..713O}.

\section{Data}
\label{sec:data_4}
\begin{figure}
    \centering
    \includegraphics[width=\columnwidth]{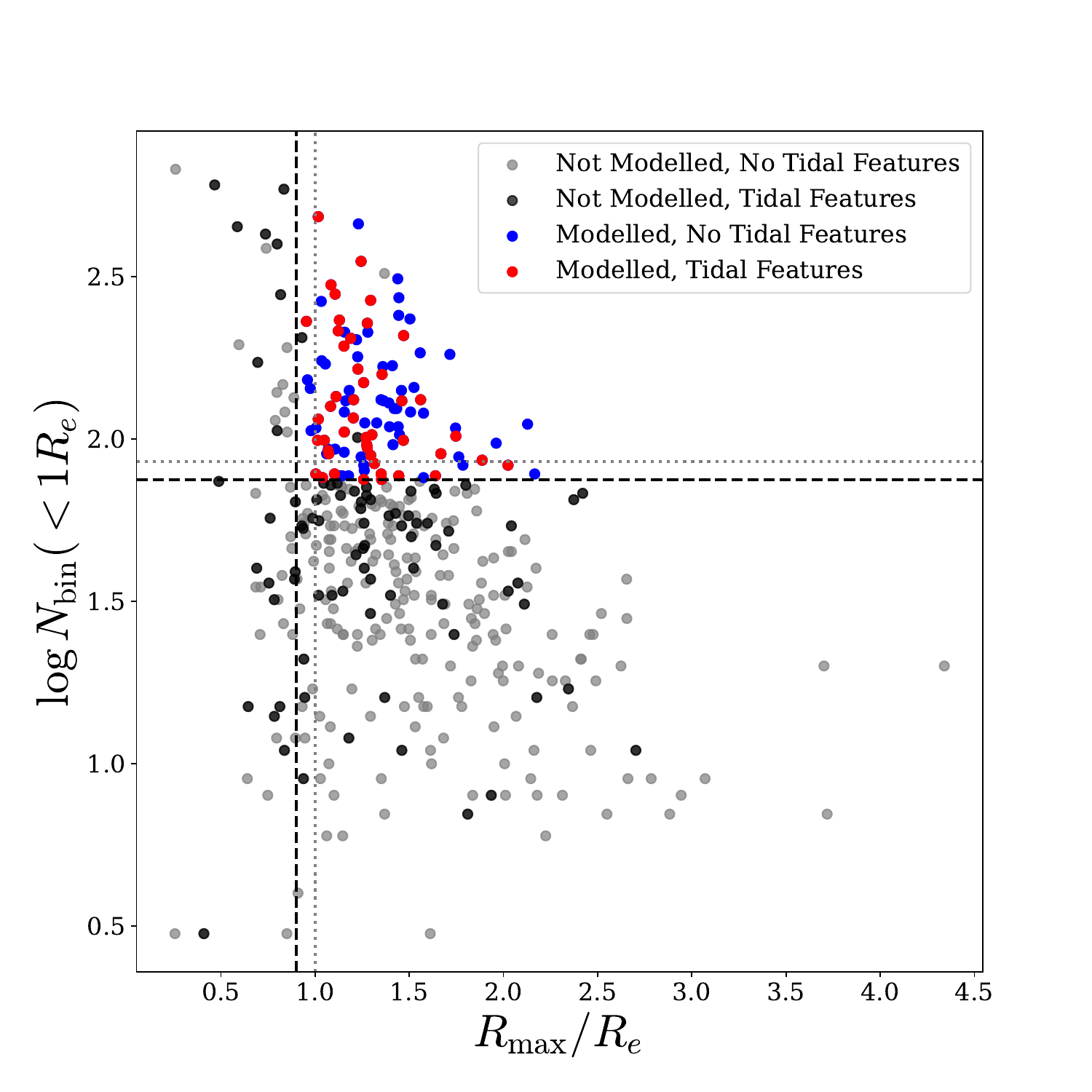}
    \caption[SAMI Galaxies Plotted in Radial Coverage vs Number of Voronoi Bins]{SAMI ETGs with total stellar mass $\rm{M}_*>10^{10}\rm{M}_\odot$, plotted in $\log N_{\text{bin}}(<1R_e)$ versus $R_{\text{max}}/R_e$. $\log N_{\text{bin}}(<1R_e)$ is the log of the number of Voronoi bins within $1R_e$, and $R_{\text{max}}/R_e$ is the maximum radius reached by a galaxy's kinematic maps, divided by $R_e$. We expand the selection criteria from \cite{2022ApJ...930..153S} from $\log N_{\text{bin}}(<1R_e)\geq 85$ and $R_{\text{max}}/R_e\geq 1$ (black dashed lines), to $\log N_{\text{bin}}(<1R_e)\geq 75$ and $R_{\text{max}}/R_e\geq 0.9$ (grey dotted lines). Galaxies shown as black and grey dots are not within our selection criteria, and do and do not show tidal features, respectively. Galaxies shown as red and blue dots are within our selection criteria, and do and do not show tidal features, respectively. There are two galaxies that lie within our selection criteria, but the model fitting failed to converge, and they were thus excluded from our sample.}
    \label{fig:schwarzschild_sample_definition}
\end{figure}
\begin{figure*}
    \centering
    \includegraphics[width=\textwidth]{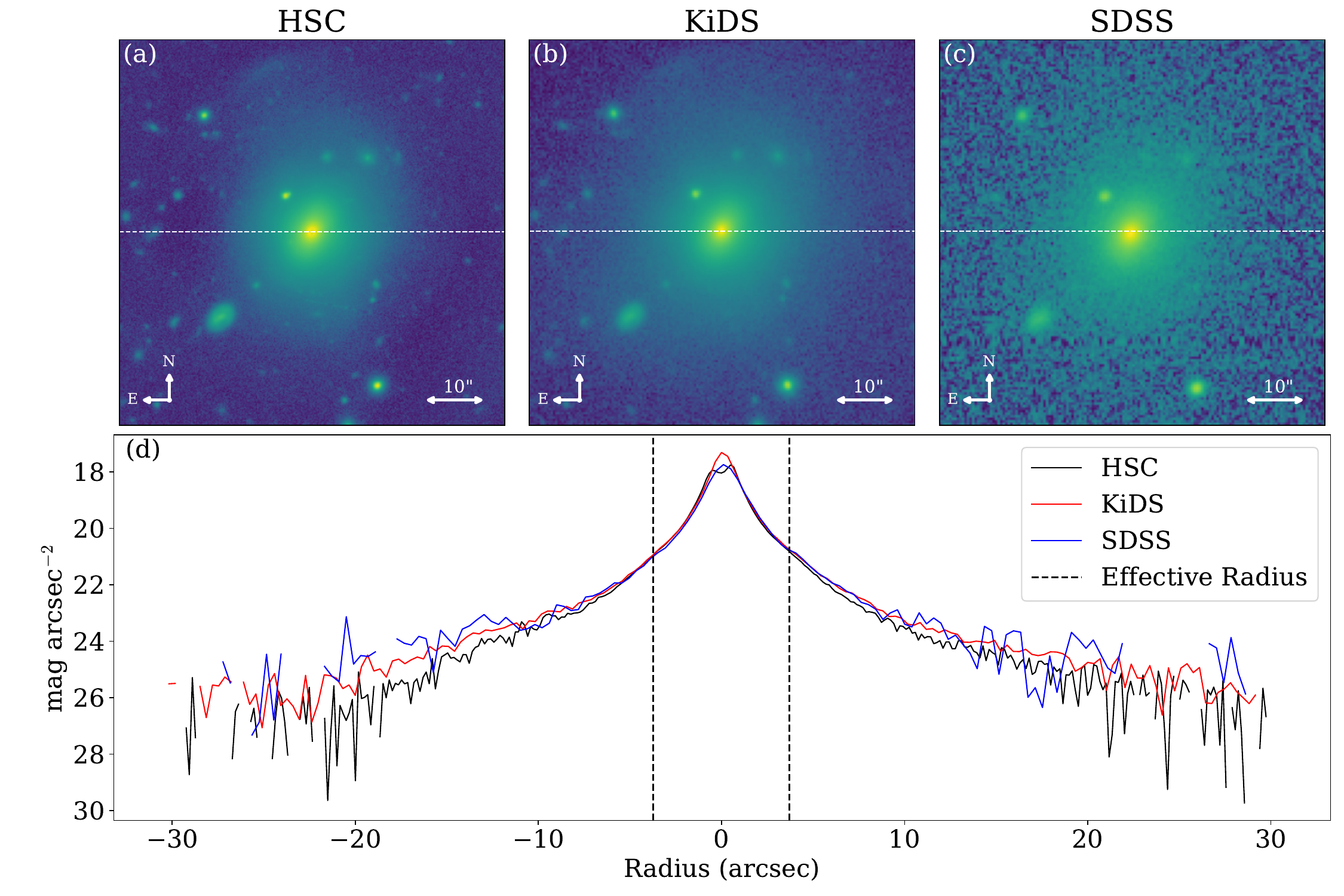}
    \caption[Imaging Comparison for 70802]{A comparison of imaging and surface brightness profiles for HSC DR2, KiDS DR4, and SDSS DR12 imaging for the SAMI galaxy with CATID = 70802. \update{This galaxy was identified as having a shell in \cite{2024MNRAS.529..810R}.} In panels (a), (b), and (c) we show a cutout for the galaxy in HSC, KiDS, and SDSS imaging, respectively. Each cutout is 60" $\times$ 60". We also show a dashed horizontal white line through the centre, from which we take a one-dimensional surface brightness profile. In panel (d), we show this surface brightness profile in mag arcsec$^{-2}$ for HSC (black), KiDS (red), and SDSS (blue). We also show the effective radius of the galaxy in vertical dashed lines. A clear non-physical decrease in peak surface brightness in the centre is clear in HSC that is not present in KiDS or SDSS, a result of saturation in HSC. SDSS peaks at a lower surface brightness than KiDS due to the poorer seeing conditions of SDSS (1.4" vs 0.7").}
    \label{fig:hsc_kids_saturation_comparison}
\end{figure*}
\begin{figure}
    \centering
    \includegraphics[width=\columnwidth]{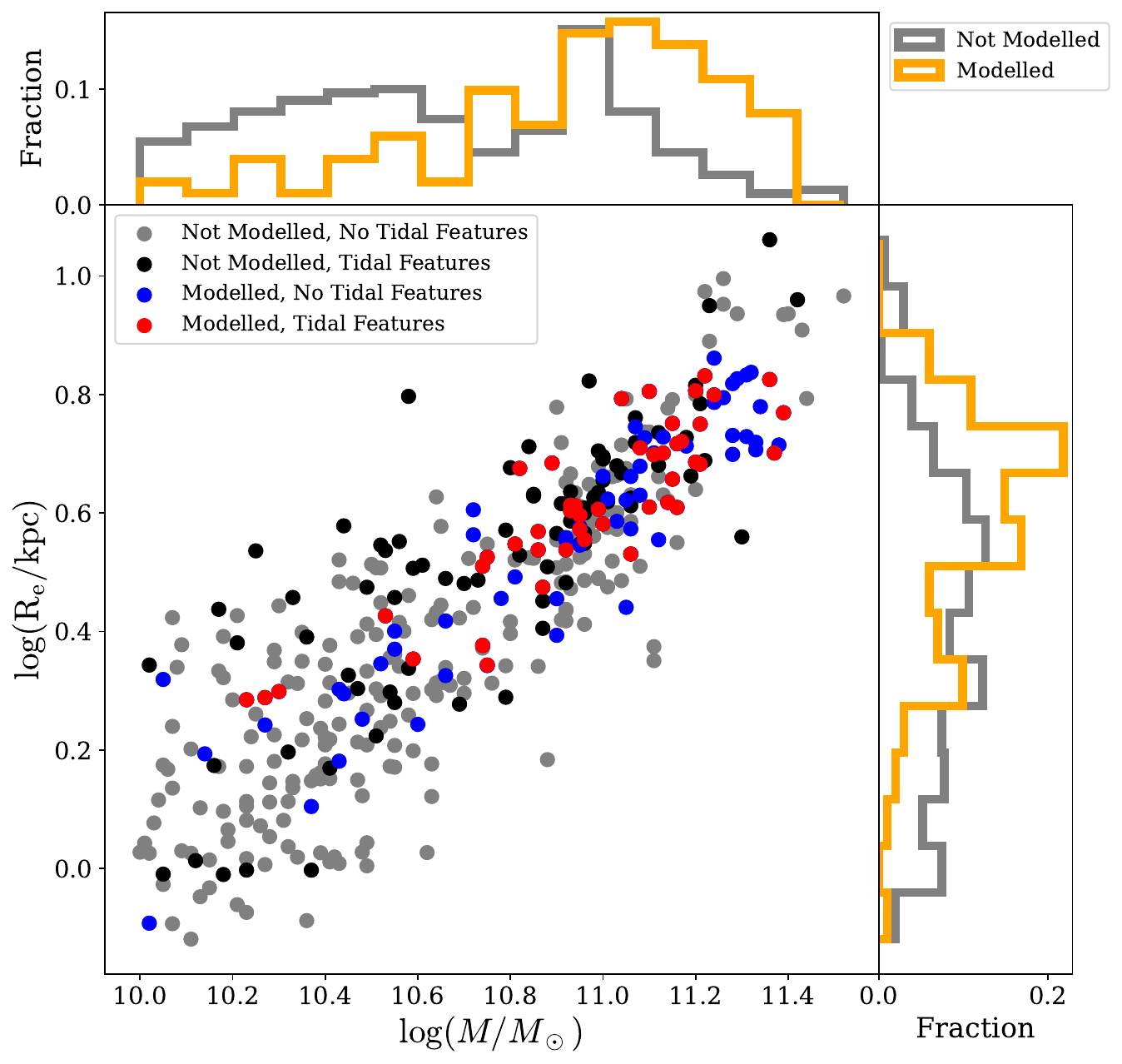}
    \caption[SAMI Galaxies Plotted in Mass vs Size]{SAMI ETGs with $\rm{M}_*>10^{10}\rm{M}_\odot$, plotted in \logm versus $\log(\rm{R}_{\rm{e}}/\rm{kpc})$. $\log(\rm{R}_{\rm{e}}/\rm{kpc})$ is the circular radius of the galaxy in kpc. Galaxies shown as black and grey dots are not within our selection criteria, and do and do not show tidal features, respectively. Galaxies shown as red and blue dots are within our selection criteria, and do and do not show tidal features, respectively. The fractions of galaxies as a function of \logm and $\log(\rm{R}_{\rm{e}}/\rm{kpc})$ are also shown as marginalised histograms for galaxies modelled and galaxies not modelled in orange and grey respectively. We observe that the galaxies that fall into our selection criteria are larger and more massive on average than the full sample. This is due to the selection requirement of at least 75 Voronoi bins within 1\re. However, the galaxies with the largest $\log(\rm{R}_{\rm{e}}/\rm{kpc})$ values are not modelled, as the stellar kinematics do not meet the required value of $R_{\text{max}}/R_e\geq 0.9$.}
    \label{fig:schwarzschild_sample_mass_size}
\end{figure}
\subsection{The SAMI Galaxy Survey}
The SAMI instrument \citep{2012MNRAS.421..872C} was installed on the Anglo-Australian Telescope and offered a 1 degree diameter field of view. SAMI utilised 13 fibre bundles \citep[Hexabundles;][]{2011OExpr..19.2649B,2014MNRAS.438..869B}, each with a 75\% fill factor. Within each bundle, there were 61 fibres, each with a diameter of 1.6\arcsec, resulting in a 15\arcsec diameter for each IFU. These IFUs, along with 26 sky fibres, were connected to the AAOmega spectrograph \citep{2006SPIE.6269E..0GS}. The spectrograph employed the 580V grating covering the wavelength range of $\sim$3700-5750\AA, providing a resolution of R=1808 ($\sigma$=70.4 \kms). The 1000R grating was additionally used, covering 6300-7400\AA, offering a resolution of R=4304 ($\sigma$=29.6 \kms) \citep{2017ApJ...835..104V}. 

The SAMI Galaxy Survey \citep{2012MNRAS.421..872C,2015MNRAS.447.2857B} targeted galaxies selected from the GAMA survey \citep{2011MNRAS.413..971D} and eight low-redshift clusters \citep{2017MNRAS.468.1824O}. Reduced data cubes \citep{2015MNRAS.446.1551S} are accessible through SAMI Galaxy Survey data releases on Data Central \citep{2015MNRAS.446.1567A,2018MNRAS.475..716G,2018MNRAS.481.2299S,2021MNRAS.505..991C}, including stellar kinematic maps. The \lre\ values used in our sample are derived from spatially resolved kinematic measurements \citep{2017ApJ...835..104V}, incorporating corrections for aperture \citep{2017MNRAS.472.1272V} and seeing \citep{2020MNRAS.497.2018H, 2021MNRAS.505.3078V}.

Light-weighted stellar ages are derived from full spectrum fitting of SAMI data, as described in \cite{2022MNRAS.516.2971V}. The average light-weighted stellar age for a galaxy was calculated by first summing the spectrum for all pixels within one effective radius (\re). The \textsc{pPXF} code \citep{2004PASP..116..138C, 2017MNRAS.466..798C} was then used to fit this full spectrum using single-age, single-metallicity stellar populations (SSP) models \citep{2015MNRAS.449.1177V}. The environmental metric used in this work is the fifth nearest neighbour surface density, $\Sigma_5$. This density is defined as $\Sigma_5=5/\pi d^2$, where $d$ is the projected comoving distance to fifth nearest neighbour galaxy, as described in \cite{2017ApJ...844...59B}. We adopt stellar mass estimates from \citet{2015MNRAS.447.2857B}, calculated from Milky Way–extinction–corrected apparent $g$ and $i$ magnitudes, following the technique of \citet{2011MNRAS.418.1587T}.
\subsection{The Kilo-Degree Survey (KiDS)}
In this paper we use the Kilo-Degree Survey (KiDS) $r$-band data from KiDS data release 4, which achieves a limiting surface brightness of $25.02 \pm 0.13$ mag arcsec$^{-2}$ \citep[$5\sigma$ in a 2" aperture, ][]{2019A&A...625A...2K}. KiDS is an optical imaging survey covering 1500 degrees$^2$ \citep{2013ExA....35...25D}, and utilises the OmegaCAM instrument \citep{2011Msngr.146....8K} on the VLT Survey Telescope (VST), which has a 2.6m diameter and is located at the Paranal Observatory in Chile. OmegaCAM has a 268 Megapixel camera, providing a $1^{\circ}\times1^{\circ}$ field of view. A focal plane array of 32 CCDs, each with $2048\times4096$ pixels, gives a total of $16,000\times16,000$ pixels, and a scale of 0.214 arcseconds/pixel. 

We investigated various imaging surveys to determine the most appropriate for our work: KiDS, SDSS and HSC-SSP. The motivation for using KiDS or HSC-SSP imaging over SDSS imaging to model the luminosity density of the galaxies in our sample is improved imaging depth and seeing. A direct comparison of surface brightness limits derived under the same assumptions is not readily available. However, KiDS $r$-band observations were obtained with an exposure time of 1800 seconds on a 2.6m telescope under a mean seeing of 0.70" \citep{2019A&A...625A...2K}. In comparison, SDSS DR7 achieved an effective exposure time of 54.1 seconds on a 2.5m telescope, with a median seeing of 1.40" \citep{2009ApJS..182..543A}. KiDS is thus significantly deeper than SDSS, and the advantage of deep imaging is that reaching the outermost isophotes of a galaxy provides the most accurate model of its light, and hence stellar mass distribution.

The HSC-SSP Wide PDR2 survey \citep{2018PASJ...70S...4A, 2015ApJ...807...22M} achieves a high surface brightness limit of $27.8\pm0.5$ mag arcsec$^{-2}$ in the $r$-band \citep[$1\sigma$ noise in 1" apertures, ][]{2024MNRAS.529..810R}, but experiences point source saturation at $\mu<17.5$ mag. Given our sample of massive early-types, this results in a significant proportion of galaxies displaying saturation in the centre. 

In Figure \ref{fig:hsc_kids_saturation_comparison} we show an example of this saturation, and compare the HSC light profile to the KiDS and SDSS light profile for a galaxy in our sample with CATID = 70802. In panels (a), (b), and (c) we show this galaxy in HSC, KiDS, and SDSS imaging, respectively, with a horizontal dashed line across the centre. In panel (d), we show the surface brightness profile along this horizontal cut for HSC in black, KiDS in red, and SDSS in blue. A clear non-physical decrease in peak surface brightness is seen for HSC, absent in KiDS or SDSS, which is a result of saturation in the HSC imaging. The surface brightness of SDSS peaks lower than KiDS, because of the poorer seeing conditions of SDSS (1.4" vs 0.7"). Outside of the centre, the profiles agree well, with HSC going deeper than KiDS, which itself goes deeper than SDSS. Finally, HSC imaging has been shown to over-estimate the sky brightness in sky-subtraction around bright, extended sources such as galaxies \citep[e.g.][]{2022ApJS..262...39H,2022MNRAS.515.5335L}. Although this is primarily an issue with HSC PDR3, it is visible when comparing panels (a) and (b), with the HSC PDR2 flux decreasing with radius faster than KiDS. As a result of the over sky-subtraction and clear saturation in HSC, we choose to use KiDS imaging, as it remains a substantial improvement in depth and seeing over SDSS.

\subsection{Sample Selection}
To investigate the relations between stellar age, stellar kinematics, and tidal features, we adopt the galaxy sample of \cite{2024MNRAS.529..810R}. This sample comprises early-type galaxies from the GAMA fields, taken from SAMI DR3 \citep{2021MNRAS.505..991C}, with a total stellar mass cut of $\logm>=10$. Tidal features were identified through visual inspection of model subtracted images. Shells or streams were identified if two out of the three people performing inspections agreed on the classification \cite[for more details, see ][]{2024MNRAS.529..810R}. Galaxy morphologies were taken from \citet{2016MNRAS.463..170C}, obtained using the classification approach of \citet{2014MNRAS.439.1245K} on SDSS DR9 $gri$ imaging \citep{2012ApJS..203...21A}. This mass cut was performed as the primary focus of \cite{2024MNRAS.529..810R} was slow rotators, whose fraction increases strongly with stellar mass \citep[e.g.][]{2017MNRAS.472.1272V}. Additionally, the stellar kinematic completeness in the SAMI Galaxy Survey falls below 50\% for $\logm<9.5$ \citep{2022ApJ...930..153S}, necessitating a total stellar mass cut for robust dynamical modelling. This sample contains 411 galaxies.

\update{We take kinematic maps of the four velocity moments $V$, $\sigma$, $h_3$, and $h_4$, along with uncertainties in each moment, from the spatially resolved kinematic measurements derived by \cite{2017ApJ...835..104V}. The velocity moments were calculated using the penalised pixel fitting code \citep[\textsc{ppxf},][]{2004PASP..116..138C}, with uncertainties estimated from 150 simulated spectra. We adopt the spaxel quality criteria for our kinematic maps used by \cite{2017ApJ...835..104V} to ensure reliability in the kinematic measurements,} but increase the minimum signal to noise per spaxel from $S/N>3$ \AA$^{-1}$ to $S/N>10$ \AA$^{-1}$. This higher threshold provides more robust measurements while allowing us to work with a reduced minimum number of spaxels within one effective radius. These criteria are therefore:
\begin{align}
    Q_1)&\;\;S/N>10\text{ \AA}^{-1} \;\;\&\;\;\sigma>35\text{ km/s}\\
    Q_2)&\;\;V_{\text{error}}<30 \text{ km/s}\;\;\&\;\;\sigma_{\text{error}}<\sigma\times0.1+25\text{ km/s}
\end{align}

The final step of the sample definition is to select from our sample based on the radial coverage of SAMI. In Figure \ref{fig:schwarzschild_sample_definition}, we show our initial sample, displayed as the number of Voronoi bins within 1\re vs radial coverage. We also show our sample in Figure \ref{fig:schwarzschild_sample_mass_size}, displayed in the mass-size plane. We extend the selection criteria from \cite{2022ApJ...930..153S}, and require each galaxy to have a minimum number of bins within 1\re\ of 75, and the minimum radius reached in the kinematic maps to be 0.9\re. These cuts in sampling and radial coverage are displayed as the black dashed lines in Figure \ref{fig:schwarzschild_sample_definition}, with the selection criteria of \cite{2022ApJ...930..153S} displayed as grey dotted lines. Galaxies shown as black (with tidal features) and grey (without tidal features) dots do not fall within our selection criteria. Galaxies shown as red (with tidal features) and blue (without tidal features) dots are within our selection criteria. Finally, two galaxies that satisfy our selection criteria failed to converge during model fitting, and they were consequently excluded from our sample. The final sample thus consists of 101 galaxies, of which 64 are shared with \citet{2022ApJ...930..153S}. \update{Of these 101 galaxies, \cite{2024MNRAS.529..810R} found:}
\begin{itemize}
    \item \update{28/101 (28\%) have a shell,}
    \item \update{24/101 (24\%) have a stream,}
    \item \update{6/101 (6\%) have a shell and a stream,}
    \item \update{55/101 (54\%) have neither a shell nor a stream.}
\end{itemize}

\section{Orbit-Superposition Modelling}
\label{sec:method_4}
We use the Schwarzschild's method \citep{1979ApJ...232..236S,1982ApJ...263..599S} to create orbit-superposition models for our galaxies, using the DYNAMITE \citep{2020ascl.soft11007J,2022A&A...667A..51T} implementation and code. We use DYNAMITE, as it accommodates triaxial models, provides multiple grid-search algorithms for efficient convergence of our free parameters, and offers a well-documented python wrapper.

The derivation of a Schwarzschild model can be summarised in three steps: 1) create a model for the gravitational potential, 2) derive a library of stellar orbits allowed by this potential, and 3) calculate which combination of orbits best reproduces the observed kinematic maps. These steps are described in detail in \cite{2008MNRAS.385..647V}, and we summarise them in Sections \ref{sec:grav_pot}-\ref{sec:orbit_weighting}.
\subsection{Gravitational Potential}
\label{sec:grav_pot}
We model the gravitational potential as consisting of a stellar mass, dark halo, and supermassive black hole component. 
\subsubsection{Stellar Potential}
The stellar mass distribution is modelled with the Multi-Gaussian Expansion (MGE) formalism \citep{1992A&A...253..366M,1994A&A...285..723E,2002MNRAS.333..400C}. This method is adopted by the DYNAMITE code that we use in this paper to construct our Schwarzschild models. MGE assumes that a galaxy's projected luminosity on the sky, $I$, can be represented by a sum of $N$ two-dimensional Gaussians:
\begin{equation}
    I(x',y') = \sum^N_{k=1}\frac{L_k}{2\pi\sigma_k'^2q_k'}\exp\bigg[-\frac{1}{2\sigma_k'^2}\bigg(x'^2+\frac{y'^2}{q_k'^2}\bigg)\bigg]
\end{equation}
where $x'$ and $y'$ are the projected coordinates on the sky, and each Gaussian component has total luminosity $L_k$, observed axial ratio $q_k'$, and dispersion $\sigma_k'$ along the major axis. \update{We note that the parameters $q_k'$ and $\sigma_k'$ are the observed quantities, whereas $q_k$ and $\sigma_k$ are the intrinsic quantities.}

We create MGE models of KiDS $r$-band images for each galaxy in our sample. The galaxy mask and PSF for each image are derived using the source-extraction and star-fitting procedures of \textsc{profound} and \textsc{profuse} respectively \citep{2018MNRAS.476.3137R,2022MNRAS.513.2985R}. The PSF, provided by \textsc{profuse} as a Moffat profile image \citep{1969A&A.....3..455M}, is then modelled with MGE so it may be analytically convolved with the MGE model of the relevant galaxy during fitting. 

\begin{figure*}
    \centering
    \includegraphics[width=\textwidth]{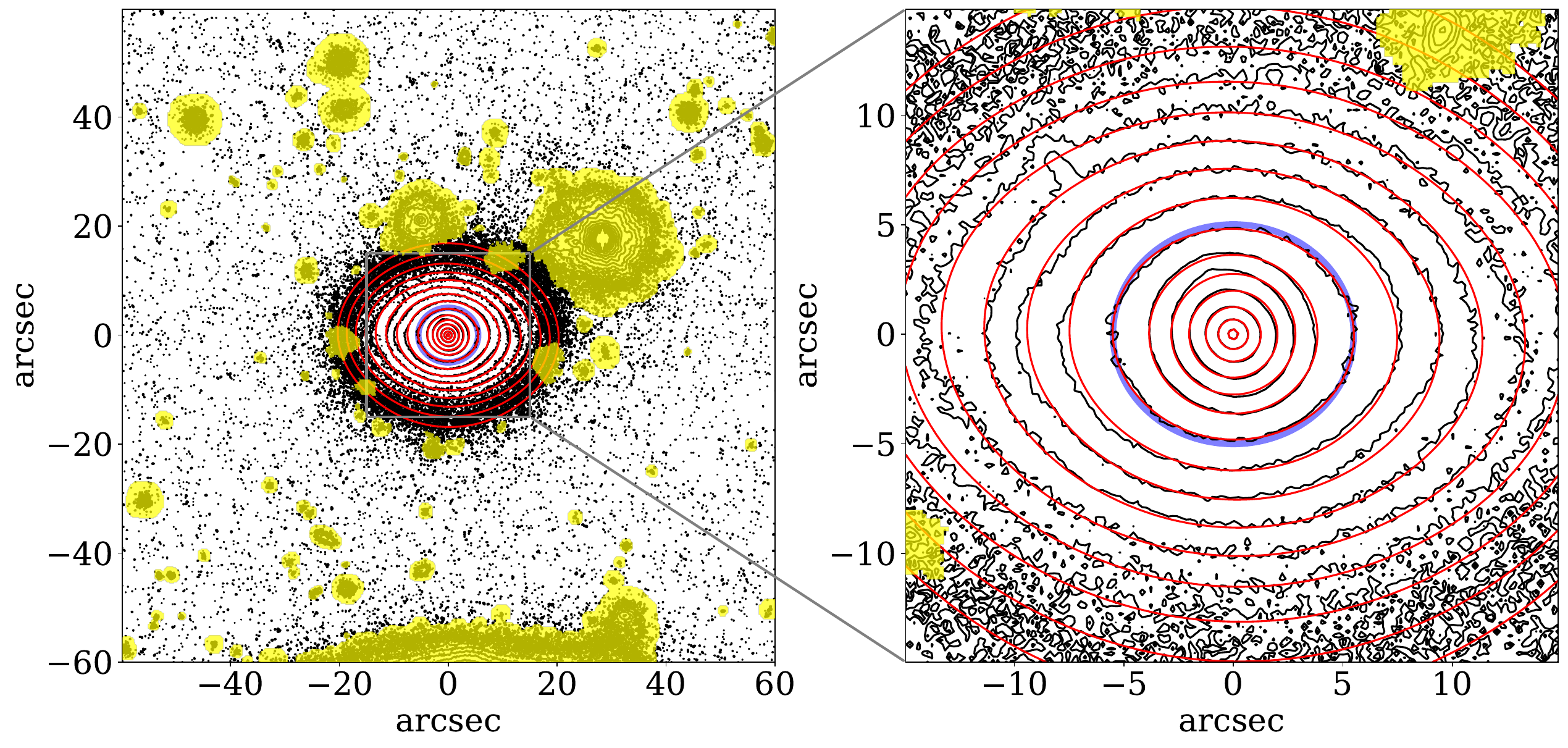}
    \caption[MGE Model for 220394]{MGE fit to the galaxy with CATAID = 220394. KiDS $r$-band imaging contours are shown in black, the MGE model contours are shown in red, and masked regions (other sources) are shown in yellow. The effective radius of the galaxy is shown in blue. The contours are spaced by 0.5 mag arcsec$^{-2}$. The right side panel is a zoomed-in version of the left side panel, with the scale shown by a box around the central region in the left side panel.}
    \label{fig:mge_fit_4}
\end{figure*}

Although DYNAMITE is capable of allowing small isophote twists in the MGE model, we fix the position angles of each Gaussian to be constant. We do this as these twists greatly reduce the available viewing angle parameter space for deprojection into three-dimensional luminosity densities. We show an example MGE fit in Figure \ref{fig:mge_fit_4}.

The MGE model of the projected luminosity is then deprojected into a three-dimensional luminosity density, given some set of viewing angles $(\vartheta,\phi,\psi)$. These viewing angles are fitted for by first converting them into the intrinsic shape parameters $(p_k,q_k,u_k=\sigma_k'/\sigma_k)$, using the relation for ellipsoidal bodies from \cite{1989ApJ...343..617D}, as described in \cite{2002MNRAS.333..400C}:
\begin{align}
    1-q^2&=\frac{\delta'[2\cos2\psi+\sin2\psi(\sec\vartheta\cot\phi-\cos\vartheta\tan\phi)]}{2\sin^2\vartheta[\delta'\cos\psi(\cos\psi+\cot\phi\sec\vartheta\sin\psi)-1]}\\
    p^2-q^2&=\frac{\delta'[2\cos2\psi+\sin2\psi(\cos\vartheta\cot\phi-\sec\vartheta\tan\phi)]}{2\sin^2\vartheta[\delta'\cos\psi(\cos\psi+\cot\phi\sec\vartheta\sin\psi)-1]}
\end{align}
where $\delta'=1-q'^2$, $p_k=B_k/A_k$, and $q_k=C_k/A_k$, with $A_k$, $B_k$, and $C_k$ being the major, medium, and minor axes of the $k^{\text{th}}$ 3D Gaussian component. The luminosity density $\nu$ is then given by:
\begin{equation}
    \nu(x,y,z) = \sum^N_{k=1}\frac{L_k}{(\sigma_k\sqrt{2\pi})^3q_kp_k}\exp\bigg[-\frac{1}{2\sigma_k^2}\bigg(x^2+\frac{y^2}{p_k^2}+\frac{z^2}{q_k^2}\bigg)\bigg]
\end{equation}
where $(x,y,z)$ is a set of coordinates centred on the Gaussians and aligned with the principal axes. Further, the deprojection requires that $q_k\leq p_k \leq 1$, $q_k'\leq q_k$ and $\max(q_k/q_k',p_k)\leq u_k \leq\min(p_i/q_i',1)$. In DYNAMITE, the deprojection is fit for by taking $(p_{\text{min}},q_{\text{min}},u_{\text{min}})$ as the free parameters. Finally, a globally constant mass-to-light ratio $\Upsilon_*$ multiplies the luminosity density to obtain the stellar mass distribution, with $\Upsilon_*$ the $4^{\text{th}}$ free parameter included in our fit. One advantage of using $r$-band KiDS photometry for our MGE model is that it traces the older stellar population that makes up the majority of the stellar mass. It is possible to derive a spatially varying mass-to-light ratio for a Schwarzschild mass model \citep[e.g. for an edge-on disc, ][]{2019MNRAS.487.3776P}, \update{but keeping $\Upsilon_*$ constant has been shown for CALIFA galaxies to not impact the internal dynamics of a Schwarzschild model \citep{2018MNRAS.473.3000Z}, and hence we use a constant value here.}
\subsubsection{Dark Matter Halo and Supermassive Black Hole}
The dark matter halo is taken to be a spherical NFW halo \citep{1996ApJ...462..563N}, with an enclosed mass profile given by:
\begin{equation}
    \text{M}(<r) = \text{M}_{200}[\ln(1+c)-c/(1+c)]^{-1}\bigg[\ln(1+c(r/r_{200}))-\frac{c(r/r_{200})}{1+c(r/r_{200})}\bigg]
\end{equation}
where $c$ is the dark matter halo concentration, and $\rm{M}_{200}$ is the mass within the virial radius $r_{200}$. The virial mass and virial radius are related via the relation $\rm{M}_{200} = \frac{4}{3}\pi200\rho_cr_{200}^3$, with the critical density $\rho_c=1.37\times10^{-7}\rm{M}_\odot\text{pc}^{-3}$. The concentration $c$ has been shown to be poorly constrained by kinematics only extending to $\sim 2\text{R}_e$, such as SAMI \citep[e.g.][]{2014ApJ...792...59Z}. We therefore fix $c$ according to $\rm{M}_{200}$ and the relation of \cite{2014MNRAS.441.3359D}. Finally, we define the dark matter fraction as $f=\rm{M}_{200}/\rm{M}_*$, and take $\log f$ as the sole free parameter to characterise the dark matter halo, and thus dark matter contribution to the gravitational potential.

The black hole contribution is modelled as a Plummer potential \citep{1911MNRAS..71..460P}, which is parametrised by its total mass $\rm{M}_{BH}$, and scale radius $a$. As the spatial resolution of SAMI kinematic maps is coarser than the size of the region where the black hole's mass dominates the potential, we fix $a = 0.001$" and the potential becomes essentially point-like. The black hole is therefore represented solely by $\rm{M}_{BH}$, which we choose to parametrise as $\log(\rm{M}_{BH})$. As this spatial resolution may also result in no meaningful constraint on $\log(\rm{M}_{BH})$, we derive an initial value from the stellar mass scaling relation of \citet{2015ApJ...813...82R}. Combining all components of the potential, we have six free parameters characterising it: $p_{\text{min}},q_{\text{min}},u_{\text{min}},\Upsilon_*,\log f,\text{ and }\log(\rm{M}_{BH})$.
\subsection{Orbit Library}
The construction of a Schwarzschild model requires a library of all allowable stellar orbits within the gravitational potential, a linear combination of which will reconstruct the observed kinematics. Although an MGE potential is not strictly separable (i.e. the Hamilton-Jacobi equation is not always separable) and thus not all orbits are regular, the non-regular orbits are usually negligible \citep{2008MNRAS.385..647V}. Regular orbits conserve three integrals of motion: Energy $E$, $I_2$, and $I_3$ \citep[e.g.][]{1985MNRAS.216..273D,2008gady.book.....B}. There are four families of orbits that exist in this potential: three types of tube orbits (\update{long, short, and intermediate axis}), and box orbits. We initiate our orbits by sampling from the three integrals of motion.

The number of points sampled across the three integrals is $n_E\times n_\theta\times n_R$, where $n_E,n_\theta,n_R$ denote the number of intervals taken across the energy $E$, azimuthal angle $\theta$ and radius $R$ on the $(x,z)$ plane. The $E$ values are sampled logarithmically in radius, with $E$ calculated as the potential at $x=r$. This grid spans $0.5\sigma'_{\rm{min}}$ to $5\sigma'_{\rm{max}}$, where $\sigma'_{\rm{min}}$ and $\sigma'_{\rm{max}}$ are the minimum and maximum observed dispersions of the MGE model. $R$ and $\theta$ are sampled in a linear open polar grid. Each initial condition sampled is then used to seed three orbits: a box orbit, a tube orbit, and a counter-rotating tube orbit. Finally, DYNAMITE allows each orbit to be dithered $n_{\rm{dither}}$ times \citep[as done in e.g.][]{2008MNRAS.385..647V,2018MNRAS.473.3000Z,2022ApJ...930..153S}, where there are $n_{\rm{dither}}^3$ orbits seeded in a grid around each set of initial conditions. This is done to smooth the model, and results in a "bundle" of orbits for each integral value. Overall, the total number of orbits generated in an orbit library is $3\times n_E\times n_\theta\times n_R \times n_{\rm{dither}}^3$.

We calculate two libraries of orbits for each galaxy. For the first library, we add $3\times n_E\times n_\theta\times n_R=3\times21\times18\times9=10\space206$ orbits. This library is used to coarsely sample the parameter space to get initial estimates for the free parameters of the model, and we refer to it as our coarse library. The second library contains $3\times n_E\times n_\theta\times n_R\times 3^3=3\times30\times20\times10\times 3^3=486\space000$ orbits, where we have dithered each orbit 3 times for each integral value. We use this library to fit for and build our final, best-fit model, and refer to it as our "fine" library.
\subsection{Orbit Weighting}
\label{sec:orbit_weighting}
To find the best-fit model, we define a goodness-of-fit, $\chi^2$, and find values for our free parameters ($p_{\text{min}},q_{\text{min}},u_{\text{min}},\Upsilon_*,\log f,\rm{M}_{BH}$) such that $\chi^2$ is minimised. In DYNAMITE, the best-fit model is taken to be that with the minimum {\it kinematic} $\chi^2$:
\begin{align*}
    \label{eq:kin_chi2}
    \chi^2 = \sum^{N_{\text{kin}}}_{n=1}\bigg[&\bigg(\frac{V^n_{\text{mod}}-V^n_{\text{obs}}}{V^n_{\text{obserr}}}\bigg)^2+\bigg(\frac{\sigma^n_{\text{mod}}-\sigma^n_{\text{obs}}}{\sigma^n_{\text{obserr}}}\bigg)^2 + \bigg(\frac{h^n_{3,\text{ mod}}-h^n_{3,\text{ obs}}}{h^n_{3,\text{ obserr}}}\bigg)^2 + \\
    &\bigg(\frac{h^n_{4,\text{ mod}}-h^n_{4,\text{ obs}}}{h^n_{4,\text{ obserr}}}\bigg)^2\bigg]
\end{align*}
\begin{figure*}
    \centering
    \includegraphics[width=\linewidth]{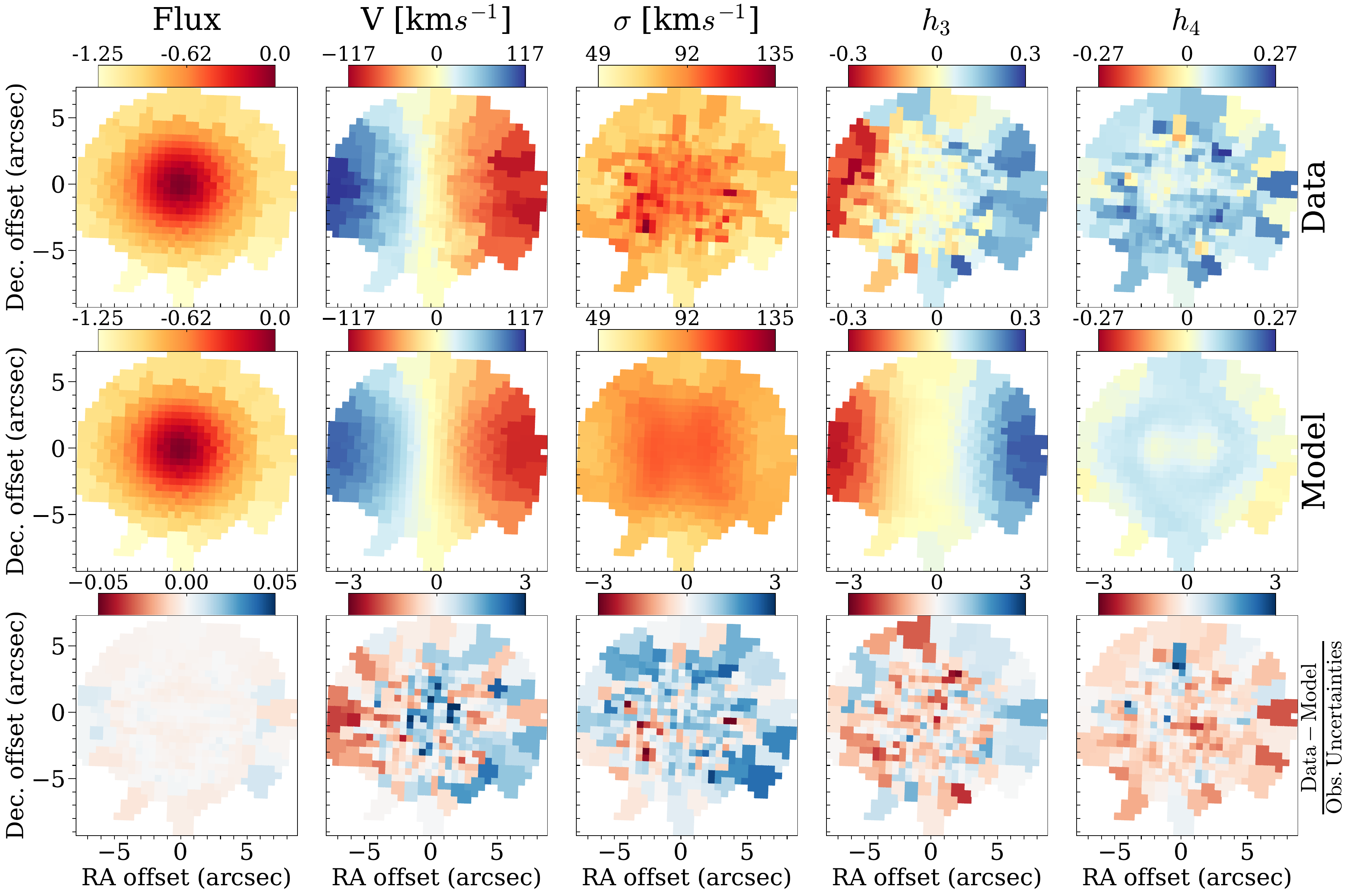}
    \caption[Schwarzschild Model for 220394]{The best-fitting Schwarzschild model for SAMI galaxy with CATID = 220394, and $\chi^2_{\text{red}}=0.94$. This galaxy displays a good fit at high spatial sampling, with a stellar mass of $\logm =10.3$, $\lre = 0.39$, $R_{\rm{max}}/R_{\rm{e}}=1.28$, and $N_{\rm{bins}} = 227$. From top to bottom, we show the input data, the model, and the data minus model divided by observational errors respectively. For the flux column, however, we show data minus model divided by data. We do this as we do not have observational errors on the flux, and it is not fit to in DYNAMITE. Some spaxels approach relative residual values of $\sim10$, or 1000\%, but due to the large number of well-fit spaxels, $\chi^2_{\text{red}}$ remains close to unity and indicates a good fit to the data. From left to right, we show the flux, velocity, velocity dispersion $\sigma$, $h_3$, and $h_4$ respectively.}
    \label{fig:220394_velmaps}
\end{figure*}
\begin{figure*}
    \centering
    \includegraphics[width=\linewidth]{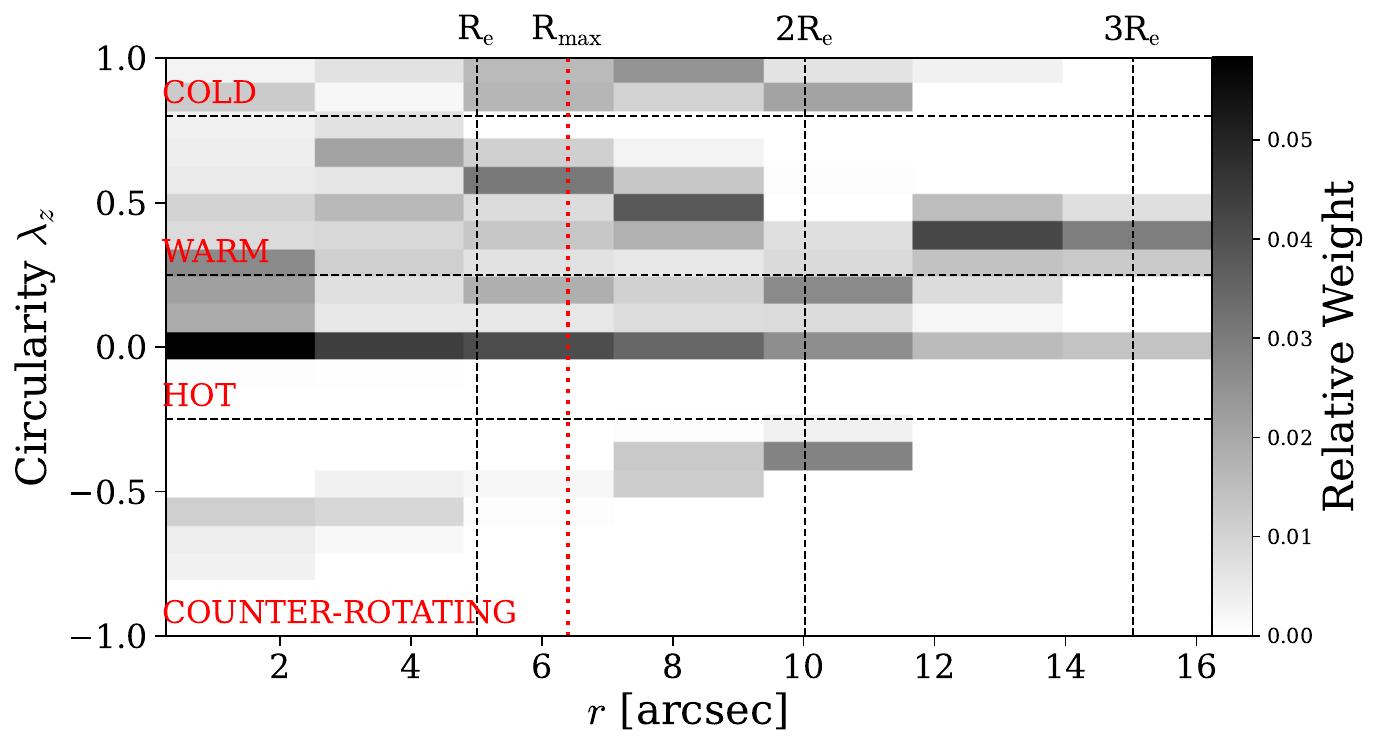}
    \caption[$\lambda_z$ vs radius for Schwarzschild Model of 220394]{The $\lambda_z-r$ phase space distribution of the best fit Schwarzschild model for the galaxy with CATID = 220394. This plot shows the probability density of the orbits used to reconstruct the velocity maps (see Figure \ref{fig:289102_velmaps}). $\lambda_z$ is defined in Equation \ref{eq:circularity}. The divisions between cold, warm, hot, and counter-rotating orbits are shown in horizontal dashed lines, and labelled. \re, 2\re\ and 3\re\ are shown as vertical dashed black lines, and the maximum radius reached by the SAMI kinematics, R$_{\rm{max}}$ is shown in a vertical dotted red line. Within 1\re, the model predicts a cold orbit fraction of 0.02, warm orbit fraction of 0.18, hot orbit fraction of 0.83, and counter-rotating orbit fraction of 0.03.}
    \label{fig:220394_orbitplot}
\end{figure*}
\begin{figure*}
    \centering
    \includegraphics[width=\linewidth]{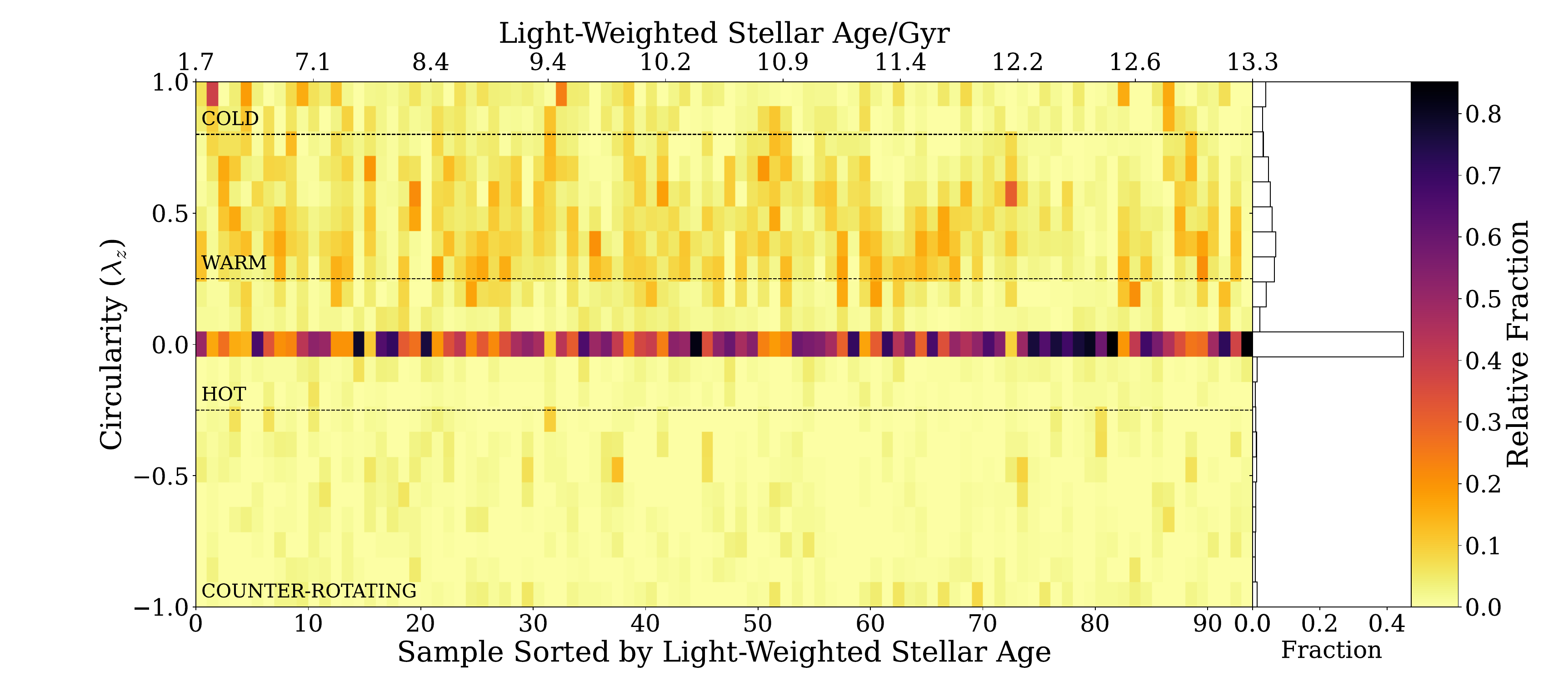}
    \caption[$\lambda_z$ for Every Galaxy Ordered by Stellar Age]{In the left panel, we show the orbit circularity ($\lambda_z$) distribution within one effective radius for each galaxy in our sample, ordered by average light-weighted stellar age within one effective radius. The definitions of cold ($\lambda_z>0.8$), warm ($0.25<\lambda_z<0.8$), hot ($-0.25<\lambda_z<0.25$), and counter-rotating ($\lambda_z<-0.25$) orbits are labelled and separasted by dashed black lines. Darker colours indicate a higher density of orbits, and lighter colours show a lower density. In the right panel, we show the marginalised distribution of $\lambda_z$ for our full sample. \update{Overall, a trend of more cold orbits is seen in the youngest galaxies, and more hot orbits in the oldest galaxies. There is a significant amount of scatter in these relationships, which can be seen more clearly in Figure \ref{fig:correlation_all_4}.}}
    \label{fig:orbital_distributions_all}
\end{figure*}
where $V^n_{\text{mod}}$, $\sigma^n_{\text{mod}}$, $h^n_{3,\text{ mod}}$, $h^n_{4,\text{ mod}}$ are the model velocities for bin $n$ in the velocity maps, $V^n_{\text{obs}}$, $\sigma^n_{\text{obs}}$, $h^n_{3,\text{ obs}}$, $h^n_{4,\text{ obs}}$ are the observed velocities, $V^n_{\text{obserr}}$, $\sigma^n_{\text{obserr}}$, $h^n_{3,\text{ obserr}}$, $h^n_{4,\text{ obserr}}$ are the observational errors and $N_{\text{kin}}$ is the number of bins. Each model velocity is convolved with the same PSF as the observations prior to calculating $\chi^2$. The reduced $\chi^2$, $\chi^2_{\text{red}}$ is defined as $\chi^2_{\text{red}}=\frac{\chi^2}{4N_{\text{kin}}-N_{\text{par}}}$, where $N_{\text{par}}$ is the number of free parameters, with $N_{\text{par}}=6$ in this work. We expect $\chi^2_{\text{red}}$ to approach 1 for the best-fit model. For each model fitting step, we use DYNAMITE's "LegacyWeightSolver" implementation of Lawson and Hanson's non-negative least-squares algorithm \citep{1974slsp.book.....L}. 

We begin with our coarse library of orbits. Similarly to \cite{2024MNRAS.533.1300D}, we use this library to first run a coarse grid search over the structural parameters $p_{\text{min}},q_{\text{min}},u_{\text{min}}$, with minimum steps of 0.03, 0.03, and 0.01 respectively. \update{The minimum and maximum values for $p_{\text{min}}$ and $u_{\text{min}}$ were set to 0.01 and 0.99. Although these can range between 0 and 1 in principle, numerical errors can arise at these values and thus we constrain our parameter search to just within them. $u_{\text{min}}$ is similarly constrained from below, but must also be flatter than the flattest observed Gaussian component, such that $0.01<u_{\text{min}}<\sigma'_{\text{min}}$.} We use DYNAMITE's "FullGrid" parameter search setting, which creates a Cartesian grid for all free parameters, and searches through it with the minimum step size. We then use the coarse library of orbits again, to run another coarse grid search over the parameters $\log f,\Upsilon_*$, with minimum steps of 0.01 each. We again run this search with the "FullGrid" parameter search setting.

For the final step, we use our fine library of orbits, and run a grid search over the full parameter space. For $p_{\text{min}},q_{\text{min}},u_{\text{min}},\Upsilon_*$, and $\log f$ we use a minimum step size of 0.01, whereas for $\log(\rm{M}_{BH})$ we use 0.1. We limited $\Upsilon_*$ between 0.1 and 10, $\log f$ between -3 and 3, and $\log(\rm{M}_{BH})$ between 1 and 10. We use the best-fit values from the first two coarse grid searches as the starting values for each parameter, and the relation from \cite{2015ApJ...813...82R} for black hole masses from stellar mass for the starting value of $\rm{M}_{BH}$. We use the "LegacyGridSearch" search setting, which takes smaller and smaller steps towards the best-fit solution.
\section{Results}
\label{sec:results_4}
\subsection{Schwarzschild Models}

We fit Schwarzschild orbit-superposition models to the 101 galaxies in our sample. For each galaxy, we explored up to 2000 models in the coarse parameter searches, and 2000 models in the final grid search. This is comparable to previous works with large sample sizes (e.g. 1000-2000 models for \citealt{2019MNRAS.486.4753J,2020MNRAS.491.1690J}, 1250 for \citealt{2022ApJ...930..153S}), and lower than previous works with smaller sample sizes (e.g. up to 8000 models for \citealt{2024MNRAS.533.1300D}). From each model we extract the velocity moment maps $V$, $\sigma$, $h_3$, and $h_4$. 

We present the velocity moment maps of an example galaxy with high spatial sampling and CATID 220394 in Figure \ref{fig:220394_velmaps}. This galaxy has 227 spatial bins within 1\re, and is well fit with $\chi^2_{\text{red}}=0.94$. \update{We show 3 further galaxies, representative of our sample, in Figures \ref{fig:511892_velmaps}-\ref{fig:238922_velmaps} of Section \ref{sec:extra_gals}. We additionally show and test the sampling of our parameter space in Section \ref{sec:parameter_search}.}

For Figure \ref{fig:220394_velmaps} (and Figures \ref{fig:511892_velmaps}-\ref{fig:238922_velmaps}), the first row shows the observed flux, $V$, $\sigma$, $h_3$, and $h_4$ from left to right. The middle row shows the same, but for the model. The bottom row shows the relative residual, where we subtract the model from the data and divide by the observational uncertainties. 

\update{We present a comparison of the models derived in this work to those found by \cite{2022ApJ...930..153S} in Section \ref{sec:model_comparison}.}

\subsection{Orbit Fractions}
The Schwarzschild model of a galaxy additionally gives information about its intrinsic angular momentum, rather than the projected quantities derived from observations. One useful quantity that we calculate for each orbit in a galaxy model is the circularity parameter, $\lambda_z$ \citep{2003ApJ...597...21A,2018MNRAS.473.3000Z}:
\begin{equation}
    \label{eq:circularity}
    \lambda_z=\frac{\overline{L_z}}{\overline{r}\cdot\overline{V_c}}
\end{equation}
where:
\begin{align*}
    \overline{L_z}&=\overline{xV_y-yV_x}\\
    \overline{r}&=\sqrt{\overline{x^2+y^2+z^2}}\\
    \overline{V_c}&=\sqrt{\overline{V_x^2+V_y^2+V_z^2+2V_xV_y+2V_xV_z+2V_yV_z}}
\end{align*}
The bar notation denotes that these values $(x,y,z,V_x,V_y,V_z)$ are averaged over time for the entire orbit's path. $\overline{V_c}$ is equivalent to the angular momentum of a purely circular orbit at $\overline{r}$, which is the maximum possible value of $\overline{L_z}$, and hence $\lambda_z$ varies between -1 and 1. $\lambda_z$ can be considered an intrinsic equivalent of \lre, as it separates rotation and dispersion supported orbits, with hot/box orbits having $\lambda_z\lessapprox0.3$, and cold/circular orbits having $\lambda_z\gtrapprox0.5$.

We show the $\lambda_z-r$ distribution for SAMI Galaxy with CATID = 220394 in Figure \ref{fig:220394_orbitplot}. \cite{2018MNRAS.473.3000Z} defined four classes of orbits within this distribution according to an orbit's $\lambda_z$: hot orbits $(-0.25\leq\lambda_z\leq0.25)$, warm orbits $(0.25\leq\lambda_z\leq0.8)$, cold orbits $(\lambda_z\geq 0.8)$, and counter-rotating orbits $(\lambda_z\leq-0.25)$. We show these classes in Figure \ref{fig:220394_orbitplot}, as well as the maximum radial extent of the SAMI kinematics, and the effective radius. 

We show the orbital distribution within 1\re\ for our full sample, ordered by stellar age, in Figure \ref{fig:orbital_distributions_all}. All stellar orbits are integrated over radii within 1\re, and their relative weights are plotted in bins of $\lambda_z$. The $x$-axis then orders the galaxies by average light-weighted stellar age within 1\re, as this has been shown to be the parameter that correlates best with \lre, and hence the angular momentum of each star as traced by $\lambda_z$. We divide $\lambda_z$ into cold, warm, hot, and counter-rotating, and show and label these groups accordingly. \update{The overall trend is that younger galaxies tend to contain higher fractions of cold orbits, whereas older galaxies tend to more hot orbits. There is, however, considerable scatter in this relationship.}

\begin{figure*}
    \centering
    \includegraphics[width=\linewidth]{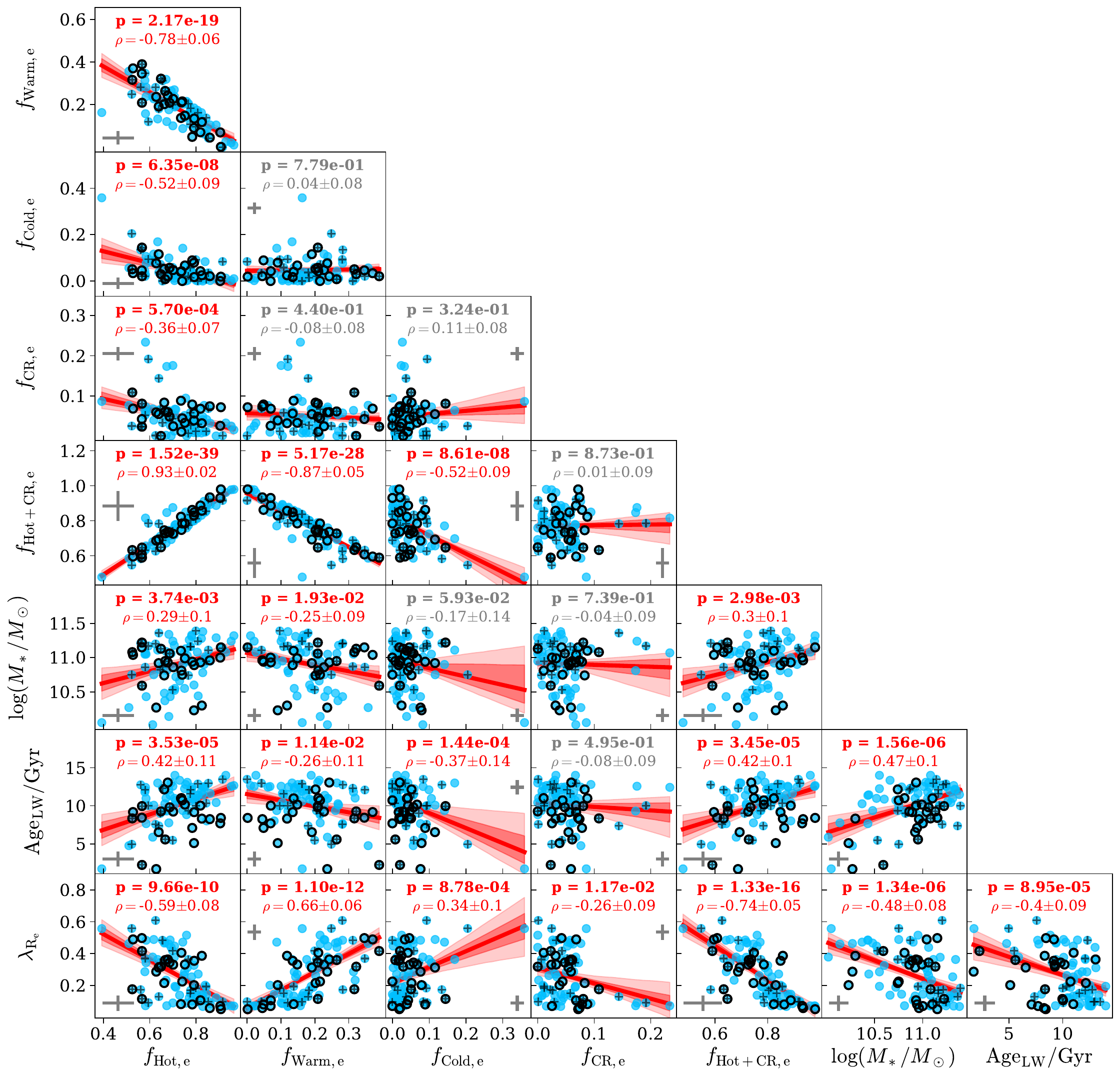}
    \caption[Correlations Between Orbital Fractions and Global Properties]{Correlations between orbital fractions and global galaxy properties for our sample. \hot, \warm, \cold, and \counter\ are the hot orbit fraction, warm orbit fraction, cold orbit fraction, and counter-rotating orbit fraction respectively, all within 1\re. \logm, $\rm{Age}_{\rm{LW}}$/Gyr, and \lre\ are stellar mass, average light-weighted stellar age within 1\re, and the spin proxy within 1\re, respectively.  \hotcr\ is the combined fraction of hot and counter-rotating orbits. Galaxies classified as having a tidal shell by \cite{2024MNRAS.529..810R} are circled in black, \update{and those with a stream are marked with a black cross}. \update{Additionally, the typical uncertainty of of each parameter is represented by a grey cross in each panel.} The best-fitting linear relationship for each correlation is shown in red, with the 68\% and 95\% confidence intervals, derived using bootstrap resampling \update{of the galaxies}, shown in lighter red shading. \update{The Pearson correlation coefficient $\rho$, its bootstrapped uncertainty, and its p-value $p$ are given for each panel. These are highlighted in red when $p<0.05$, indicating a statistically significant correlation.}}
    \label{fig:correlation_all_4}
\end{figure*}
We derive orbit fractions (cold, warm, hot, counter-rotating) within 1\re\ for each galaxy in our sample. \update{We note that uncertainties for orbit fractions are not returned by DYNAMITE, as a full Monte Carlo treatment is required to derive them. For example, \cite{2022ApJ...930..153S} derived orbit fraction uncertainties in their Appendix C for Schwarzschild models of SAMI galaxies. For a subset of their galaxies, they perturbed their best-fit model by adding Gaussian noise with standard deviation equal to the mean error of each observed kinematic moment ($V$, $\sigma$, $h_3$, $h_4$). From 50 realisations per galaxy, re-deriving Schwarzschild models for each, they found that the orbit fractions showed a typical uncertainty of $10\%-15\%$. Alternatively, \cite{2018NatAs...2..233Z} compared the intrinsic orbital distribution of simulated CALIFA-like galaxies to the derived distribution from Schwarzschild models. They found that the combination of statistical errors, systematic biases, and systemic errors resulted in a typical uncertainty of $\sim 20\%$ in orbit fractions. Given that we use the same kinematic data as \cite{2022ApJ...930..153S}, we adopt a constant $10\%$ uncertainty in each orbital fraction.}

We show the correlations between these fractions and other global galaxy properties in Figure \ref{fig:correlation_all_4}. This figure presents the hot, warm, cold, and counter-rotating fractions within 1\re\, labelled as \hot, \warm, \cold, and \counter\ respectively. We additionally show the stellar mass, \logm, the mean light-weighted stellar age within 1\re, $\rm{Age}_{\rm{LW}}$/Gyr, the spin parameter proxy, \lre, and the sum of hot and counter-rotating orbit fractions, \hotcr. Galaxies with tidal shells, as identified in \cite{2024MNRAS.529..810R}, are circled in black, and those with tidal streams are marked with a black cross. \update{Additionally, the typical uncertainty of of each parameter is represented by a grey cross in each panel. The uncertainty in stellar mass is taken to be 0.1 dex \citep{2011MNRAS.418.1587T}, the uncertainty in \lre\ is taken as 10.9\% \citep{2021MNRAS.505.3078V}, and the uncertainty in light-weighted stellar age is estimated as 0.97 Gyr \citep{2022MNRAS.516.2971V}.} A best-fitting linear relation is shown for each panel in red, with $68\%$ and $95\%$ confidence intervals, derived using bootstrap resampling \update{of the galaxies}, shown in lighter red shading. \update{The bootstrap resampling was performed by redrawing our sample 1000 times with replacement, and calculating the linear relation for each realisation. This method was used for all bootstrap resampling in this work. The Pearson correlation coefficient $\rho$, its bootstrapped uncertainty, and its p-value $p$ are given for each panel. These are highlighted in red when $p<0.05$, indicating a statistically significant correlation.} 

There are strong correlations between each respective orbit fraction, and we note that these are partially driven by the fact that all orbit fractions must sum to one for each galaxy. \lre\ shows a significant correlation with each orbit fraction, as expected since it traces the balance between rotation and dispersion set by the stellar orbits. Among these relations, \lre\ shows the strongest and tightest correlation with the combined hot plus counter-rotating fraction, \hotcr,  followed by the warm fraction, \warm, and then the hot fraction, \hot. The quantity \hotcr\ is mathematically equivalent to $1-f_{\rm{Cold+Warm, e}}$, reflecting the requirement for all fractions to sum to unity. The light-weighted stellar age, $\rm{Age}_{\rm{LW}}$, correlates most strongly with \logm, followed by, at roughly equal significance, \hotcr, \hot, \lre and \cold. While \warm\ and \hot\ both show a strong positive and negative correlation with \lre respectively, only \hot\ shows a strong correlation with $\rm{Age}_{\rm{LW}}$. The $\rm{Age}_{\rm{LW}}$-\logm\ and $\rm{Age}_{\rm{LW}}$/Gyr-\lre\ trends have previously been shown for the full SAMI sample by \cite{2024MNRAS.529.3446C}.

\subsection{Correlation Analysis}
\begin{figure*}
    \centering
    \includegraphics[width=\linewidth]{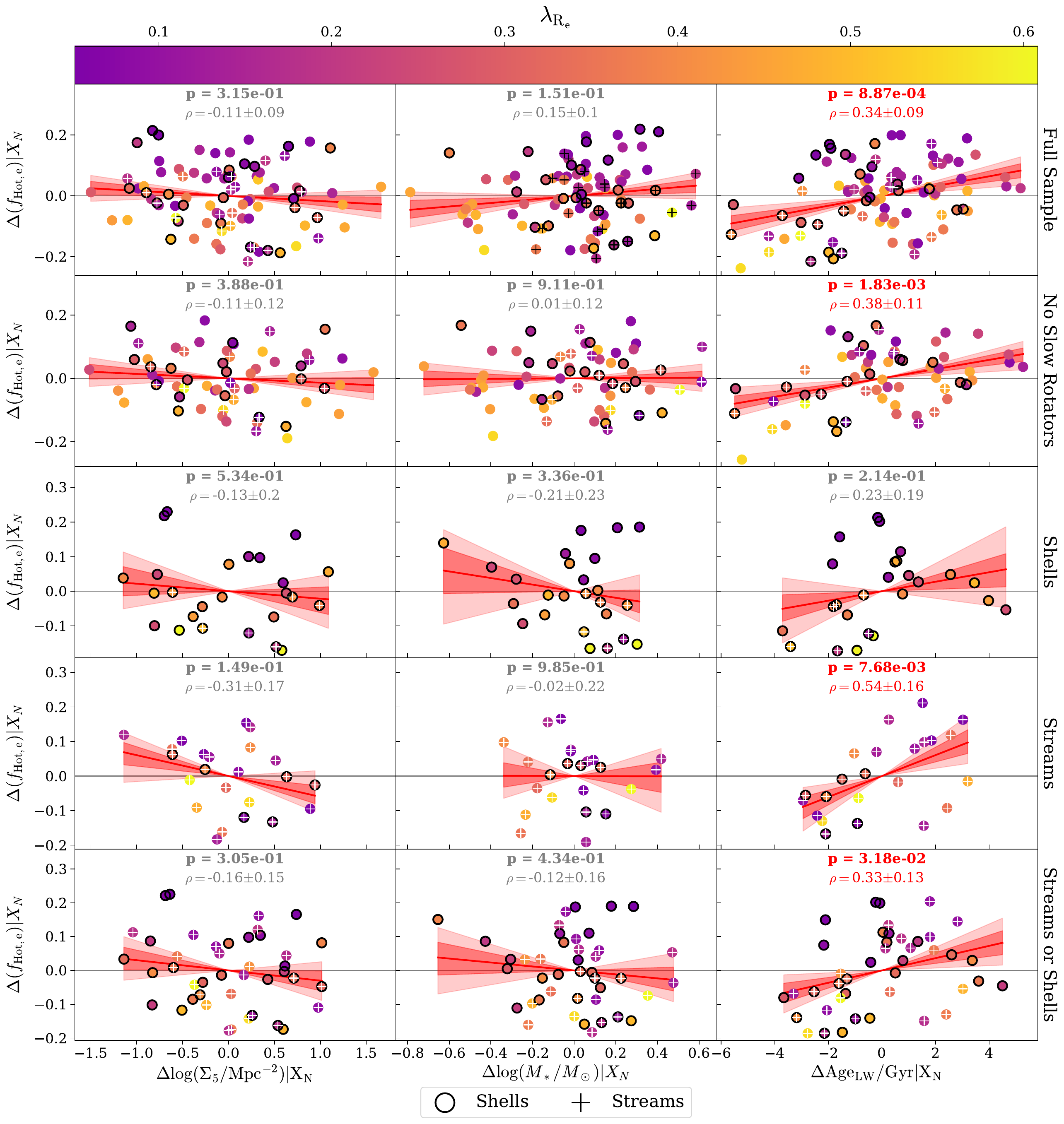}
    \caption[Partial Correlations Between Hot Orbits and Global Properties]{Correlations between the hot orbital fraction \hot, and \age, \env, and \logm. For each correlation, the linear correlations of the other variables is controlled for. The top row displays the full sample, and the second row shows the full sample excluding slow rotators (as defined for the SAMI sample in \citealt{2021MNRAS.505.3078V}). Subsequent rows display galaxies with shells, galaxies with streams, and galaxies with streams or shells. Galaxies are coloured by \lre. Galaxies with tidal shells are circled in black, and galaxies with tidal streams display a white cross. The best-fitting linear relation for each correlation is shown in red, with 68\% and 95\% confidence intervals shaded. \update{The Pearson correlation coefficient $\rho$, its bootstrapped uncertainty, and its p-value $p$ are given for each panel. These are highlighted in red when $p<0.05$, indicating a statistically significant correlation.}}
    \label{fig:partial_correlation_hot_basic}
\end{figure*}
In Figure \ref{fig:partial_correlation_hot_basic}, we present correlations between \hot\ and average light-weighted stellar age (\age), environment (\env), and stellar mass (\logm). We choose to begin this analysis with \hot\ as it shows a strong correlation with \age\ in Figure \ref{fig:correlation_all_4}. We perform this analysis for several samples: the full sample, the full sample without slow rotators (as defined for SAMI in \citealt{2021MNRAS.505.3078V}), galaxies with shells, galaxies with streams, and galaxies with shells or streams. For each correlation, we account for the linear correlation of the other variables. For example, the correlation between \hot\ and \age is shown after controlling for the correlations with \env\ and \logm. This is known as a partial correlation \citep[e.g.][]{1979ats..book.....K}. Galaxies are coloured by \lre, with tidal shells (black circles) and streams (white crosses) indicated too. For each panel, the best-fitting linear relation is shown in red, with $68\%$ and $95\%$ confidence intervals, derived using bootstrap resampling \update {of the galaxies}, shaded in lighter red. \update{The Pearson correlation coefficient $\rho$, its bootstrapped uncertainty, and its p-value $p$ are given for each panel. These are highlighted in red when $p<0.05$, indicating a statistically significant correlation. We acknowledge that a cut of significance at $p=0.05$ is somewhat arbitrary. However, it corresponds to a $\sim2\sigma$ result, and we are more interested in which orbits correlate with age and less interested in the exact strength of the correlation, so we choose to leave it as our point of significance.}

The choice of parameters for this partial correlation analysis is inspired by \cite{2024MNRAS.529.3446C}, who used partial correlations between \age, \logm, \env, and \lre\ for the full SAMI sample to determine which parameter \lre\ is most strongly correlated with. They found that \lre\ only showed a significant partial correlation with \age. We aim to determine which orbits are driving this relationship. We note here that we are not examining the full SAMI sample, and our sample is biased, only considering massive ($\logm>10$) early-type galaxies with sufficient radial kinematic coverage to build an orbit-superposition dynamical model. Figure \ref{fig:partial_correlation_hot_basic} shows that \age\ and \hot\ are positively correlated even once mass and environment are accounted for, i.e. galaxies with a large proportion of hot orbits within 1\re\ have older mean stellar ages within 1\re. This correlation remains when slow rotators are removed, and when only galaxies with tidal streams are considered. Galaxies with tidal shells do not show this correlation, however. Additionally, there is no partial correlation between \hot\ and \logm\ or \env\ in any sample. Given that hot orbits are dispersion supported and drive low \lre\ values, and we see that \hot\ is more correlated with \age\ than \logm\ or \env\ in our full sample, this result is in agreement with the findings of \cite{2024MNRAS.529.3446C}.

\begin{figure*}
    \centering
    \includegraphics[width=\linewidth]{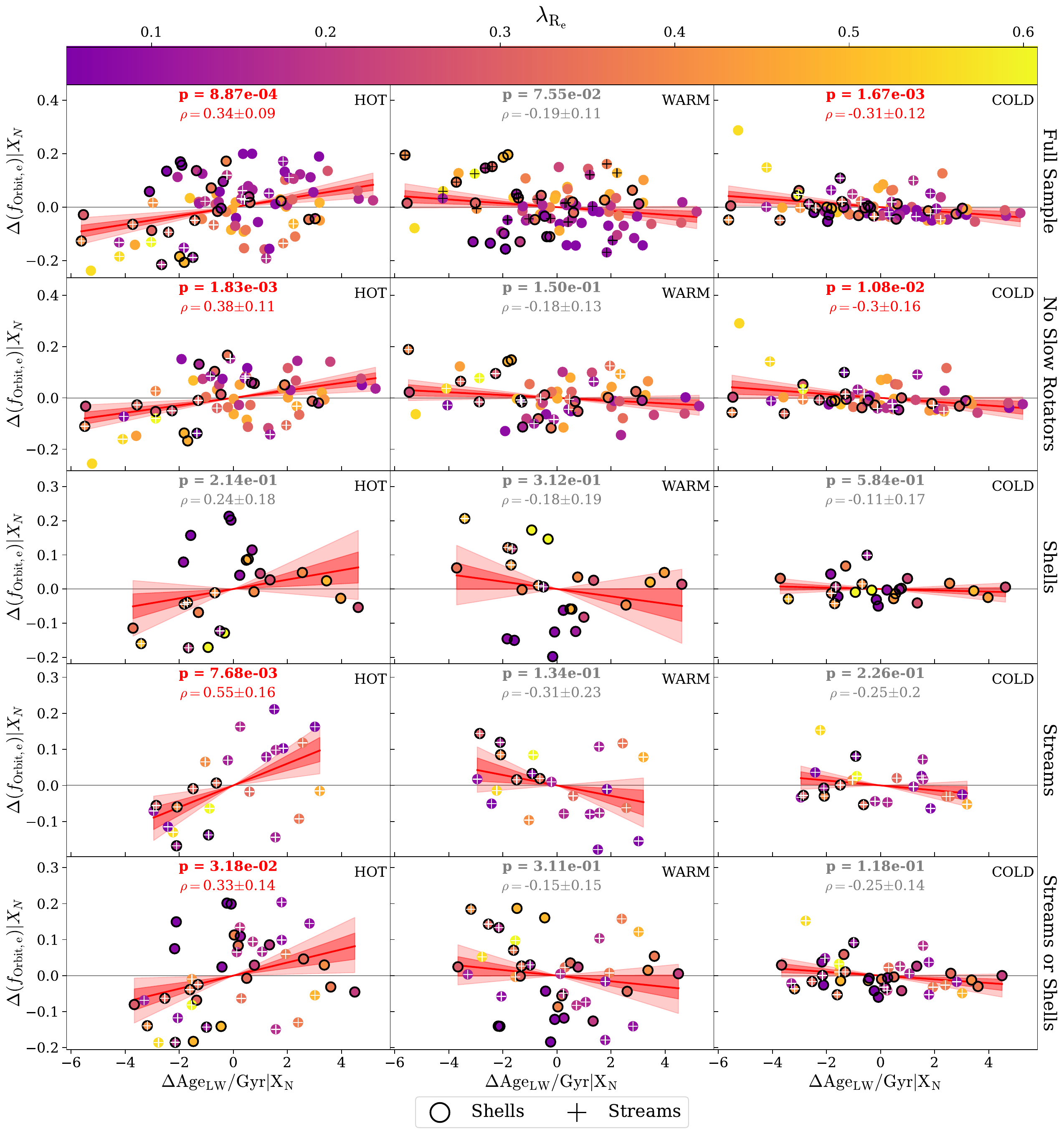}
    \caption[Partial Correlations Between All Orbits and Global Properties]{Correlations between the hot, warm, and cold orbital fractions, and \age. For each correlation, the linear correlations with \logm\ and \env\ are controlled for. The points are displayed and rows are ordered as in Figure \ref{fig:partial_correlation_hot_basic}. The left, middle, and right columns show the partial correlations for \hot, \warm, and \cold\ respectively. \update{The Pearson correlation coefficient $\rho$, its bootstrapped uncertainty, and its p-value $p$ are given for each panel. These are highlighted in red when $p<0.05$, indicating a statistically significant correlation.}}
    \label{fig:partial_correlation_all}
\end{figure*}

In Figure \ref{fig:partial_correlation_all}, we extend the analysis to show how \age\ correlates with the orbital components \hot, \warm, and \cold, again controlling for \logm\ and \env\ as in Figure \ref{fig:partial_correlation_hot_basic}. \update{We choose not to show the partial correlations between orbit fractions and \logm\ or \env\ in this figure, as \age\ seems to be the primary driver of kinematics. However, the effects of \logm\ and \env\ are still controlled for. Additionally, we do not show partial correlations with the counter-rotating fraction, as this shows no significant results.} The left, middle and right columns show the partial correlations of \age\ with \hot, \warm, and \cold\ respectively. The rows follow the same layout as in Figure \ref{fig:partial_correlation_hot_basic}, showing the full sample, the full sample excluding slow rotators, galaxies with shells, galaxies with streams, and galaxies with shells or streams. \update{We note that partial correlations with significant p-values additionally have significant Pearson correlation coefficients. Due to our assumed 10\% uncertainty in each orbital fraction for each galaxy \citep[as found by][]{2022ApJ...930..153S}, we perform an additional test to confirm the significance of our correlations. When drawing galaxies for our bootstrap resampling, we perturb each orbital fraction by adding Gaussian noise with standard deviation equal to 10\% of our derived value. We then perform the same Pearson test, and find that each partial correlation that displays significance in Figure \ref{fig:partial_correlation_all} maintains significance. We acknowledge that this is not as thorough as a full MCMC derivation of uncertainties for every galaxy's orbital fraction, but this would be very computationally expensive, and would not impact our result.}

\update{Figure \ref{fig:partial_correlation_all}} shows that hot and cold orbits are correlated with \age, but warm orbits are not. Given that the relative orbital fractions set the observed $V$, $\sigma$, and thus \lre\ \citep[e.g.][]{2008gady.book.....B,2008MNRAS.385..647V,2016ARA&A..54..597C}, the \age–\lre\ relation can, in principle, be re-stated as correlations between \age\ and the orbital fractions. Figure \ref{fig:partial_correlation_all} shows that the \age-\lre\ relationship is driven by hot and cold orbits, with warm orbits not playing a significant role. Additionally, given the correlations remain when we remove slow rotators, this result is not solely a result of the slow/fast rotator dichotomy. \update{We again note that this is a conclusion based on which correlations are significant, and not a reflection of the strength of each correlation.}

\update{The lack of significant correlation between hot orbits and age for galaxies with shells is likely due to both low number statistics and the observed trends amongst shell galaxies, namely lower \lre and significantly lower mean light-weighted stellar ages \citep[e.g.][]{2024MNRAS.529..810R,2024MNRAS.529.3446C}. A possible scenario to explain the lower stellar ages in shell galaxies comes from the work by \cite{2018MNRAS.473.4956L}, who showed that in order for a galaxy to become a slow rotator, it needs to quench before merging. Given that mean stellar age should approximately increase with time since quenching, it is natural that slow rotators that quenched recently are more likely to have merged recently, resulting in a population of galaxies with lower stellar ages and still-visible tidal shells. In this second scenario, it is further unsurprising that there is a large population of old, low-spin galaxies with no visible tidal features \citep{2024MNRAS.529.3446C}, as they formed through quenching and merging in the early Universe and their shells have since faded. Overall, the mergers identified through tidal shells represent the subset of galaxies that have both quenched and merged recently, and therefore deviate from the broader trend in which hot orbit fraction increases and \lre\ decreases with stellar age.} 

\update{We note that galaxies which have experienced a merger sufficiently recently could challenge the steady-state assumption implicit in Schwarzschild models. This could lead to, for example, an artificially enhanced orbit fraction in galaxies that have experienced a very recent radial merger. As our sample is drawn from \cite{2024MNRAS.529..810R} who excluded ongoing mergers, we are confident that the vast majority of galaxies in our sample have achieved sufficient phase-mixing post merger and can be reliably modelled.}

\subsection{Mergers and Orbit Fractions}

To analyse the impact of mergers on orbital distributions and galaxy shapes, we begin by calculating the triaxiality at 1\re, $\rm{T}_{\rm{R_{\rm{e}}}}$, defined as:
\begin{equation}
    \rm{T}_{\rm{R}_{\rm{e}}} = \frac{1-p_{\rm{R}_{\rm{e}}}^2}{1-q_{\rm{R}_{\rm{e}}}^2}
\end{equation}
where $p_{\rm{R}_{\rm{e}}}$ and $q_{\rm{R}_{\rm{e}}}$ are the intrinsic shape parameters at 1\re.

\begin{figure*}
    \centering
    \includegraphics[width=\linewidth]{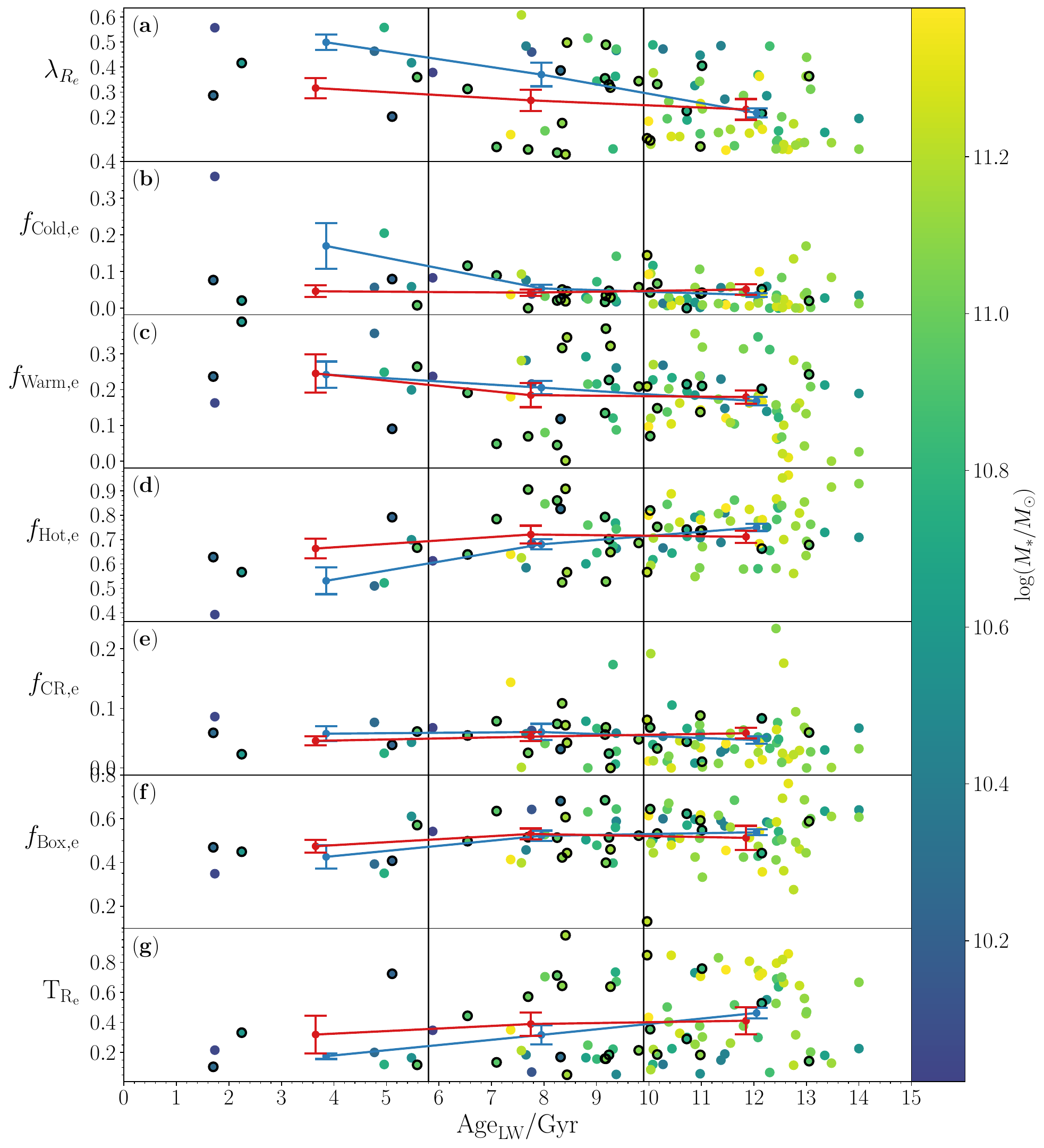}
    \caption[SAMI Galaxies in Light-Weighted Age versus Orbital and Shape Parameters]{The distribution of our sample in \age\ versus various orbital and shape parameters. Galaxies are coloured by \logm, and circled in black if they have a tidal shell. \update{We only show shells, as there are no correlations for streams.} Each panel is divided into three equally spaced age bins, denoted by vertical black lines. In panel (a), galaxies are represented by their mean light-weighted stellar age within 1\re\, \age, against the spin parameter proxy \lre. Panels (b)-(g) are displayed by \age\ versus \cold, \warm, \hot, \counter, \boxorbit, and $\rm{T}_{\rm{R_{\rm{e}}}}$, respectively. \update{\boxorbit\ is the fraction of box orbits within 1\re.} In each age bin, the average value of the y-axis parameter for galaxies without shells is shown in blue, and for galaxies with shells in red.}
    \label{fig:age_spin_shell}
\end{figure*}

In panel (a) of Figure \ref{fig:age_spin_shell}, we show the distribution of galaxies in \age\ versus \lre, coloured by \logm. Galaxies with shells identified are circled in black. \update{We only show shells in black, as there are no correlations for streams.} The panel is further divided into three equally spaced age bins, denoted by black vertical lines\update{, similarly to  \cite{2024MNRAS.529..810R}}. In each age bin, the average \lre\ value for galaxies without shells is shown in blue, and for galaxies with shells in red. \update{Uncertainties are the uncertainties of the mean within each age bin.} Panels (b)-(f) are constructed analogously to panel (a), but they show our sample in \age\ versus \cold, \warm, \hot, \counter, \boxorbit\ and $\rm{T}_{\rm{R_{\rm{e}}}}$, respectively. \update{\boxorbit\ represents the fraction of box orbits within 1\re. These orbits are not defined on the $\lambda_z-r$ plane, but are rather a family of orbit sampled by DYNAMITE.}

We find that, consistent with \cite{2024MNRAS.529..810R}, young and intermediate-age galaxies (the lower two age bins) show significantly lower \lre\ values when they have a shell feature compared with galaxies that do not, at $3.6\sigma$ and $1.6\sigma$ significance for the young and intermediate bins, respectively. We do note, however, that there are only eight galaxies in the lowest age bin. Although \cite{2024MNRAS.529..810R} did not show this trend in the lowest age bin, the sample in this work is different, and the broad trend remains that younger galaxies with shells display lower values of \lre\ than those without shells. We find that this relationship is primarily driven by changes in the hot and cold orbital fractions (panels (b) and (d)), where the youngest galaxies with shells exhibit moderately higher hot orbit fractions and lower cold orbit fractions than young galaxies without shells, both with a 1.9$\sigma$ significance. Warm and counter-rotating orbits, shown in panels (c) and (e), do not have any significant difference in their distributions for galaxies with and without shells, regardless of age. There is a weak trend towards the youngest galaxies with shells being more triaxial, but the difference is not statistically significant. \update{There is no significant correlation of box orbits with shells.} Finally, the result seen in Figure \ref{fig:partial_correlation_all} that galaxies without shells increase their hot orbits with age and galaxies with shells do not, can be clearly seen in panel (d) of Figure \ref{fig:age_spin_shell}. \update{We note that in the lowest age bin there are only four galaxies with shells and four without, but the mean hot orbit fraction still differs significantly ($1.9\sigma$).}

\section{Discussion}
\label{sec:discussion_4}

\subsection{Spin Transformation is Driven by Hot and Cold Orbits}

We have constructed orbit-superposition dynamical models of a subsample of the SAMI galaxy survey to investigate the transformation of galaxy kinematics in massive galaxies as a function of both age and merger history. By deriving the circularity parameter $\lambda_z$ for the orbits in our models, we find that cold ($\lambda_z>0.8$) and warm ($0.25<\lambda_z<0.8$) orbit fractions correlate positively with \lre, while hot ($-0.25<\lambda_z<0.25$) and counter-rotating ($\lambda_z<-0.25$) orbit fractions correlate negatively with \lre. Correlations between orbital fractions and \lre\ are expected, as \lre\ traces the dynamical contribution of different orbits. While \cite{2022ApJ...930..153S} demonstrated the same trends for SAMI passive galaxies, warm orbits are intermediate by definition, and could trace \lre\ differently in rotation-dominated late-type discs. A major advantage of an orbit fraction driven analysis is that it provides a means to disentangle the orbital structure of galaxies, where mergers and secular evolution are expected to leave different imprints in the spin transformation of galaxies \citep[e.g.][]{2006ApJ...645..209D,2010ApJ...721.1878S,2020MNRAS.498..940M}. We note here, with care, that while the contributions of different orbital families and \lre\ evolve over time in a galaxy, stellar angular momentum is always conserved.

The question of what drives spin in galaxies has been investigated in many previous studies. Motivated by the morphology-density relation of \cite{1980ApJ...236..351D}, the role of environment was first identified as a key factor correlated with \lre\ by \cite{ATLAS3DVII}, with further studies supporting this trend \citep[e.g.][]{2013MNRAS.429.1258D, 2013MNRAS.436...19H,2014MNRAS.441..274S,2014MNRAS.443..485F}. Stellar mass has also been proposed as a primary driver of galaxy spin \citep[e.g.][]{2017ApJ...844...59B,2017MNRAS.471.1428V,2017ApJ...851L..33G}, although there is evidence from more recent works that mass and environment are independent influences on \lre\ \citep[e.g.][]{2017MNRAS.464.3850L,2019arXiv191005139G,2021ApJ...918...84R,2021MNRAS.508.2307V}. Further complicating the picture, stellar age has been shown to strongly correlate with stellar kinematics \citep{2018NatAs...2..483V}, with \cite{2024MNRAS.529.3446C} finding that stellar age is more strongly correlated with \lre\ than either mass or environment within the SAMI sample.

We take advantage of our derived orbit fractions to further investigate the relationships between \lre, \age, \env, and \logm. Using partial correlations, we find that the fraction of hot orbits increases with stellar age ($p=8.87\times10^{-4}$), and the fraction of cold orbits decreases with stellar age ($p=1.67\times10^{-3}$), independent of mass and environment. In contrast, the warm orbit fraction shows no significant correlation with age ($p=7.55\times10^{-2}$) once mass and environment are controlled for. These results are broadly consistent with the \lre-stellar age relation reported by \cite{2024MNRAS.529.3446C}, as the relative fractions of orbits determine the observed $V$, $\sigma$, and \lre\ \citep[e.g.][]{2008gady.book.....B,2008MNRAS.385..647V,2016ARA&A..54..597C}. However, the lack of any significant trend between warm orbits and age suggests that the transformation of \lre\ with age occurs with stars transitioning directly from cold to hot orbits. Secular processes such as bar-driven instabilities \citep[e.g.][]{2005AIPC..804..333A,2016MNRAS.462.1697A} and disc/spiral arm torques \citep[e.g.][]{2014RvMP...86....1S,2021MNRAS.503.5826A,2024ApJS..271....1Y}, as well as environmental/group effects \citep[e.g.][]{2011MNRAS.415.1783B,2017ApJ...837...68C} are capable of redistributing angular momentum in galaxies, however in these scenarios the orbital transformation is expected to happen smoothly, with stars heated progressively from cold, to warm, and finally to hot orbits \citep{2006ApJ...645..209D,2010ApJ...721.1878S}. The absence of any significant warm orbit correlation with age implies that these pathways do not dominate the spin evolution of SAMI galaxies in our sample. Rather, the transition from cold to hot orbits appears to happen more stochastically. \update{We do not suggest that warm orbits do not contribute substantially to the total orbital distribution, but rather that their fraction stays consistent across stellar age, while hot and cold orbits do not.}  We note that warm orbits are represented in thick discs, spheroids and bars \citep{2018MNRAS.479..945Z}, \update{and the warm orbits contained within any of these components could show a trend with the age of the component that is not apparent when the galaxy is viewed as a whole. For example, \cite{2026MNRAS.547ag471J}  showed that weak bars in the SAMI sample display higher \lre\ and lower \age\ as compared to strong bars, in which bar strength could be strongly tied to intrinsic orbital distributions. Future work with simulations or Schwarzschild models explicitly including a bar \citep[e.g.][]{2024MNRAS.534..861T} could help disentangle this.}

\subsection{Mergers as the Driver of Orbit Transformations}

The fact that warm orbits do not correlate with age gives valuable insight into the processes that drive the reduction of \lre\ with stellar age. Stochastic events such as major mergers are well known to strongly reduce \lre\ \citep[e.g.][]{2014MNRAS.444.3357N,2018MNRAS.473.4956L}, but this also strongly depends on the gas content of the merging galaxies \citep[e.g.][]{2020IAUFM..30A.208L}, as high gas fractions may result in the re-formation of a spinning disc post-merger. Numerical simulations further demonstrate that mergers generate hot and counter-rotating orbits \citep[e.g.][]{2020MNRAS.498..940M}, with radial major mergers in particular producing a substantial population of box orbits \citep{1990dig..book..186B,2003ApJ...597..893N,2005MNRAS.360.1185J}. Our initial results \update{somewhat }support this scenario, as the combined fraction of hot and counter-rotating orbital components provides the strongest correlation with \lre\ ($p=1.33\times10^{-16}$), consistent with a merger-driven origin for the reduction of galaxy spin. \update{However, as shown in panel (f) of Figure \ref{fig:age_spin_shell}, we find no relation between box orbits and shells.}

Galaxy mergers leave behind distinct stellar debris structures in the form of tidal streams and shells. As shown by \cite{2024MNRAS.529..810R}, however, the connection between merger history, identifiable features, and kinematic transformation is not straightforward. They concluded that i) tidal shells trace radial mergers which impact the inner stellar kinematics more significantly than the circularly infalling mergers that form streams, and ii) tidal features can only reliably trace mergers in galaxies with young stellar ages, as such features fade after $\sim2-4$ Gyr \citep[e.g.][]{2008MNRAS.391.1137L, 2010MNRAS.404..575L,2010MNRAS.404..590L,2017MNRAS.465.2895L,2019A&A...632A.122M,2021ApJ...912...45N,2023MNRAS.523.4381D}. A high gas fraction in mergers can increase the detectability timescale of tidal features \citep[e.g.][]{2018ApJ...857..144H,2010MNRAS.404..575L}, however this also suppresses the generation of box orbits in central regions \citep{1996ApJ...471..115B,2006MNRAS.372..839N,2010ApJ...723..818H}. Consequently, any correlation between tidal features and orbital structure is not perfectly analogous to tracing merger history.

Mergers are expected to influence both stellar ages and kinematics. While any merger can funnel cold gas to the centre of the galaxy, driving star formation and reducing mean stellar age \citep[e.g.,][]{2014MNRAS.437.2137S,2015MNRAS.454.1742K,2019MNRAS.482L..55T,2022MNRAS.514.3294B}, the impact on stellar kinematics depends on the merger geometry. Galaxies with shells are produced as the result of radial, typically major mergers \citep[e.g.][]{1984ApJ...279..596Q,1989ApJ...342....1H,2019MNRAS.487..318K,2024MNRAS.530.4422K,2025MNRAS.543.3391K} that strongly heat the inner stellar regions and affect the observed stellar kinematics \citep[e.g.][]{2003ApJ...597..893N,2005MNRAS.360.1185J,2014MNRAS.445.1065R}. In contrast, circular or tangential mergers that form streams tend to only affect the outer regions \citep{2019MNRAS.487..318K}, and preserve the inner rotational support \citep{2022arXiv220808443V,2024MNRAS.529..810R}. This framework predicts that while all merger galaxies should show young mean stellar ages, galaxies with shells should show signatures of dynamically hot inner regions, and galaxies with streams should maintain their pre-merger stellar kinematics.

We divide our sample into three equally spaced age bins, and find that for the youngest galaxies ($\age\lessapprox6$ Gyr), the galaxies that have shells (as identified by \citealt{2024MNRAS.529..810R}) show moderately higher hot orbit fractions ($1.9\sigma$) and lower cold orbit fractions ($1.9\sigma$) than galaxies without shells. No significant difference is found for warm orbits, however. For the intermediate and oldest galaxies, we find no difference in any orbital fraction for galaxies with shells. As discussed by \cite{2024MNRAS.529..810R} and \cite{2024MNRAS.529.3446C}, this does not necessarily imply that these galaxies did not build their hot orbits through mergers, but rather that the relevant mergers may have occurred sufficiently long ago that any identifiable features have since faded with time. 

\update{In this analysis thus far we have interpreted shells as evidence of radial merger events that have dynamically heated the central regions of a galaxy and increased the hot orbit fraction. Now, we consider the potential of merger scenarios to contribute to warm orbit populations. In particular, given the intermediate geometries of mergers that result in stellar streams, the stars in these features are likely to phase mix into warm orbits. When performing a similar analysis as in Figure \ref{fig:age_spin_shell}, we find that for galaxies with intermediate ages, those with streams display a significantly higher warm orbit fraction within 1\re\ than those without. This supports the scenario of the phase mixing of stellar material from streams contributing to the build up of the warm orbit population. However, this does not appear to dominate the global redistribution of angular momentum in our sample. In particular, the strong correlations between stellar age, \hot, and \cold\ indicate that the primary driver of the spin-age relation is the direct transformation of cold orbits into hot orbits through radial merger events, rather than the intermediate formation of warm orbits.}

While these results support the scenario that mergers transform cold orbits to hot orbits, the galaxies that currently display shells differ systematically from the overall population. Although partial correlations show that the fraction of hot orbits increases with age across the full sample ($p=8.87\times10^{-4}$), galaxies with shells do not show this trend ($p=2.14\times10^{-1}$). Conversely, galaxies with streams maintain a significant increase of hot orbits with age ($p=7.68\times10^{-3}$). This again may reflect the short visibility lifetime of shells. In Figure \ref{fig:age_spin_shell}, the average \lre, orbit fraction and triaxiality for galaxies with shells is consistent between the intermediate age bin and the oldest age bin, suggesting that older galaxies may have experienced the same mergers that leave shells, but their shells have already faded.

\subsection{Disentangling the Role of Slow Rotators and Warm Orbits in Angular Momentum Evolution}
\begin{figure}
    \centering
    \includegraphics[width=\linewidth]{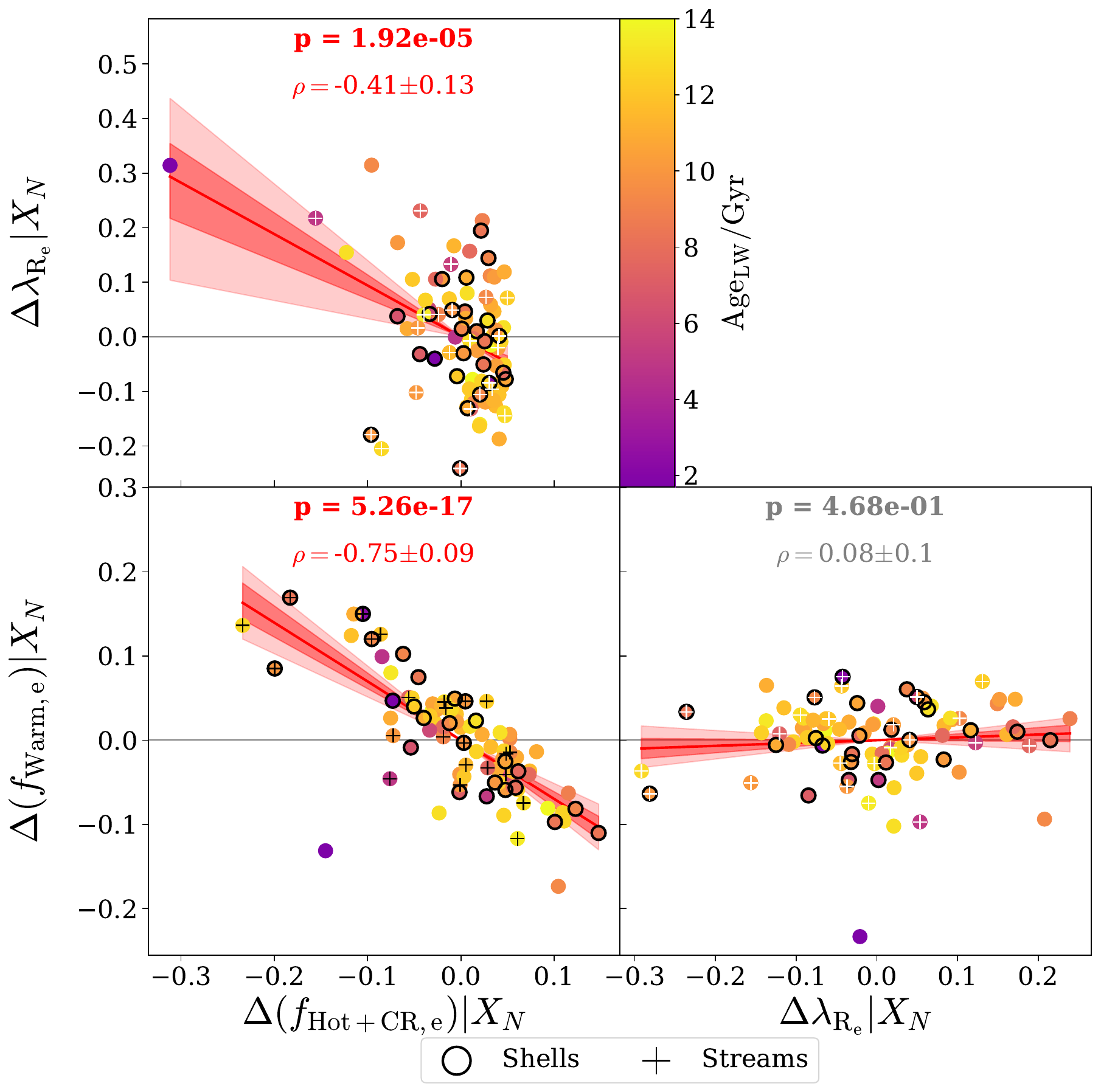}
    \caption[Partial Correlation Between Warm and Hot + Counter-Rotating Orbits]{Correlations between the hot + counter rotating orbital fraction (\hotcr), warm orbital fraction (\warm), and \lre. For each correlation between two variables, the linear correlation with the third variable is controlled for. Points are coloured by stellar age. Galaxies with shells are circles in black, and galaxies with streams display a white cross. The best-fitting linear relation for each correlation is shown in red, with 68\% and 95\% intervals shaded. \update{The Pearson correlation coefficient $\rho$, its bootstrapped uncertainty, and its p-value $p$ are given for each panel. These are highlighted in red when $p<0.05$, indicating a statistically significant correlation.}}
    \label{fig:warm_hotcr_test}
\end{figure}

The discussion thus far has supported the scenario in which massive SAMI early-type galaxies redistribute angular momentum from ordered to random motion through stochastic, merger-driven heating, transforming cold orbits directly into hot ones, and resulting in no significant correlations of the transitionary warm orbits with stellar age. Here we discuss several aspects of the data and methods that warrant further consideration. Firstly, slow rotators are very low spin galaxies \citep[e.g.][]{2011MNRAS.414..888E} that increase in fraction in the high-mass regime \citep{2017MNRAS.472.1272V,2020MNRAS.491..773G} and are primarily formed through mergers \citep{2018MNRAS.473.4956L}. A large proportion of slow rotators in our sample could bias results, particularly if fast rotators experience angular momentum evolution through different physical processes. Secondly, the behaviour of warm orbits remains ambiguous. They correlate more strongly with \lre\ than hot orbits ($p=1.10\times10^{-12}$ versus $p=9.66\times10^{-10}$), but unlike hot orbits, show no residual correlation with stellar age once mass and environment are controlled for. This further suggests that warm orbits are driven by secular evolution, and hence have no residual relation once stellar mass is controlled for.

Slow rotators present a distinct kinematic class of galaxy \citep[e.g.][]{2018MNRAS.477.4711G,2021MNRAS.505.3078V}, characterised by low \lre, high stellar mass and old stellar age \citep{2011MNRAS.414..888E}, typically attributed to the remnants of major, gas-poor mergers \citep[e.g.][]{2011MNRAS.416.1654B,2014MNRAS.444.3357N,2016ARA&A..54..597C,2018MNRAS.473.4956L}. Given their prevalence in the high-mass regime, which we sample from, it is possible that the correlations observed in our full sample are simply reflecting the increasing fraction of slow rotators at high stellar mass and age. To test this, we remove all slow rotator galaxies from our sample (i.e. all galaxies with $\varepsilon<0.428$ and $\lre<0.12+0.25\varepsilon$, as defined for SAMI by \citealt{2021MNRAS.505.3078V}). This consists of 34 galaxies, or $33\%$ of our sample. After this cut, the broad trends remain the same: controlling for mass and environment, the fraction of hot orbits increases with stellar age ($p=1.83\times10^{-3}$), the fraction of cold orbits decreases with stellar age ($p=1.08\times10^{-2}$), and the fraction of warm orbit shows no significant correlation with stellar age ($p=1.50\times10^{-1}$). Moreover, mass and environment show no significant correlation with any orbit fraction once the other variables are controlled for. These results indicate that stellar age remains the dominant predictor of internal dynamics across the slow and fast rotator dichotomy, and the observed correlations of hot and cold orbits with age are not not solely a property of slow rotators. Rather, merger-driven kinematic heating appears to operate across the full galaxy population, including the fast rotators.

The correlation of warm orbits with \lre\ is complicated by the absence of a corresponding correlation between warm orbits and stellar age once stellar mass and environment are controlled for. This is unexpected, as \lre\ is known to correlate most strongly with age in the SAMI sample \citep{2024MNRAS.529.3446C}, and we find that older galaxies in our sample have significantly more hot orbits and fewer cold orbits than younger systems. The lack of an age trend for warm orbits therefore suggests that they do not represent a distinct stage in the transformation from high to low spin. One plausible explanation is that "warm" classification traces too large a range in $\lambda_z$  ($0.25<\lambda_z<0.8$), encompassing both near-circular and moderately radial orbits. Such a wide range may mix multiple physical regimes and wash out any underlying age dependence. We tested this by dividing the warm orbit $\lambda_z$ range in half, creating "cold warm" and "hot warm" fractions. Both subsets independently correlate positively with \lre\ however, and continue to show no trend with age. This reinforces the idea that warm orbits do not represent a distinct stage in spin evolution. 

A further possibility is that warm orbits simply represent the largest residual fraction of orbits that are not created by mergers (i.e hot and counter-rotating). Given that the total orbit fraction must sum to unity, the correlation of warm orbits with \lre\ could be partially or entirely driven by the correlation of hot and counter-rotating orbits with \lre. We test this in Figure \ref{fig:warm_hotcr_test}, by showing the partial correlation of \warm, \hotcr, and \lre. Presented similarly to Figure \ref{fig:partial_correlation_hot_basic} but with points coloured by stellar age, we show in each panel the partial correlation of two variables with one another, once the linear correlation of the third variable has been accounted for. We see that when \warm\ is accounted for, there is still a strong negative correlation between \hotcr\ and \lre\ ($p=1.92\times10^{-5}$). Further, there exists a strong correlation between \hotcr\ and \warm\ independently of \lre ($p=5.26\times10^{-17}$). However, once correlations with \hotcr\ are removed, there is no residual trend between \warm\ and \lre ($p=4.68\times10^{-1}$). These trends persist even with the most extreme outliers removed. This suggests that the relationship between warm orbits and \lre\ may be an indirect consequence of the correlation between merger-driven orbits and \lre, but further work is needed to establish this.
\section{Conclusions}
\label{sec:conclusion_4}
In this paper, we investigated the evolution of angular momentum in massive SAMI early-type galaxies through detailed dynamical modelling. The light distribution of each galaxy was modelled with a Multi-Gaussian Expansion (MGE) model, based on KiDS $r$-band imaging. This MGE model was then converted to a total mass model by applying a global mass-to-light ratio and the inclusion of an NFW dark matter halo. We then used these mass models to construct orbit-superposition Schwarzschild models, and derived the fractions of cold, warm, hot, and counter-rotating orbits within one effective radius (\re) for each galaxy.

Within our sample, we find that the \lre-age relation reported by \cite{2024MNRAS.529.3446C} is driven primarily by a transition from cold to hot orbits as galaxies age. A partial correlation analysis shows that the fraction of hot orbits increases significantly with stellar age ($p=8.87\times10^{-4}$), while the cold orbit fraction decreases ($p=1.67\times10^{-3}$). Warm orbits show no significant correlation with stellar age, and neither stellar mass nor environment show any residual trend with \lre. The absence of a strong warm orbit correlation with age suggests that secular heating processes do not dominate spin evolution in massive early-type galaxies.

Our results suggest that mergers are likely the dominant mechanism driving the transformation from cold to hot orbits. Using the tidal feature classifications from \citet{2024MNRAS.529..810R}, we find that the youngest galaxies ($\age\lessapprox6$ Gyr) with shell features exhibit moderately higher hot orbit fractions ($1.9\sigma$) and lower cold orbit fractions ($1.9\sigma$) compared to galaxies without shells. However, shell galaxies do not show an increase in hot orbits with age as the larger sample does, which is consistent with the short visibility timescale of tidal shells. The similarity in average orbit fractions and triaxialities between intermediate-age shell systems and old-age systems with or without shells suggests that older galaxies may have experienced the same mergers, but their shells have simply faded with time.

We find that merger-driven heating of orbits operates across the full galaxy population, and is not driven solely by slow rotators. After removing slow rotators from our sample, partial correlation analysis again shows that the fraction of hot orbits increases with stellar age ($p=1.83\times10^{-3}$), the fraction of cold orbits decreases ($p=1.08\times10^{-2}$), and warm orbits show no significant trend. Furthermore, the apparent correlation between warm orbits and \lre\ can be fully explained by the combined fraction of hot and counter-rotating orbits, which are likely produced by merger-driven kinematic transformation. Stellar age remains the dominant predictor of stellar kinematic parameters across both slow and fast rotators, while warm orbits appear to play little role in the spin-down of massive early-type galaxies.

Future studies incorporating late-type star-forming discs into the sample \citep[e.g.][]{2023A&A...672A..84D} would allow for an understanding of kinematic transformation across the full local galaxy population, rather than just early-types. Further, more detailed Schwarzschild implementations that explicitly include bar modelling and orbits \citep[now available in the publicly available DYNAMITE code, based on][]{2024MNRAS.534..861T} may help to disentangle the contributions of different orbital families beyond the broad cold, warm, hot and counter-rotating classifications used in this work. The combination of orbit-superposition methods with stellar population properties, i.e. population-orbital superposition method or "Extragalactic Archaeology" \citep{2025ARA&A..63..259V}, has also shown promise as a pathway to uncover the assembly history of galaxies \citep[e.g.][]{2024MNRAS.534..861T}.

Upcoming instruments such as HARMONI on the ELT \citep{2010SPIE.7735E..2IT} will deliver high spatial resolution observations at cosmic noon ($z\sim2-3$), allowing for the study of galaxies at the peak of the star-formation epoch. BlueMUSE \citep{2019arXiv190601657R} on the VLT will provide high spatial and spectral resolution in a spectral range that will deliver the stellar population properties needed for population-orbital superposition methods, while the Hector Survey \citep{2025PASA...42..150O} will provide observations in diverse environments. Finally, similar analyses have been performed for simulated galaxies \citep[e.g.][]{2024MNRAS.528.2326S}, demonstrating that Schwarzschild models can recover the dynamical imprints of mergers. Extending this work to higher-resolution simulations would allow finer orbital structures to be resolved and compared, offering a test for how reliably the Schwarzschild technique can reconstruct the true dynamical evolution of galaxies.

We conclude that mergers are the dominant driver of angular momentum evolution in massive early-type galaxies. These mergers reshape the internal orbital structures of present day fast and slow rotators, establishing the observed kinematic spin-down with stellar age.

\section*{Acknowledgements}

The SAMI Galaxy Survey is based on observations made at the Anglo-Australian Telescope. SAMI was developed jointly by the University of Sydney and the Australian Astronomical Observatory (AAO). The SAMI input catalogue is based on data taken from the Sloan Digital Sky Survey, the GAMA Survey and the VST ATLAS Survey. The SAMI Galaxy Survey is supported by the Australian Research Council (ARC) Centre of Excellence ASTRO 3D (CE170100013) and CAASTRO (CE110001020), and other participating institutions. The SAMI Galaxy Survey website is \url{http://sami-survey.org/}.

JJB acknowledges support of an Australian Research Council Future Fellowship (FT180100231). 
JBH is supported by an ARC Laureate Fellowship FL140100278. The SAMI instrument was funded by Bland-Hawthorn's former Federation Fellowship FF0776384, an ARC LIEF grant LE130100198 (PI Bland-Hawthorn) and funding from the Anglo-Australian Observatory. SMS acknowledges funding from the Australian Research Council (DE220100003). Parts of this research were conducted by the Australian Research Council Centre of Excellence for All Sky Astrophysics in 3 Dimensions (ASTRO 3D), through project number CE170100013. 

Based on observations made with ESO Telescopes at the La Silla Paranal Observatory under programme IDs 177.A-3016, 177.A-3017, 177.A-3018 and 179.A-2004, and on data products produced by the KiDS consortium. The KiDS production team acknowledges support from: Deutsche Forschungsgemeinschaft, ERC, NOVA and NWO-M grants; Target; the University of Padova, and the University Federico II (Naples).

The Hyper Suprime-Cam (HSC) collaboration includes the astronomical communities of Japan and Taiwan, and Princeton University. The HSC instrumentation and software were developed by the National Astronomical Observatory of Japan (NAOJ), the Kavli Institute for the Physics and Mathematics of the Universe (Kavli IPMU), the University of Tokyo, the High Energy Accelerator Research Organization (KEK), the Academia Sinica Institute for Astronomy and Astrophysics in Taiwan (ASIAA), and Princeton University. Funding was contributed by the FIRST program from Japanese Cabinet Office, the Ministry of Education, Culture, Sports, Science and Technology (MEXT), the Japan Society for the Promotion of Science (JSPS), Japan Science and Technology Agency (JST), the Toray Science Foundation, NAOJ, Kavli IPMU, KEK, ASIAA, and Princeton University.

This paper makes use of software developed for the Large Synoptic Survey Telescope. We thank the LSST Project for making their code available as free software at \url{http://dm.lsst.org}.

The Pan-STARRS1 Surveys (PS1) have been made possible through contributions of the Institute for Astronomy, the University of Hawaii, the Pan-STARRS Project Office, the Max-Planck Society and its participating institutes, the Max Planck Institute for Astronomy, Heidelberg and the Max Planck Institute for Extraterrestrial Physics, Garching, The Johns Hopkins University, Durham University, the University of Edinburgh, Queen’s University Belfast, the Harvard-Smithsonian Center for Astrophysics, the Las Cumbres Observatory Global Telescope Network Incorporated, the National Central University of Taiwan, the Space Telescope Science Institute, the National Aeronautics and Space Administration under Grant No. NNX08AR22G issued through the Planetary Science Division of the NASA Science Mission Directorate, the National Science Foundation under Grant No. AST-1238877, the University of Maryland, and Eotvos Lorand University (ELTE) and the Los Alamos National Laboratory.

Based [in part] on data collected at the Subaru Telescope and retrieved from the HSC data archive system, which is operated by Subaru Telescope and Astronomy Data Center at National Astronomical Observatory of Japan.

\section*{Data Availability}

The results of the data displayed in the figures can be obtained by contacting the corresponding author upon request. The images originate from the Kilo-Degree Survey Data Release 4 \citep{2019A&A...625A...2K}. The data utilized in this paper, including kinematic measurements, originate from SAMI Data Release 3 \citep{2021MNRAS.505..991C}. This data is accessible through Australian Astronomical Optics’ Data Central at \url{https://datacentral.org.au/}.



\bibliographystyle{mnras}
\bibliography{mybib} 

@ARTICLE{1994A&A...285..723E,
       author = {{Emsellem}, E. and {Monnet}, G. and {Bacon}, R.},
        title = "{The multi-gaussian expansion method: a tool for building realistic photometric and kinematical models of stellar systems I. The formalism}",
      journal = {\aap},
         year = 1994,
        month = may,
       volume = {285},
        pages = {723-738},
       adsurl = {https://ui.adsabs.harvard.edu/abs/1994A&A...285..723E}
}

@ARTICLE{2002MNRAS.333..400C,
       author = {{Cappellari}, Michele},
        title = "{Efficient multi-Gaussian expansion of galaxies}",
      journal = {\mnras},
         year = 2002,
        month = jun,
       volume = {333},
       number = {2},
        pages = {400-410},
          doi = {10.1046/j.1365-8711.2002.05412.x},
archivePrefix = {arXiv},
       eprint = {astro-ph/0201430},
 primaryClass = {astro-ph},
       adsurl = {https://ui.adsabs.harvard.edu/abs/2002MNRAS.333..400C}
}

@ARTICLE{2017ApJ...835..104V,
       author = {{van de Sande}, Jesse and {Bland-Hawthorn}, Joss and {Fogarty}, Lisa M.~R. and {Cortese}, Luca and {d'Eugenio}, Francesco and {Croom}, Scott M. and {Scott}, Nicholas and {Allen}, James T. and {Brough}, Sarah and {Bryant}, Julia J. and {Cecil}, Gerald and {Colless}, Matthew and {Couch}, Warrick J. and {Davies}, Roger and {Elahi}, Pascal J. and {Foster}, Caroline and {Goldstein}, Gregory and {Goodwin}, Michael and {Groves}, Brent and {Ho}, I. -Ting and {Jeong}, Hyunjin and {Jones}, D. Heath and {Konstantopoulos}, Iraklis S. and {Lawrence}, Jon S. and {Leslie}, Sarah K. and {L{\'o}pez-S{\'a}nchez}, {\'A}ngel R. and {McDermid}, Richard M. and {McElroy}, Rebecca and {Medling}, Anne M. and {Oh}, Sree and {Owers}, Matt S. and {Richards}, Samuel N. and {Schaefer}, Adam L. and {Sharp}, Rob and {Sweet}, Sarah M. and {Taranu}, Dan and {Tonini}, Chiara and {Walcher}, C. Jakob and {Yi}, Sukyoung K.},
        title = "{The SAMI Galaxy Survey: Revisiting Galaxy Classification through High-order Stellar Kinematics}",
      journal = {\apj},
         year = 2017,
        month = jan,
       volume = {835},
       number = {1},
          eid = {104},
        pages = {104},
          doi = {10.3847/1538-4357/835/1/104},
archivePrefix = {arXiv},
       eprint = {1611.07039},
 primaryClass = {astro-ph.GA},
       adsurl = {https://ui.adsabs.harvard.edu/abs/2017ApJ...835..104V}
}

@ARTICLE{2004PASP..116..138C,
       author = {{Cappellari}, Michele and {Emsellem}, Eric},
        title = "{Parametric Recovery of Line-of-Sight Velocity Distributions from Absorption-Line Spectra of Galaxies via Penalized Likelihood}",
      journal = {\pasp},
         year = 2004,
        month = feb,
       volume = {116},
       number = {816},
        pages = {138-147},
          doi = {10.1086/381875},
archivePrefix = {arXiv},
       eprint = {astro-ph/0312201},
 primaryClass = {astro-ph},
       adsurl = {https://ui.adsabs.harvard.edu/abs/2004PASP..116..138C}
}

@ARTICLE{2017MNRAS.466..798C,
       author = {{Cappellari}, Michele},
        title = "{Improving the full spectrum fitting method: accurate convolution with Gauss-Hermite functions}",
      journal = {\mnras},
         year = 2017,
        month = apr,
       volume = {466},
       number = {1},
        pages = {798-811},
          doi = {10.1093/mnras/stw3020},
archivePrefix = {arXiv},
       eprint = {1607.08538},
 primaryClass = {astro-ph.GA},
       adsurl = {https://ui.adsabs.harvard.edu/abs/2017MNRAS.466..798C}
}

@ARTICLE{2024MNRAS.529..810R,
       author = {{Rutherford}, Tomas H. and {van de Sande}, Jesse and {Croom}, Scott M. and {Valenzuela}, Lucas M. and {Remus}, Rhea-Silvia and {D'Eugenio}, Francesco and {Vaughan}, Sam P. and {Zovaro}, Henry R.~M. and {Casura}, Sarah and {Barsanti}, Stefania and {Bland-Hawthorn}, Joss and {Brough}, Sarah and {Bryant}, Julia J. and {Goodwin}, Michael and {Lorente}, Nuria and {Oh}, Sree and {Ristea}, Andrei},
        title = "{The SAMI Galaxy Survey: using tidal streams and shells to trace the dynamical evolution of massive galaxies}",
      journal = {\mnras},
         year = 2024,
        month = apr,
       volume = {529},
       number = {2},
        pages = {810-830},
          doi = {10.1093/mnras/stae398},
archivePrefix = {arXiv},
       eprint = {2402.02728},
 primaryClass = {astro-ph.GA},
       adsurl = {https://ui.adsabs.harvard.edu/abs/2024MNRAS.529..810R}
}

@ARTICLE{2022MNRAS.514.3294B,
       author = {{Bickley}, Robert W. and {Ellison}, Sara L. and {Patton}, David R. and {Bottrell}, Connor and {Gwyn}, Stephen and {Hudson}, Michael J.},
        title = "{Star formation characteristics of CNN-identified post-mergers in the Ultraviolet Near Infrared Optical Northern Survey (UNIONS)}",
      journal = {\mnras},
         year = 2022,
        month = aug,
       volume = {514},
       number = {3},
        pages = {3294-3307},
          doi = {10.1093/mnras/stac1500},
archivePrefix = {arXiv},
       eprint = {2205.14103},
 primaryClass = {astro-ph.GA},
       adsurl = {https://ui.adsabs.harvard.edu/abs/2022MNRAS.514.3294B}
}

@ARTICLE{2019MNRAS.482L..55T,
       author = {{Thorp}, Mallory D. and {Ellison}, Sara L. and {Simard}, Luc and {S{\'a}nchez}, Sebastian F. and {Antonio}, Braulio},
        title = "{Spatially resolved star formation and metallicity profiles in post-merger galaxies from MaNGA}",
      journal = {\mnras},
         year = 2019,
        month = jan,
       volume = {482},
       number = {1},
        pages = {L55-L59},
          doi = {10.1093/mnrasl/sly185},
archivePrefix = {arXiv},
       eprint = {1810.00897},
 primaryClass = {astro-ph.GA},
       adsurl = {https://ui.adsabs.harvard.edu/abs/2019MNRAS.482L..55T}
}

@ARTICLE{2020MNRAS.491..773G,
       author = {{Guo}, Kexin and {Cortese}, Luca and {Obreschkow}, Danail and {Catinella}, Barbara and {van de Sande}, Jesse and {Croom}, Scott M. and {Brough}, Sarah and {Sweet}, Sarah and {Bryant}, Julia J. and {Medling}, Anne and {Bland-Hawthorn}, Joss and {Owers}, Matt and {Richards}, Samuel N.},
        title = "{The SAMI Galaxy Survey: the contribution of different kinematic classes to the stellar mass function of nearby galaxies}",
      journal = {\mnras},
         year = 2020,
        month = jan,
       volume = {491},
       number = {1},
        pages = {773-781},
          doi = {10.1093/mnras/stz3042},
archivePrefix = {arXiv},
       eprint = {1911.01433},
 primaryClass = {astro-ph.GA},
       adsurl = {https://ui.adsabs.harvard.edu/abs/2020MNRAS.491..773G}
}

@ARTICLE{2016ARA&A..54..597C,
       author = {{Cappellari}, Michele},
        title = "{Structure and Kinematics of Early-Type Galaxies from Integral Field Spectroscopy}",
      journal = {\araa},
         year = 2016,
        month = sep,
       volume = {54},
        pages = {597-665},
          doi = {10.1146/annurev-astro-082214-122432},
archivePrefix = {arXiv},
       eprint = {1602.04267},
 primaryClass = {astro-ph.GA},
       adsurl = {https://ui.adsabs.harvard.edu/abs/2016ARA&A..54..597C}
}

@ARTICLE{2019MNRAS.486.4753J,
       author = {{Jin}, Yunpeng and {Zhu}, Ling and {Long}, R.~J. and {Mao}, Shude and {Xu}, Dandan and {Li}, Hongyu and {van de Ven}, Glenn},
        title = "{Evaluating the ability of triaxial Schwarzschild modelling to estimate properties of galaxies from the Illustris simulation}",
      journal = {\mnras},
         year = 2019,
        month = jul,
       volume = {486},
       number = {4},
        pages = {4753-4772},
          doi = {10.1093/mnras/stz1170},
archivePrefix = {arXiv},
       eprint = {1904.12942},
 primaryClass = {astro-ph.GA},
       adsurl = {https://ui.adsabs.harvard.edu/abs/2019MNRAS.486.4753J}
}

@software{2020ascl.soft11007J,
       author = {{Jethwa}, Prashin and {Thater}, Sabine and {Maindl}, Thomas and {Van de Ven}, Glenn},
        title = "{DYNAMITE: DYnamics, Age and Metallicity Indicators Tracing Evolution}",
 howpublished = {Astrophysics Source Code Library, record ascl:2011.007},
         year = 2020,
        month = nov,
          eid = {ascl:2011.007},
       adsurl = {https://ui.adsabs.harvard.edu/abs/2020ascl.soft11007J}
}

@ARTICLE{2020ApJ...889...39V,
       author = {{Vasiliev}, Eugene and {Valluri}, Monica},
        title = "{A New Implementation of the Schwarzchild Method for Constructing Observationally Driven Dynamical Models of Galaxies of All Morphological Types}",
      journal = {\apj},
         year = 2020,
        month = jan,
       volume = {889},
       number = {1},
          eid = {39},
        pages = {39},
          doi = {10.3847/1538-4357/ab5fe0},
archivePrefix = {arXiv},
       eprint = {1912.04288},
 primaryClass = {astro-ph.GA},
       adsurl = {https://ui.adsabs.harvard.edu/abs/2020ApJ...889...39V}
}

@ARTICLE{1978MNRAS.183..341W,
       author = {{White}, S.~D.~M. and {Rees}, M.~J.},
        title = "{Core condensation in heavy halos: a two-stage theory for galaxy formation and clustering.}",
      journal = {\mnras},
         year = 1978,
        month = may,
       volume = {183},
        pages = {341-358},
          doi = {10.1093/mnras/183.3.341},
       adsurl = {https://ui.adsabs.harvard.edu/abs/1978MNRAS.183..341W}
}

@ARTICLE{2014MNRAS.444.3357N,
       author = {{Naab}, Thorsten and {Oser}, L. and {Emsellem}, E. and {Cappellari}, Michele and {Krajnovi{\'c}}, D. and {McDermid}, R.~M. and {Alatalo}, K. and {Bayet}, E. and {Blitz}, L. and {Bois}, M. and {Bournaud}, F. and {Bureau}, M. and {Crocker}, A. and {Davies}, R.~L. and {Davis}, T.~A. and {de Zeeuw}, P.~T. and {Duc}, P. -A. and {Hirschmann}, M. and {Johansson}, P.~H. and {Khochfar}, S. and {Kuntschner}, H. and {Morganti}, R. and {Oosterloo}, T. and {Sarzi}, M. and {Scott}, N. and {Serra}, P. and {van de Ven}, G. and {Weijmans}, A. and {Young}, L.~M.},
        title = "{The ATLAS$^{3D}$ project - XXV. Two-dimensional kinematic analysis of simulated galaxies and the cosmological origin of fast and slow rotators}",
      journal = {\mnras},
         year = 2014,
        month = nov,
       volume = {444},
       number = {4},
        pages = {3357-3387},
          doi = {10.1093/mnras/stt1919},
archivePrefix = {arXiv},
       eprint = {1311.0284},
 primaryClass = {astro-ph.CO},
       adsurl = {https://ui.adsabs.harvard.edu/abs/2014MNRAS.444.3357N}
}

@ARTICLE{2020MNRAS.493.3778S,
       author = {{Schulze}, Felix and {Remus}, Rhea-Silvia and {Dolag}, Klaus and {Bellstedt}, Sabine and {Burkert}, Andreas and {Forbes}, Duncan A.},
        title = "{Kinematics of simulated galaxies II: Probing the stellar kinematics of galaxies out to large radii}",
      journal = {\mnras},
         year = 2020,
        month = apr,
       volume = {493},
       number = {3},
        pages = {3778-3799},
          doi = {10.1093/mnras/staa511},
archivePrefix = {arXiv},
       eprint = {2001.02237},
 primaryClass = {astro-ph.GA},
       adsurl = {https://ui.adsabs.harvard.edu/abs/2020MNRAS.493.3778S}
}

@ARTICLE{2003ApJ...597...21A,
       author = {{Abadi}, Mario G. and {Navarro}, Julio F. and {Steinmetz}, Matthias and {Eke}, Vincent R.},
        title = "{Simulations of Galaxy Formation in a {\ensuremath{\Lambda}} Cold Dark Matter Universe. II. The Fine Structure of Simulated Galactic Disks}",
      journal = {\apj},
         year = 2003,
        month = nov,
       volume = {597},
       number = {1},
        pages = {21-34},
          doi = {10.1086/378316},
archivePrefix = {arXiv},
       eprint = {astro-ph/0212282},
 primaryClass = {astro-ph},
       adsurl = {https://ui.adsabs.harvard.edu/abs/2003ApJ...597...21A}
}

@ARTICLE{2008MNRAS.385..647V,
       author = {{van den Bosch}, R.~C.~E. and {van de Ven}, G. and {Verolme}, E.~K. and {Cappellari}, M. and {de Zeeuw}, P.~T.},
        title = "{Triaxial orbit based galaxy models with an application to the (apparent) decoupled core galaxy NGC 4365}",
      journal = {\mnras},
         year = 2008,
        month = apr,
       volume = {385},
       number = {2},
        pages = {647-666},
          doi = {10.1111/j.1365-2966.2008.12874.x},
archivePrefix = {arXiv},
       eprint = {0712.0113},
 primaryClass = {astro-ph},
       adsurl = {https://ui.adsabs.harvard.edu/abs/2008MNRAS.385..647V}
}

@ARTICLE{2019MNRAS.487.3776P,
       author = {{Poci}, Adriano and {McDermid}, Richard M. and {Zhu}, Ling and {van de Ven}, Glenn},
        title = "{Combining stellar populations with orbit-superposition dynamical modelling: the formation history of the lenticular galaxy NGC 3115}",
      journal = {\mnras},
         year = 2019,
        month = aug,
       volume = {487},
       number = {3},
        pages = {3776-3796},
          doi = {10.1093/mnras/stz1154},
archivePrefix = {arXiv},
       eprint = {1904.11605},
 primaryClass = {astro-ph.GA},
       adsurl = {https://ui.adsabs.harvard.edu/abs/2019MNRAS.487.3776P}
}

@ARTICLE{2024MNRAS.534..861T,
       author = {{Tahmasebzadeh}, Behzad and {Zhu}, Ling and {Shen}, Juntai and {Gadotti}, Dimitri A. and {Valluri}, Monica and {Thater}, Sabine and {van de Ven}, Glenn and {Jin}, Yunpeng and {Gerhard}, Ortwin and {Erwin}, Peter and {Jethwa}, Prashin and {Zocchi}, Alice and {Lilley}, Edward J. and {Fragkoudi}, Francesca and {de Lorenzo-C{\'a}ceres}, Adriana and {M{\'e}ndez-Abreu}, Jairo and {Neumann}, Justus and {Guo}, Rui},
        title = "{Schwarzschild modelling of barred s0 galaxy NGC 4371}",
      journal = {\mnras},
         year = 2024,
        month = oct,
       volume = {534},
       number = {1},
        pages = {861-882},
          doi = {10.1093/mnras/stae2109},
archivePrefix = {arXiv},
       eprint = {2310.00497},
 primaryClass = {astro-ph.GA},
       adsurl = {https://ui.adsabs.harvard.edu/abs/2024MNRAS.534..861T}
}

@BOOK{2008gady.book.....B,
       author = {{Binney}, James and {Tremaine}, Scott},
        title = "{Galactic Dynamics: Second Edition}",
         year = 2008,
       adsurl = {https://ui.adsabs.harvard.edu/abs/2008gady.book.....B}
}

@ARTICLE{2003PASP..115..763C,
       author = {{Chabrier}, Gilles},
        title = "{Galactic Stellar and Substellar Initial Mass Function}",
      journal = {\pasp},
         year = 2003,
        month = jul,
       volume = {115},
       number = {809},
        pages = {763-795},
          doi = {10.1086/376392},
archivePrefix = {arXiv},
       eprint = {astro-ph/0304382},
 primaryClass = {astro-ph},
       adsurl = {https://ui.adsabs.harvard.edu/abs/2003PASP..115..763C}
}

@ARTICLE{2009A&A...501L...9D,
       author = {{Di Matteo}, P. and {Jog}, C.~J. and {Lehnert}, M.~D. and {Combes}, F. and {Semelin}, B.},
        title = "{Generation of rotationally dominated galaxies by mergers of pressure-supported progenitors}",
      journal = {\aap},
         year = 2009,
        month = jul,
       volume = {501},
       number = {3},
        pages = {L9-L13},
          doi = {10.1051/0004-6361/200912354},
archivePrefix = {arXiv},
       eprint = {0906.0010},
 primaryClass = {astro-ph.CO},
       adsurl = {https://ui.adsabs.harvard.edu/abs/2009A&A...501L...9D}
}

@ARTICLE{2017MNRAS.464.3850L,
       author = {{Lagos}, Claudia del P. and {Theuns}, Tom and {Stevens}, Adam R.~H. and {Cortese}, Luca and {Padilla}, Nelson D. and {Davis}, Timothy A. and {Contreras}, Sergio and {Croton}, Darren},
        title = "{Angular momentum evolution of galaxies in EAGLE}",
      journal = {\mnras},
         year = 2017,
        month = feb,
       volume = {464},
       number = {4},
        pages = {3850-3870},
          doi = {10.1093/mnras/stw2610},
archivePrefix = {arXiv},
       eprint = {1609.01739},
 primaryClass = {astro-ph.GA},
       adsurl = {https://ui.adsabs.harvard.edu/abs/2017MNRAS.464.3850L}
}

@ARTICLE{2017ApJ...837...68C,
       author = {{Choi}, Hoseung and {Yi}, Sukyoung K.},
        title = "{On the Evolution of Galaxy Spin in a Cosmological Hydrodynamic Simulation of Galaxy Clusters}",
      journal = {\apj},
         year = 2017,
        month = mar,
       volume = {837},
       number = {1},
          eid = {68},
        pages = {68},
          doi = {10.3847/1538-4357/aa5e4b},
archivePrefix = {arXiv},
       eprint = {1702.00517},
 primaryClass = {astro-ph.GA},
       adsurl = {https://ui.adsabs.harvard.edu/abs/2017ApJ...837...68C}
}

@ARTICLE{1979ApJ...232..236S,
       author = {{Schwarzschild}, M.},
        title = "{A numerical model for a triaxial stellar system in dynamical equilibrium.}",
      journal = {\apj},
         year = 1979,
        month = aug,
       volume = {232},
        pages = {236-247},
          doi = {10.1086/157282},
       adsurl = {https://ui.adsabs.harvard.edu/abs/1979ApJ...232..236S}
}

@ARTICLE{2012MNRAS.421..872C,
       author = {{Croom}, Scott M. and {Lawrence}, Jon S. and {Bland-Hawthorn}, Joss and {Bryant}, Julia J. and {Fogarty}, Lisa and {Richards}, Samuel and {Goodwin}, Michael and {Farrell}, Tony and {Miziarski}, Stan and {Heald}, Ron and {Jones}, D. Heath and {Lee}, Steve and {Colless}, Matthew and {Brough}, Sarah and {Hopkins}, Andrew M. and {Bauer}, Amanda E. and {Birchall}, Michael N. and {Ellis}, Simon and {Horton}, Anthony and {Leon-Saval}, Sergio and {Lewis}, Geraint and {L{\'o}pez-S{\'a}nchez}, {\'A}. R. and {Min}, Seong-Sik and {Trinh}, Christopher and {Trowland}, Holly},
        title = "{The Sydney-AAO Multi-object Integral field spectrograph}",
      journal = {\mnras},
         year = 2012,
        month = mar,
       volume = {421},
       number = {1},
        pages = {872-893},
          doi = {10.1111/j.1365-2966.2011.20365.x},
archivePrefix = {arXiv},
       eprint = {1112.3367},
 primaryClass = {astro-ph.CO},
       adsurl = {https://ui.adsabs.harvard.edu/abs/2012MNRAS.421..872C}
}

@ARTICLE{1992A&A...253..366M,
       author = {{Monnet}, G. and {Bacon}, R. and {Emsellem}, E.},
        title = "{Modelling the stellar intensity and radial velocity fields in triaxial galaxies by sums of Gaussian functions.}",
      journal = {\aap},
         year = 1992,
        month = jan,
       volume = {253},
        pages = {366-373},
       adsurl = {https://ui.adsabs.harvard.edu/abs/1992A&A...253..366M}
}

@ARTICLE{2021ApJ...918...84R,
       author = {{Rutherford}, Tomas H. and {Croom}, Scott M. and {van de Sande}, Jesse and {Lagos}, Claudia del P. and {Bland-Hawthorn}, Joss and {Brough}, S. and {Bryant}, Julia J. and {D'Eugenio}, Francesco and {Owers}, Matt S.},
        title = "{The SAMI Galaxy Survey: Detection of Environmental Dependence of Galaxy Spin in Observations and Simulations Using Marked Correlation Functions}",
      journal = {\apj},
         year = 2021,
        month = sep,
       volume = {918},
       number = {2},
          eid = {84},
        pages = {84},
          doi = {10.3847/1538-4357/ac0e8d},
archivePrefix = {arXiv},
       eprint = {2107.00546},
 primaryClass = {astro-ph.GA},
       adsurl = {https://ui.adsabs.harvard.edu/abs/2021ApJ...918...84R}
}

@ARTICLE{2023A&A...672A..27B,
       author = {{B{\'\i}lek}, M. and {Duc}, P. -A. and {Sola}, E.},
        title = "{Origin of the differences in rotational support among early-type galaxies: The case of galaxies outside clusters}",
      journal = {\aap},
         year = 2023,
        month = apr,
       volume = {672},
          eid = {A27},
        pages = {A27},
          doi = {10.1051/0004-6361/202244749},
archivePrefix = {arXiv},
       eprint = {2210.02478},
 primaryClass = {astro-ph.GA},
       adsurl = {https://ui.adsabs.harvard.edu/abs/2023A&A...672A..27B}
}

@ARTICLE{2022ApJS..262...39H,
       author = {{Huang}, Qifeng and {Fan}, Lulu},
        title = "{Massive Early-type Galaxies in the HSC-SSP: Flux Fraction of Tidal Features and Merger Rates}",
      journal = {\apjs},
         year = 2022,
        month = oct,
       volume = {262},
       number = {2},
          eid = {39},
        pages = {39},
          doi = {10.3847/1538-4365/ac85b1},
archivePrefix = {arXiv},
       eprint = {2207.14320},
 primaryClass = {astro-ph.GA},
       adsurl = {https://ui.adsabs.harvard.edu/abs/2022ApJS..262...39H}
}

@ARTICLE{2022MNRAS.516.2971V,
       author = {{Vaughan}, Sam P. and {Barone}, Tania M. and {Croom}, Scott M. and {Cortese}, Luca and {D'Eugenio}, Francesco and {Brough}, Sarah and {Colless}, Matthew and {McDermid}, Richard M. and {van de Sande}, Jesse and {Scott}, Nicholas and {Bland-Hawthorn}, Joss and {Bryant}, Julia J. and {Lawrence}, J.~S. and {L{\'o}pez-S{\'a}nchez}, {\'A}ngel R. and {Lorente}, Nuria P.~F. and {Owers}, Matt S. and {Richards}, Samuel N.},
        title = "{The SAMI galaxy survey: Galaxy size can explain the offset between star-forming and passive galaxies in the mass-metallicity relationship}",
      journal = {\mnras},
         year = 2022,
        month = oct,
       volume = {516},
       number = {2},
        pages = {2971-2987},
          doi = {10.1093/mnras/stac2304},
archivePrefix = {arXiv},
       eprint = {2208.06939},
 primaryClass = {astro-ph.GA},
       adsurl = {https://ui.adsabs.harvard.edu/abs/2022MNRAS.516.2971V}
}

@ARTICLE{2022arXiv220808443V,
       author = {{Valenzuela}, Lucas M. and {Remus}, Rhea-Silvia},
        title = "{A Stream Come True -- Connecting tidal tails, shells, streams, and planes with galaxy kinematics and formation history}",
      journal = {arXiv e-prints},
         year = 2022,
        month = aug,
          eid = {arXiv:2208.08443},
        pages = {arXiv:2208.08443},
          doi = {10.48550/arXiv.2208.08443},
archivePrefix = {arXiv},
       eprint = {2208.08443},
 primaryClass = {astro-ph.GA},
       adsurl = {https://ui.adsabs.harvard.edu/abs/2022arXiv220808443V}
}

@ARTICLE{2022MNRAS.513.2985R,
       author = {{Robotham}, A.~S.~G. and {Bellstedt}, S. and {Driver}, S.~P.},
        title = "{ProFuse: physical multiband structural decomposition of galaxies and the mass-size-age plane}",
      journal = {\mnras},
         year = 2022,
        month = jun,
       volume = {513},
       number = {2},
        pages = {2985-3012},
          doi = {10.1093/mnras/stac1032},
archivePrefix = {arXiv},
       eprint = {2204.04897},
 primaryClass = {astro-ph.GA},
       adsurl = {https://ui.adsabs.harvard.edu/abs/2022MNRAS.513.2985R}
}

@ARTICLE{2021MNRAS.505.3078V,
       author = {{van de Sande}, Jesse and {Vaughan}, Sam P. and {Cortese}, Luca and {Scott}, Nicholas and {Bland-Hawthorn}, Joss and {Croom}, Scott M. and {Lagos}, Claudia D.~P. and {Brough}, Sarah and {Bryant}, Julia J. and {Devriendt}, Julien and {Dubois}, Yohan and {D'Eugenio}, Francesco and {Foster}, Caroline and {Fraser-McKelvie}, Amelia and {Harborne}, Katherine E. and {Lawrence}, Jon S. and {Oh}, Sree and {Owers}, Matt S. and {Poci}, Adriano and {Remus}, Rhea-Silvia and {Richards}, Samuel N. and {Schulze}, Felix and {Sweet}, Sarah M. and {Varidel}, Mathew R. and {Welker}, Charlotte},
        title = "{The SAMI Galaxy Survey: a statistical approach to an optimal classification of stellar kinematics in galaxy surveys}",
      journal = {\mnras},
         year = 2021,
        month = aug,
       volume = {505},
       number = {2},
        pages = {3078-3106},
          doi = {10.1093/mnras/stab1490},
archivePrefix = {arXiv},
       eprint = {2011.08199},
 primaryClass = {astro-ph.GA},
       adsurl = {https://ui.adsabs.harvard.edu/abs/2021MNRAS.505.3078V}
}

@ARTICLE{2021MNRAS.505..991C,
       author = {{Croom}, Scott M. and {Owers}, Matt S. and {Scott}, Nicholas and {Poetrodjojo}, Henry and {Groves}, Brent and {van de Sande}, Jesse and {Barone}, Tania M. and {Cortese}, Luca and {D'Eugenio}, Francesco and {Bland-Hawthorn}, Joss and {Bryant}, Julia and {Oh}, Sree and {Brough}, Sarah and {Agostino}, James and {Casura}, Sarah and {Catinella}, Barbara and {Colless}, Matthew and {Cecil}, Gerald and {Davies}, Roger L. and {Drinkwater}, Michael J. and {Driver}, Simon P. and {Ferreras}, Ignacio and {Foster}, Caroline and {Fraser-McKelvie}, Amelia and {Lawrence}, Jon and {Leslie}, Sarah K. and {Liske}, Jochen and {L{\'o}pez-S{\'a}nchez}, {\'A}ngel R. and {Lorente}, Nuria P.~F. and {McElroy}, Rebecca and {Medling}, Anne M. and {Obreschkow}, Danail and {Richards}, Samuel N. and {Sharp}, Rob and {Sweet}, Sarah M. and {Taranu}, Dan S. and {Taylor}, Edward N. and {Tescari}, Edoardo and {Thomas}, Adam D. and {Tocknell}, James and {Vaughan}, Sam P.},
        title = "{The SAMI Galaxy Survey: the third and final data release}",
      journal = {\mnras},
         year = 2021,
        month = jul,
       volume = {505},
       number = {1},
        pages = {991-1016},
          doi = {10.1093/mnras/stab229},
archivePrefix = {arXiv},
       eprint = {2101.12224},
 primaryClass = {astro-ph.GA},
       adsurl = {https://ui.adsabs.harvard.edu/abs/2021MNRAS.505..991C}
}

@ARTICLE{2021ApJ...912...45N,
       author = {{Nevin}, R. and {Blecha}, L. and {Comerford}, J. and {Greene}, J.~E. and {Law}, D.~R. and {Stark}, D.~V. and {Westfall}, K.~B. and {Vazquez-Mata}, J.~A. and {Smethurst}, R. and {Argudo-Fern{\'a}ndez}, M. and {Brownstein}, J.~R. and {Drory}, N.},
        title = "{Accurate Identification of Galaxy Mergers with Stellar Kinematics}",
      journal = {\apj},
         year = 2021,
        month = may,
       volume = {912},
       number = {1},
          eid = {45},
        pages = {45},
          doi = {10.3847/1538-4357/abe2a9},
archivePrefix = {arXiv},
       eprint = {2102.02208},
 primaryClass = {astro-ph.GA},
       adsurl = {https://ui.adsabs.harvard.edu/abs/2021ApJ...912...45N}
}

@ARTICLE{2020MNRAS.497.2018H,
       author = {{Harborne}, K.~E. and {van de Sande}, J. and {Cortese}, L. and {Power}, C. and {Robotham}, A.~S.~G. and {Lagos}, C.~D.~P. and {Croom}, S.},
        title = "{Recovering {\ensuremath{\lambda}}$_{R}$ and V/{\ensuremath{\sigma}} from seeing-dominated IFS data}",
      journal = {\mnras},
         year = 2020,
        month = sep,
       volume = {497},
       number = {2},
        pages = {2018-2038},
          doi = {10.1093/mnras/staa1847},
archivePrefix = {arXiv},
       eprint = {2006.12730},
 primaryClass = {astro-ph.GA},
       adsurl = {https://ui.adsabs.harvard.edu/abs/2020MNRAS.497.2018H}
}

@ARTICLE{2020IAUFM..30A.208L,
       author = {{Lagos}, Claudia del P.},
        title = "{Angular Momentum Evolution of Galaxies: the Perspective of Hydrodynamical Simulations}",
      journal = {IAU Focus Meeting},
         year = 2020,
        month = jan,
       volume = {30},
        pages = {208-214},
          doi = {10.1017/S1743921319004095},
       adsurl = {https://ui.adsabs.harvard.edu/abs/2020IAUFM..30A.208L}
}

@ARTICLE{2019A&A...632A.122M,
       author = {{Mancillas}, Brisa and {Duc}, Pierre-Alain and {Combes}, Fran{\c{c}}oise and {Bournaud}, Fr{\'e}d{\'e}ric and {Emsellem}, Eric and {Martig}, Marie and {Michel-Dansac}, Leo},
        title = "{Probing the merger history of red early-type galaxies with their faint stellar substructures}",
      journal = {\aap},
         year = 2019,
        month = dec,
       volume = {632},
          eid = {A122},
        pages = {A122},
          doi = {10.1051/0004-6361/201936320},
archivePrefix = {arXiv},
       eprint = {1909.07500},
 primaryClass = {astro-ph.GA},
       adsurl = {https://ui.adsabs.harvard.edu/abs/2019A&A...632A.122M}
}

@ARTICLE{2019arXiv191005139G,
       author = {{Graham}, Mark T. and {Cappellari}, Michele and {Bershady}, Matthew A. and {Drory}, Niv},
        title = "{SDSS-IV MaNGA: New benchmark for the connection between stellar angular momentum and environment: a study of about 900 groups/clusters}",
      journal = {arXiv e-prints},
         year = 2019,
        month = oct,
          eid = {arXiv:1910.05139},
        pages = {arXiv:1910.05139},
          doi = {10.48550/arXiv.1910.05139},
archivePrefix = {arXiv},
       eprint = {1910.05139},
 primaryClass = {astro-ph.GA},
       adsurl = {https://ui.adsabs.harvard.edu/abs/2019arXiv191005139G}
}

@ARTICLE{2019MNRAS.487..318K,
       author = {{Karademir}, Geray S. and {Remus}, Rhea-Silvia and {Burkert}, Andreas and {Dolag}, Klaus and {Hoffmann}, Tadziu L. and {Moster}, Benjamin P. and {Steinwandel}, Ulrich P. and {Zhang}, Jielai},
        title = "{The outer stellar halos of galaxies: how radial merger mass deposition, shells, and streams depend on infall-orbit configurations}",
      journal = {\mnras},
         year = 2019,
        month = jul,
       volume = {487},
       number = {1},
        pages = {318-332},
          doi = {10.1093/mnras/stz1251},
archivePrefix = {arXiv},
       eprint = {1808.10454},
 primaryClass = {astro-ph.GA},
       adsurl = {https://ui.adsabs.harvard.edu/abs/2019MNRAS.487..318K}
}

@ARTICLE{2018MNRAS.481.2299S,
       author = {{Scott}, Nicholas and {van de Sande}, Jesse and {Croom}, Scott M. and {Groves}, Brent and {Owers}, Matt S. and {Poetrodjojo}, Henry and {D'Eugenio}, Francesco and {Medling}, Anne M. and {Barat}, Dilyar and {Barone}, Tania M. and {Bland-Hawthorn}, Joss and {Brough}, Sarah and {Bryant}, Julia and {Cortese}, Luca and {Foster}, Caroline and {Green}, Andrew W. and {Oh}, Sree and {Colless}, Matthew and {Drinkwater}, Michael J. and {Driver}, Simon P. and {Goodwin}, Michael and {Gunawardhana}, Madusha L.~P. and {Federrath}, Christoph and {Harischandra}, Lloyd and {Jin}, Yifei and {Lawrence}, J.~S. and {Lorente}, Nuria P. and {Mannering}, Elizabeth and {O'Toole}, Simon and {Richards}, Samuel N. and {Sanchez}, Sebastian F. and {Schaefer}, Adam L. and {Sealey}, Katrina and {Sharp}, Rob and {Sweet}, Sarah M. and {Taranu}, Dan S. and {Varidel}, Mathew},
        title = "{The SAMI Galaxy Survey: Data Release Two with absorption-line physics value-added products}",
      journal = {\mnras},
         year = 2018,
        month = dec,
       volume = {481},
       number = {2},
        pages = {2299-2319},
          doi = {10.1093/mnras/sty2355},
archivePrefix = {arXiv},
       eprint = {1808.03365},
 primaryClass = {astro-ph.GA},
       adsurl = {https://ui.adsabs.harvard.edu/abs/2018MNRAS.481.2299S}
}

@ARTICLE{2018MNRAS.477.4711G,
       author = {{Graham}, Mark T. and {Cappellari}, Michele and {Li}, Hongyu and {Mao}, Shude and {Bershady}, Matthew A. and {Bizyaev}, Dmitry and {Brinkmann}, Jonathan and {Brownstein}, Joel R. and {Bundy}, Kevin and {Drory}, Niv and {Law}, David R. and {Pan}, Kaike and {Thomas}, Daniel and {Wake}, David A. and {Weijmans}, Anne-Marie and {Westfall}, Kyle B. and {Yan}, Renbin},
        title = "{SDSS-IV MaNGA: stellar angular momentum of about 2300 galaxies: unveiling the bimodality of massive galaxy properties}",
      journal = {\mnras},
         year = 2018,
        month = jul,
       volume = {477},
       number = {4},
        pages = {4711-4737},
          doi = {10.1093/mnras/sty504},
archivePrefix = {arXiv},
       eprint = {1802.08213},
 primaryClass = {astro-ph.GA},
       adsurl = {https://ui.adsabs.harvard.edu/abs/2018MNRAS.477.4711G}
}

@ARTICLE{2018ApJ...857..144H,
       author = {{Hood}, Callie E. and {Kannappan}, Sheila J. and {Stark}, David V. and {Dell'Antonio}, Ian P. and {Moffett}, Amanda J. and {Eckert}, Kathleen D. and {Norris}, Mark A. and {Hendel}, David},
        title = "{The Origin of Faint Tidal Features around Galaxies in the RESOLVE Survey}",
      journal = {\apj},
         year = 2018,
        month = apr,
       volume = {857},
       number = {2},
          eid = {144},
        pages = {144},
          doi = {10.3847/1538-4357/aab719},
archivePrefix = {arXiv},
       eprint = {1803.05447},
 primaryClass = {astro-ph.GA},
       adsurl = {https://ui.adsabs.harvard.edu/abs/2018ApJ...857..144H}
}

@ARTICLE{2018NatAs...2..483V,
       author = {{van de Sande}, Jesse and {Scott}, Nicholas and {Bland-Hawthorn}, Joss and {Brough}, Sarah and {Bryant}, Julia J. and {Colless}, Matthew and {Cortese}, Luca and {Croom}, Scott M. and {d'Eugenio}, Francesco and {Foster}, Caroline and {Goodwin}, Michael and {Konstantopoulos}, Iraklis S. and {Lawrence}, Jon S. and {McDermid}, Richard M. and {Medling}, Anne M. and {Owers}, Matt S. and {Richards}, Samuel N. and {Sharp}, Rob},
        title = "{A relation between the characteristic stellar ages of galaxies and their intrinsic shapes}",
      journal = {Nature Astronomy},
         year = 2018,
        month = apr,
       volume = {2},
        pages = {483-488},
          doi = {10.1038/s41550-018-0436-x},
archivePrefix = {arXiv},
       eprint = {1804.07769},
 primaryClass = {astro-ph.GA},
       adsurl = {https://ui.adsabs.harvard.edu/abs/2018NatAs...2..483V}
}

@ARTICLE{2018MNRAS.475..716G,
       author = {{Green}, Andrew W. and {Croom}, Scott M. and {Scott}, Nicholas and {Cortese}, Luca and {Medling}, Anne M. and {D'Eugenio}, Francesco and {Bryant}, Julia J. and {Bland-Hawthorn}, Joss and {Allen}, J.~T. and {Sharp}, Rob and {Ho}, I. -Ting and {Groves}, Brent and {Drinkwater}, Michael J. and {Mannering}, Elizabeth and {Harischandra}, Lloyd and {van de Sande}, Jesse and {Thomas}, Adam D. and {O'Toole}, Simon and {McDermid}, Richard M. and {Vuong}, Minh and {Sealey}, Katrina and {Bauer}, Amanda E. and {Brough}, S. and {Catinella}, Barbara and {Cecil}, Gerald and {Colless}, Matthew and {Couch}, Warrick J. and {Driver}, Simon P. and {Federrath}, Christoph and {Foster}, Caroline and {Goodwin}, Michael and {Hampton}, Elise J. and {Hopkins}, A.~M. and {Jones}, D. Heath and {Konstantopoulos}, Iraklis S. and {Lawrence}, J.~S. and {Leon-Saval}, Sergio G. and {Liske}, Jochen and {L{\'o}pez-S{\'a}nchez}, {\'A}ngel R. and {Lorente}, Nuria P.~F. and {Mould}, Jeremy and {Obreschkow}, Danail and {Owers}, Matt S. and {Richards}, Samuel N. and {Robotham}, Aaron S.~G. and {Schaefer}, Adam L. and {Sweet}, Sarah M. and {Taranu}, Dan S. and {Tescari}, Edoardo and {Tonini}, Chiara and {Zafar}, T.},
        title = "{The SAMI Galaxy Survey: Data Release One with emission-line physics value-added products}",
      journal = {\mnras},
         year = 2018,
        month = mar,
       volume = {475},
       number = {1},
        pages = {716-734},
          doi = {10.1093/mnras/stx3135},
archivePrefix = {arXiv},
       eprint = {1707.08402},
 primaryClass = {astro-ph.GA},
       adsurl = {https://ui.adsabs.harvard.edu/abs/2018MNRAS.475..716G}
}

@ARTICLE{2018MNRAS.473.4956L,
       author = {{Lagos}, Claudia del P. and {Stevens}, Adam R.~H. and {Bower}, Richard G. and {Davis}, Timothy A. and {Contreras}, Sergio and {Padilla}, Nelson D. and {Obreschkow}, Danail and {Croton}, Darren and {Trayford}, James W. and {Welker}, Charlotte and {Theuns}, Tom},
        title = "{Quantifying the impact of mergers on the angular momentum of simulated galaxies}",
      journal = {\mnras},
         year = 2018,
        month = feb,
       volume = {473},
       number = {4},
        pages = {4956-4974},
          doi = {10.1093/mnras/stx2667},
archivePrefix = {arXiv},
       eprint = {1701.04407},
 primaryClass = {astro-ph.GA},
       adsurl = {https://ui.adsabs.harvard.edu/abs/2018MNRAS.473.4956L}
}

@ARTICLE{2018PASJ...70S...4A,
       author = {{Aihara}, Hiroaki and {Arimoto}, Nobuo and {Armstrong}, Robert and {Arnouts}, St{\'e}phane and {Bahcall}, Neta A. and {Bickerton}, Steven and {Bosch}, James and {Bundy}, Kevin and {Capak}, Peter L. and {Chan}, James H.~H. and {Chiba}, Masashi and {Coupon}, Jean and {Egami}, Eiichi and {Enoki}, Motohiro and {Finet}, Francois and {Fujimori}, Hiroki and {Fujimoto}, Seiji and {Furusawa}, Hisanori and {Furusawa}, Junko and {Goto}, Tomotsugu and {Goulding}, Andy and {Greco}, Johnny P. and {Greene}, Jenny E. and {Gunn}, James E. and {Hamana}, Takashi and {Harikane}, Yuichi and {Hashimoto}, Yasuhiro and {Hattori}, Takashi and {Hayashi}, Masao and {Hayashi}, Yusuke and {He{\l}miniak}, Krzysztof G. and {Higuchi}, Ryo and {Hikage}, Chiaki and {Ho}, Paul T.~P. and {Hsieh}, Bau-Ching and {Huang}, Kuiyun and {Huang}, Song and {Ikeda}, Hiroyuki and {Imanishi}, Masatoshi and {Inoue}, Akio K. and {Iwasawa}, Kazushi and {Iwata}, Ikuru and {Jaelani}, Anton T. and {Jian}, Hung-Yu and {Kamata}, Yukiko and {Karoji}, Hiroshi and {Kashikawa}, Nobunari and {Katayama}, Nobuhiko and {Kawanomoto}, Satoshi and {Kayo}, Issha and {Koda}, Jin and {Koike}, Michitaro and {Kojima}, Takashi and {Komiyama}, Yutaka and {Konno}, Akira and {Koshida}, Shintaro and {Koyama}, Yusei and {Kusakabe}, Haruka and {Leauthaud}, Alexie and {Lee}, Chien-Hsiu and {Lin}, Lihwai and {Lin}, Yen-Ting and {Lupton}, Robert H. and {Mandelbaum}, Rachel and {Matsuoka}, Yoshiki and {Medezinski}, Elinor and {Mineo}, Sogo and {Miyama}, Shoken and {Miyatake}, Hironao and {Miyazaki}, Satoshi and {Momose}, Rieko and {More}, Anupreeta and {More}, Surhud and {Moritani}, Yuki and {Moriya}, Takashi J. and {Morokuma}, Tomoki and {Mukae}, Shiro and {Murata}, Ryoma and {Murayama}, Hitoshi and {Nagao}, Tohru and {Nakata}, Fumiaki and {Niida}, Mana and {Niikura}, Hiroko and {Nishizawa}, Atsushi J. and {Obuchi}, Yoshiyuki and {Oguri}, Masamune and {Oishi}, Yukie and {Okabe}, Nobuhiro and {Okamoto}, Sakurako and {Okura}, Yuki and {Ono}, Yoshiaki and {Onodera}, Masato and {Onoue}, Masafusa and {Osato}, Ken and {Ouchi}, Masami and {Price}, Paul A. and {Pyo}, Tae-Soo and {Sako}, Masao and {Sawicki}, Marcin and {Shibuya}, Takatoshi and {Shimasaku}, Kazuhiro and {Shimono}, Atsushi and {Shirasaki}, Masato and {Silverman}, John D. and {Simet}, Melanie and {Speagle}, Joshua and {Spergel}, David N. and {Strauss}, Michael A. and {Sugahara}, Yuma and {Sugiyama}, Naoshi and {Suto}, Yasushi and {Suyu}, Sherry H. and {Suzuki}, Nao and {Tait}, Philip J. and {Takada}, Masahiro and {Takata}, Tadafumi and {Tamura}, Naoyuki and {Tanaka}, Manobu M. and {Tanaka}, Masaomi and {Tanaka}, Masayuki and {Tanaka}, Yoko and {Terai}, Tsuyoshi and {Terashima}, Yuichi and {Toba}, Yoshiki and {Tominaga}, Nozomu and {Toshikawa}, Jun and {Turner}, Edwin L. and {Uchida}, Tomohisa and {Uchiyama}, Hisakazu and {Umetsu}, Keiichi and {Uraguchi}, Fumihiro and {Urata}, Yuji and {Usuda}, Tomonori and {Utsumi}, Yousuke and {Wang}, Shiang-Yu and {Wang}, Wei-Hao and {Wong}, Kenneth C. and {Yabe}, Kiyoto and {Yamada}, Yoshihiko and {Yamanoi}, Hitomi and {Yasuda}, Naoki and {Yeh}, Sherry and {Yonehara}, Atsunori and {Yuma}, Suraphong},
        title = "{The Hyper Suprime-Cam SSP Survey: Overview and survey design}",
      journal = {\pasj},
         year = 2018,
        month = jan,
       volume = {70},
          eid = {S4},
        pages = {S4},
          doi = {10.1093/pasj/psx066},
archivePrefix = {arXiv},
       eprint = {1704.05858},
 primaryClass = {astro-ph.IM},
       adsurl = {https://ui.adsabs.harvard.edu/abs/2018PASJ...70S...4A}
}

@ARTICLE{2018PASJ...70S...8A,
       author = {{Aihara}, Hiroaki and {Armstrong}, Robert and {Bickerton}, Steven and {Bosch}, James and {Coupon}, Jean and {Furusawa}, Hisanori and {Hayashi}, Yusuke and {Ikeda}, Hiroyuki and {Kamata}, Yukiko and {Karoji}, Hiroshi and {Kawanomoto}, Satoshi and {Koike}, Michitaro and {Komiyama}, Yutaka and {Lang}, Dustin and {Lupton}, Robert H. and {Mineo}, Sogo and {Miyatake}, Hironao and {Miyazaki}, Satoshi and {Morokuma}, Tomoki and {Obuchi}, Yoshiyuki and {Oishi}, Yukie and {Okura}, Yuki and {Price}, Paul A. and {Takata}, Tadafumi and {Tanaka}, Manobu M. and {Tanaka}, Masayuki and {Tanaka}, Yoko and {Uchida}, Tomohisa and {Uraguchi}, Fumihiro and {Utsumi}, Yousuke and {Wang}, Shiang-Yu and {Yamada}, Yoshihiko and {Yamanoi}, Hitomi and {Yasuda}, Naoki and {Arimoto}, Nobuo and {Chiba}, Masashi and {Finet}, Francois and {Fujimori}, Hiroki and {Fujimoto}, Seiji and {Furusawa}, Junko and {Goto}, Tomotsugu and {Goulding}, Andy and {Gunn}, James E. and {Harikane}, Yuichi and {Hattori}, Takashi and {Hayashi}, Masao and {He{\l}miniak}, Krzysztof G. and {Higuchi}, Ryo and {Hikage}, Chiaki and {Ho}, Paul T.~P. and {Hsieh}, Bau-Ching and {Huang}, Kuiyun and {Huang}, Song and {Imanishi}, Masatoshi and {Iwata}, Ikuru and {Jaelani}, Anton T. and {Jian}, Hung-Yu and {Kashikawa}, Nobunari and {Katayama}, Nobuhiko and {Kojima}, Takashi and {Konno}, Akira and {Koshida}, Shintaro and {Kusakabe}, Haruka and {Leauthaud}, Alexie and {Lee}, Chien-Hsiu and {Lin}, Lihwai and {Lin}, Yen-Ting and {Mandelbaum}, Rachel and {Matsuoka}, Yoshiki and {Medezinski}, Elinor and {Miyama}, Shoken and {Momose}, Rieko and {More}, Anupreeta and {More}, Surhud and {Mukae}, Shiro and {Murata}, Ryoma and {Murayama}, Hitoshi and {Nagao}, Tohru and {Nakata}, Fumiaki and {Niida}, Mana and {Niikura}, Hiroko and {Nishizawa}, Atsushi J. and {Oguri}, Masamune and {Okabe}, Nobuhiro and {Ono}, Yoshiaki and {Onodera}, Masato and {Onoue}, Masafusa and {Ouchi}, Masami and {Pyo}, Tae-Soo and {Shibuya}, Takatoshi and {Shimasaku}, Kazuhiro and {Simet}, Melanie and {Speagle}, Joshua and {Spergel}, David N. and {Strauss}, Michael A. and {Sugahara}, Yuma and {Sugiyama}, Naoshi and {Suto}, Yasushi and {Suzuki}, Nao and {Tait}, Philip J. and {Takada}, Masahiro and {Terai}, Tsuyoshi and {Toba}, Yoshiki and {Turner}, Edwin L. and {Uchiyama}, Hisakazu and {Umetsu}, Keiichi and {Urata}, Yuji and {Usuda}, Tomonori and {Yeh}, Sherry and {Yuma}, Suraphong},
        title = "{First data release of the Hyper Suprime-Cam Subaru Strategic Program}",
      journal = {\pasj},
         year = 2018,
        month = jan,
       volume = {70},
          eid = {S8},
        pages = {S8},
          doi = {10.1093/pasj/psx081},
archivePrefix = {arXiv},
       eprint = {1702.08449},
 primaryClass = {astro-ph.IM},
       adsurl = {https://ui.adsabs.harvard.edu/abs/2018PASJ...70S...8A}
}

@ARTICLE{2017MNRAS.472.1272V,
       author = {{van de Sande}, Jesse and {Bland-Hawthorn}, Joss and {Brough}, Sarah and {Croom}, Scott M. and {Cortese}, Luca and {Foster}, Caroline and {Scott}, Nicholas and {Bryant}, Julia J. and {d'Eugenio}, Francesco and {Tonini}, Chiara and {Goodwin}, Michael and {Konstantopoulos}, Iraklis S. and {Lawrence}, Jon S. and {Medling}, Anne M. and {Owers}, Matt S. and {Richards}, Samuel N. and {Schaefer}, Adam L. and {Yi}, Sukyoung K.},
        title = "{The SAMI Galaxy Survey: revising the fraction of slow rotators in IFS galaxy surveys}",
      journal = {\mnras},
         year = 2017,
        month = dec,
       volume = {472},
       number = {2},
        pages = {1272-1285},
          doi = {10.1093/mnras/stx1751},
archivePrefix = {arXiv},
       eprint = {1707.03402},
 primaryClass = {astro-ph.GA},
       adsurl = {https://ui.adsabs.harvard.edu/abs/2017MNRAS.472.1272V}
}

@ARTICLE{2017ApJ...851L..33G,
       author = {{Greene}, J.~E. and {Leauthaud}, A. and {Emsellem}, E. and {Goddard}, D. and {Ge}, J. and {Andrews}, B.~H. and {Brinkman}, J. and {Brownstein}, J.~R. and {Greco}, J. and {Law}, D. and {Lin}, Y. -T. and {Masters}, K.~L. and {Merrifield}, M. and {More}, S. and {Okabe}, N. and {Schneider}, D.~P. and {Thomas}, D. and {Wake}, D.~A. and {Yan}, R. and {Drory}, N.},
        title = "{SDSS-IV MaNGA: Probing the Kinematic Morphology-Density Relation of Early-type Galaxies with MaNGA}",
      journal = {\apjl},
         year = 2017,
        month = dec,
       volume = {851},
       number = {2},
          eid = {L33},
        pages = {L33},
          doi = {10.3847/2041-8213/aa8ace},
archivePrefix = {arXiv},
       eprint = {1708.07843},
 primaryClass = {astro-ph.GA},
       adsurl = {https://ui.adsabs.harvard.edu/abs/2017ApJ...851L..33G}
}

@ARTICLE{2017MNRAS.471.1428V,
       author = {{Veale}, Melanie and {Ma}, Chung-Pei and {Greene}, Jenny E. and {Thomas}, Jens and {Blakeslee}, John P. and {McConnell}, Nicholas and {Walsh}, Jonelle L. and {Ito}, Jennifer},
        title = "{The MASSIVE Survey - VII. The relationship of angular momentum, stellar mass and environment of early-type galaxies}",
      journal = {\mnras},
         year = 2017,
        month = oct,
       volume = {471},
       number = {2},
        pages = {1428-1445},
          doi = {10.1093/mnras/stx1639},
archivePrefix = {arXiv},
       eprint = {1703.08573},
 primaryClass = {astro-ph.GA},
       adsurl = {https://ui.adsabs.harvard.edu/abs/2017MNRAS.471.1428V}
}

@ARTICLE{2017MNRAS.468.3883P,
       author = {{Penoyre}, Zephyr and {Moster}, Benjamin P. and {Sijacki}, Debora and {Genel}, Shy},
        title = "{The origin and evolution of fast and slow rotators in the Illustris simulation}",
      journal = {\mnras},
         year = 2017,
        month = jul,
       volume = {468},
       number = {4},
        pages = {3883-3906},
          doi = {10.1093/mnras/stx762},
archivePrefix = {arXiv},
       eprint = {1703.00545},
 primaryClass = {astro-ph.GA},
       adsurl = {https://ui.adsabs.harvard.edu/abs/2017MNRAS.468.3883P}
}

@ARTICLE{2017ApJ...844...59B,
       author = {{Brough}, Sarah and {van de Sande}, Jesse and {Owers}, Matt S. and {d'Eugenio}, Francesco and {Sharp}, Rob and {Cortese}, Luca and {Scott}, Nicholas and {Croom}, Scott M. and {Bassett}, Rob and {Bekki}, Kenji and {Bland-Hawthorn}, Joss and {Bryant}, Julia J. and {Davies}, Roger and {Drinkwater}, Michael J. and {Driver}, Simon P. and {Foster}, Caroline and {Goldstein}, Gregory and {L{\'o}pez-S{\'a}nchez}, {\'A}. R. and {Medling}, Anne M. and {Sweet}, Sarah M. and {Taranu}, Dan S. and {Tonini}, Chiara and {Yi}, Sukyoung K. and {Goodwin}, Michael and {Lawrence}, J.~S. and {Richards}, Samuel N.},
        title = "{The SAMI Galaxy Survey: Mass as the Driver of the Kinematic Morphology-Density Relation in Clusters}",
      journal = {\apj},
         year = 2017,
        month = jul,
       volume = {844},
       number = {1},
          eid = {59},
        pages = {59},
          doi = {10.3847/1538-4357/aa7a11},
archivePrefix = {arXiv},
       eprint = {1704.01169},
 primaryClass = {astro-ph.GA},
       adsurl = {https://ui.adsabs.harvard.edu/abs/2017ApJ...844...59B}
}

@ARTICLE{2017MNRAS.468.1824O,
       author = {{Owers}, M.~S. and {Allen}, J.~T. and {Baldry}, I. and {Bryant}, J.~J. and {Cecil}, G.~N. and {Cortese}, L. and {Croom}, S.~M. and {Driver}, S.~P. and {Fogarty}, L.~M.~R. and {Green}, A.~W. and {Helmich}, E. and {de Jong}, J.~T.~A. and {Kuijken}, K. and {Mahajan}, S. and {McFarland}, J. and {Pracy}, M.~B. and {Robotham}, A.~G.~S. and {Sikkema}, G. and {Sweet}, S. and {Taylor}, E.~N. and {Verdoes Kleijn}, G. and {Bauer}, A.~E. and {Bland-Hawthorn}, J. and {Brough}, S. and {Colless}, M. and {Couch}, W.~J. and {Davies}, R.~L. and {Drinkwater}, M.~J. and {Goodwin}, M. and {Hopkins}, A.~M. and {Konstantopoulos}, I.~S. and {Foster}, C. and {Lawrence}, J.~S. and {Lorente}, N.~P.~F. and {Medling}, A.~M. and {Metcalfe}, N. and {Richards}, S.~N. and {van de Sande}, J. and {Scott}, N. and {Shanks}, T. and {Sharp}, R. and {Thomas}, A.~D. and {Tonini}, C.},
        title = "{The SAMI Galaxy Survey: the cluster redshift survey, target selection and cluster properties}",
      journal = {\mnras},
         year = 2017,
        month = jun,
       volume = {468},
       number = {2},
        pages = {1824-1849},
          doi = {10.1093/mnras/stx562},
archivePrefix = {arXiv},
       eprint = {1703.00997},
 primaryClass = {astro-ph.GA},
       adsurl = {https://ui.adsabs.harvard.edu/abs/2017MNRAS.468.1824O}
}

@ARTICLE{2017MNRAS.465.2895L,
       author = {{Lofthouse}, E.~K. and {Kaviraj}, S. and {Conselice}, C.~J. and {Mortlock}, A. and {Hartley}, W.},
        title = "{Major mergers are not significant drivers of star formation or morphological transformation around the epoch of peak cosmic star formation}",
      journal = {\mnras},
         year = 2017,
        month = mar,
       volume = {465},
       number = {3},
        pages = {2895-2900},
          doi = {10.1093/mnras/stw2895},
archivePrefix = {arXiv},
       eprint = {1608.03892},
 primaryClass = {astro-ph.GA},
       adsurl = {https://ui.adsabs.harvard.edu/abs/2017MNRAS.465.2895L}
}

@ARTICLE{2016MNRAS.463..170C,
       author = {{Cortese}, L. and {Fogarty}, L.~M.~R. and {Bekki}, K. and {van de Sande}, J. and {Couch}, W. and {Catinella}, B. and {Colless}, M. and {Obreschkow}, D. and {Taranu}, D. and {Tescari}, E. and {Barat}, D. and {Bland-Hawthorn}, J. and {Bloom}, J. and {Bryant}, J.~J. and {Cluver}, M. and {Croom}, S.~M. and {Drinkwater}, M.~J. and {d'Eugenio}, F. and {Konstantopoulos}, I.~S. and {Lopez-Sanchez}, A. and {Mahajan}, S. and {Scott}, N. and {Tonini}, C. and {Wong}, O.~I. and {Allen}, J.~T. and {Brough}, S. and {Goodwin}, M. and {Green}, A.~W. and {Ho}, I. -T. and {Kelvin}, L.~S. and {Lawrence}, J.~S. and {Lorente}, N.~P.~F. and {Medling}, A.~M. and {Owers}, M.~S. and {Richards}, S. and {Sharp}, R. and {Sweet}, S.~M.},
        title = "{The SAMI Galaxy Survey: the link between angular momentum and optical morphology}",
      journal = {\mnras},
         year = 2016,
        month = nov,
       volume = {463},
       number = {1},
        pages = {170-184},
          doi = {10.1093/mnras/stw1891},
archivePrefix = {arXiv},
       eprint = {1608.00291},
 primaryClass = {astro-ph.GA},
       adsurl = {https://ui.adsabs.harvard.edu/abs/2016MNRAS.463..170C}
}

@ARTICLE{2015MNRAS.454.1742K,
       author = {{Knapen}, Johan H. and {Cisternas}, Mauricio and {Querejeta}, Miguel},
        title = "{Interacting galaxies in the nearby Universe: only moderate increase of star formation}",
      journal = {\mnras},
         year = 2015,
        month = dec,
       volume = {454},
       number = {2},
        pages = {1742-1750},
          doi = {10.1093/mnras/stv2135},
archivePrefix = {arXiv},
       eprint = {1509.05164},
 primaryClass = {astro-ph.GA},
       adsurl = {https://ui.adsabs.harvard.edu/abs/2015MNRAS.454.1742K}
}

@ARTICLE{2015ApJ...807...22M,
       author = {{Miyazaki}, Satoshi and {Oguri}, Masamune and {Hamana}, Takashi and {Tanaka}, Masayuki and {Miller}, Lance and {Utsumi}, Yousuke and {Komiyama}, Yutaka and {Furusawa}, Hisanori and {Sakurai}, Junya and {Kawanomoto}, Satoshi and {Nakata}, Fumiaki and {Uraguchi}, Fumihiro and {Koike}, Michitaro and {Tomono}, Daigo and {Lupton}, Robert and {Gunn}, James E. and {Karoji}, Hiroshi and {Aihara}, Hiroaki and {Murayama}, Hitoshi and {Takada}, Masahiro},
        title = "{Properties of Weak Lensing Clusters Detected on Hyper Suprime-Cam{\textquoteright}s 2.3 deg$^{2}$ field}",
      journal = {\apj},
         year = 2015,
        month = jul,
       volume = {807},
       number = {1},
          eid = {22},
        pages = {22},
          doi = {10.1088/0004-637X/807/1/22},
archivePrefix = {arXiv},
       eprint = {1504.06974},
 primaryClass = {astro-ph.CO},
       adsurl = {https://ui.adsabs.harvard.edu/abs/2015ApJ...807...22M}
}

@ARTICLE{2015MNRAS.449.1177V,
       author = {{Vazdekis}, A. and {Coelho}, P. and {Cassisi}, S. and {Ricciardelli}, E. and {Falc{\'o}n-Barroso}, J. and {S{\'a}nchez-Bl{\'a}zquez}, P. and {La Barbera}, F. and {Beasley}, M.~A. and {Pietrinferni}, A.},
        title = "{Evolutionary stellar population synthesis with MILES - II. Scaled-solar and {\ensuremath{\alpha}}-enhanced models}",
      journal = {\mnras},
         year = 2015,
        month = may,
       volume = {449},
       number = {2},
        pages = {1177-1214},
          doi = {10.1093/mnras/stv151},
archivePrefix = {arXiv},
       eprint = {1504.08032},
 primaryClass = {astro-ph.GA},
       adsurl = {https://ui.adsabs.harvard.edu/abs/2015MNRAS.449.1177V}
}

@ARTICLE{2015MNRAS.447.2857B,
       author = {{Bryant}, J.~J. and {Owers}, M.~S. and {Robotham}, A.~S.~G. and {Croom}, S.~M. and {Driver}, S.~P. and {Drinkwater}, M.~J. and {Lorente}, N.~P.~F. and {Cortese}, L. and {Scott}, N. and {Colless}, M. and {Schaefer}, A. and {Taylor}, E.~N. and {Konstantopoulos}, I.~S. and {Allen}, J.~T. and {Baldry}, I. and {Barnes}, L. and {Bauer}, A.~E. and {Bland-Hawthorn}, J. and {Bloom}, J.~V. and {Brooks}, A.~M. and {Brough}, S. and {Cecil}, G. and {Couch}, W. and {Croton}, D. and {Davies}, R. and {Ellis}, S. and {Fogarty}, L.~M.~R. and {Foster}, C. and {Glazebrook}, K. and {Goodwin}, M. and {Green}, A. and {Gunawardhana}, M.~L. and {Hampton}, E. and {Ho}, I. -T. and {Hopkins}, A.~M. and {Kewley}, L. and {Lawrence}, J.~S. and {Leon-Saval}, S.~G. and {Leslie}, S. and {McElroy}, R. and {Lewis}, G. and {Liske}, J. and {L{\'o}pez-S{\'a}nchez}, {\'A}. R. and {Mahajan}, S. and {Medling}, A.~M. and {Metcalfe}, N. and {Meyer}, M. and {Mould}, J. and {Obreschkow}, D. and {O'Toole}, S. and {Pracy}, M. and {Richards}, S.~N. and {Shanks}, T. and {Sharp}, R. and {Sweet}, S.~M. and {Thomas}, A.~D. and {Tonini}, C. and {Walcher}, C.~J.},
        title = "{The SAMI Galaxy Survey: instrument specification and target selection}",
      journal = {\mnras},
         year = 2015,
        month = mar,
       volume = {447},
       number = {3},
        pages = {2857-2879},
          doi = {10.1093/mnras/stu2635},
archivePrefix = {arXiv},
       eprint = {1407.7335},
 primaryClass = {astro-ph.GA},
       adsurl = {https://ui.adsabs.harvard.edu/abs/2015MNRAS.447.2857B}
}

@ARTICLE{2015MNRAS.446.1551S,
       author = {{Sharp}, R. and {Allen}, J.~T. and {Fogarty}, L.~M.~R. and {Croom}, S.~M. and {Cortese}, L. and {Green}, A.~W. and {Nielsen}, J. and {Richards}, S.~N. and {Scott}, N. and {Taylor}, E.~N. and {Barnes}, L.~A. and {Bauer}, A.~E. and {Birchall}, M. and {Bland-Hawthorn}, J. and {Bloom}, J.~V. and {Brough}, S. and {Bryant}, J.~J. and {Cecil}, G.~N. and {Colless}, M. and {Couch}, W.~J. and {Drinkwater}, M.~J. and {Driver}, S. and {Foster}, C. and {Goodwin}, M. and {Gunawardhana}, M.~L.~P. and {Ho}, I. -T. and {Hampton}, E.~J. and {Hopkins}, A.~M. and {Jones}, H. and {Konstantopoulos}, I.~S. and {Lawrence}, J.~S. and {Leslie}, S.~K. and {Lewis}, G.~F. and {Liske}, J. and {L{\'o}pez-S{\'a}nchez}, {\'A}. R. and {Lorente}, N.~P.~F. and {McElroy}, R. and {Medling}, A.~M. and {Mahajan}, S. and {Mould}, J. and {Parker}, Q. and {Pracy}, M.~B. and {Obreschkow}, D. and {Owers}, M.~S. and {Schaefer}, A.~L. and {Sweet}, S.~M. and {Thomas}, A.~D. and {Tonini}, C. and {Walcher}, C.~J.},
        title = "{The SAMI Galaxy Survey: cubism and covariance, putting round pegs into square holes}",
      journal = {\mnras},
         year = 2015,
        month = jan,
       volume = {446},
       number = {2},
        pages = {1551-1566},
          doi = {10.1093/mnras/stu2055},
archivePrefix = {arXiv},
       eprint = {1407.5237},
 primaryClass = {astro-ph.IM},
       adsurl = {https://ui.adsabs.harvard.edu/abs/2015MNRAS.446.1551S}
}

@ARTICLE{2015MNRAS.446.1567A,
       author = {{Allen}, J.~T. and {Croom}, S.~M. and {Konstantopoulos}, I.~S. and {Bryant}, J.~J. and {Sharp}, R. and {Cecil}, G.~N. and {Fogarty}, L.~M.~R. and {Foster}, C. and {Green}, A.~W. and {Ho}, I. -T. and {Owers}, M.~S. and {Schaefer}, A.~L. and {Scott}, N. and {Bauer}, A.~E. and {Baldry}, I. and {Barnes}, L.~A. and {Bland-Hawthorn}, J. and {Bloom}, J.~V. and {Brough}, S. and {Colless}, M. and {Cortese}, L. and {Couch}, W.~J. and {Drinkwater}, M.~J. and {Driver}, S.~P. and {Goodwin}, M. and {Gunawardhana}, M.~L.~P. and {Hampton}, E.~J. and {Hopkins}, A.~M. and {Kewley}, L.~J. and {Lawrence}, J.~S. and {Leon-Saval}, S.~G. and {Liske}, J. and {L{\'o}pez-S{\'a}nchez}, {\'A}. R. and {Lorente}, N.~P.~F. and {McElroy}, R. and {Medling}, A.~M. and {Mould}, J. and {Norberg}, P. and {Parker}, Q.~A. and {Power}, C. and {Pracy}, M.~B. and {Richards}, S.~N. and {Robotham}, A.~S.~G. and {Sweet}, S.~M. and {Taylor}, E.~N. and {Thomas}, A.~D. and {Tonini}, C. and {Walcher}, C.~J.},
        title = "{The SAMI Galaxy Survey: Early Data Release}",
      journal = {\mnras},
         year = 2015,
        month = jan,
       volume = {446},
       number = {2},
        pages = {1567-1583},
          doi = {10.1093/mnras/stu2057},
archivePrefix = {arXiv},
       eprint = {1407.6068},
 primaryClass = {astro-ph.GA},
       adsurl = {https://ui.adsabs.harvard.edu/abs/2015MNRAS.446.1567A}
}

@ARTICLE{2014MNRAS.443..485F,
       author = {{Fogarty}, L.~M.~R. and {Scott}, Nicholas and {Owers}, Matt S. and {Brough}, S. and {Croom}, Scott M. and {Pracy}, Michael B. and {Houghton}, R.~C.~W. and {Bland-Hawthorn}, Joss and {Colless}, Matthew and {Davies}, Roger L. and {Jones}, D. Heath and {Allen}, J.~T. and {Bryant}, Julia J. and {Goodwin}, Michael and {Green}, Andrew W. and {Konstantopoulos}, Iraklis S. and {Lawrence}, J.~S. and {Richards}, Samuel and {Cortese}, Luca and {Sharp}, Rob},
        title = "{The SAMI Pilot Survey: the kinematic morphology-density relation in Abell 85, Abell 168 and Abell 2399}",
      journal = {\mnras},
         year = 2014,
        month = sep,
       volume = {443},
       number = {1},
        pages = {485-503},
          doi = {10.1093/mnras/stu1165},
archivePrefix = {arXiv},
       eprint = {1406.3899},
 primaryClass = {astro-ph.GA},
       adsurl = {https://ui.adsabs.harvard.edu/abs/2014MNRAS.443..485F}
}

@ARTICLE{2014MNRAS.439.1245K,
       author = {{Kelvin}, Lee S. and {Driver}, Simon P. and {Robotham}, Aaron S.~G. and {Graham}, Alister W. and {Phillipps}, Steven and {Agius}, Nicola K. and {Alpaslan}, Mehmet and {Baldry}, Ivan and {Bamford}, Steven P. and {Bland-Hawthorn}, Joss and {Brough}, Sarah and {Brown}, Michael J.~I. and {Colless}, Matthew and {Conselice}, Christopher J. and {Hopkins}, Andrew M. and {Liske}, Jochen and {Loveday}, Jon and {Norberg}, Peder and {Pimbblet}, Kevin A. and {Popescu}, Cristina C. and {Prescott}, Matthew and {Taylor}, Edward N. and {Tuffs}, Richard J.},
        title = "{Galaxy And Mass Assembly (GAMA): ugrizYJHK S{\'e}rsic luminosity functions and the cosmic spectral energy distribution by Hubble type}",
      journal = {\mnras},
         year = 2014,
        month = apr,
       volume = {439},
       number = {2},
        pages = {1245-1269},
          doi = {10.1093/mnras/stt2391},
archivePrefix = {arXiv},
       eprint = {1401.1817},
 primaryClass = {astro-ph.CO},
       adsurl = {https://ui.adsabs.harvard.edu/abs/2014MNRAS.439.1245K}
}

@ARTICLE{2014MNRAS.438..869B,
       author = {{Bryant}, J.~J. and {Bland-Hawthorn}, J. and {Fogarty}, L.~M.~R. and {Lawrence}, J.~S. and {Croom}, S.~M.},
        title = "{Focal ratio degradation in lightly fused hexabundles}",
      journal = {\mnras},
         year = 2014,
        month = feb,
       volume = {438},
       number = {1},
        pages = {869-877},
          doi = {10.1093/mnras/stt2254},
archivePrefix = {arXiv},
       eprint = {1311.6865},
 primaryClass = {astro-ph.IM},
       adsurl = {https://ui.adsabs.harvard.edu/abs/2014MNRAS.438..869B}
}

@ARTICLE{2014MNRAS.437.2137S,
       author = {{Scott}, Caroline and {Kaviraj}, Sugata},
        title = "{Star formation and AGN activity in interacting galaxies: a near-UV perspective}",
      journal = {\mnras},
         year = 2014,
        month = jan,
       volume = {437},
       number = {3},
        pages = {2137-2145},
          doi = {10.1093/mnras/stt2014},
archivePrefix = {arXiv},
       eprint = {1310.5148},
 primaryClass = {astro-ph.GA},
       adsurl = {https://ui.adsabs.harvard.edu/abs/2014MNRAS.437.2137S}
}

@ARTICLE{2013MNRAS.436...19H,
       author = {{Houghton}, R.~C.~W. and {Davies}, Roger L. and {D'Eugenio}, F. and {Scott}, N. and {Thatte}, N. and {Clarke}, F. and {Tecza}, M. and {Salter}, G.~S. and {Fogarty}, L.~M.~R. and {Goodsall}, T.},
        title = "{Fast and slow rotators in the densest environments: a SWIFT IFS study of the Coma cluster}",
      journal = {\mnras},
         year = 2013,
        month = nov,
       volume = {436},
       number = {1},
        pages = {19-33},
          doi = {10.1093/mnras/stt1399},
archivePrefix = {arXiv},
       eprint = {1308.6581},
 primaryClass = {astro-ph.CO},
       adsurl = {https://ui.adsabs.harvard.edu/abs/2013MNRAS.436...19H}
}

@ARTICLE{2013MNRAS.429.1258D,
       author = {{D'Eugenio}, F. and {Houghton}, R.~C.~W. and {Davies}, R.~L. and {Dalla Bont{\`a}}, E.},
        title = "{Fast and slow rotators in the densest environments: a FLAMES/GIRAFFE integral field spectroscopy study of galaxies in A1689 at z = 0.183}",
      journal = {\mnras},
         year = 2013,
        month = feb,
       volume = {429},
       number = {2},
        pages = {1258-1266},
          doi = {10.1093/mnras/sts406},
archivePrefix = {arXiv},
       eprint = {1205.5545},
 primaryClass = {astro-ph.CO},
       adsurl = {https://ui.adsabs.harvard.edu/abs/2013MNRAS.429.1258D}
}

@ARTICLE{2012ApJS..203...21A,
       author = {{Ahn}, Christopher P. and {Alexandroff}, Rachael and {Allende Prieto}, Carlos and {Anderson}, Scott F. and {Anderton}, Timothy and {Andrews}, Brett H. and {Aubourg}, {\'E}ric and {Bailey}, Stephen and {Balbinot}, Eduardo and {Barnes}, Rory and {Bautista}, Julian and {Beers}, Timothy C. and {Beifiori}, Alessandra and {Berlind}, Andreas A. and {Bhardwaj}, Vaishali and {Bizyaev}, Dmitry and {Blake}, Cullen H. and {Blanton}, Michael R. and {Blomqvist}, Michael and {Bochanski}, John J. and {Bolton}, Adam S. and {Borde}, Arnaud and {Bovy}, Jo and {Brandt}, W.~N. and {Brinkmann}, J. and {Brown}, Peter J. and {Brownstein}, Joel R. and {Bundy}, Kevin and {Busca}, N.~G. and {Carithers}, William and {Carnero}, Aurelio R. and {Carr}, Michael A. and {Casetti-Dinescu}, Dana I. and {Chen}, Yanmei and {Chiappini}, Cristina and {Comparat}, Johan and {Connolly}, Natalia and {Crepp}, Justin R. and {Cristiani}, Stefano and {Croft}, Rupert A.~C. and {Cuesta}, Antonio J. and {da Costa}, Luiz N. and {Davenport}, James R.~A. and {Dawson}, Kyle S. and {de Putter}, Roland and {De Lee}, Nathan and {Delubac}, Timoth{\'e}e and {Dhital}, Saurav and {Ealet}, Anne and {Ebelke}, Garrett L. and {Edmondson}, Edward M. and {Eisenstein}, Daniel J. and {Escoffier}, S. and {Esposito}, Massimiliano and {Evans}, Michael L. and {Fan}, Xiaohui and {Femen{\'\i}a Castell{\'a}}, Bruno and {Fern{\'a}ndez Alvar}, Emma and {Ferreira}, Leticia D. and {Filiz Ak}, N. and {Finley}, Hayley and {Fleming}, Scott W. and {Font-Ribera}, Andreu and {Frinchaboy}, Peter M. and {Garc{\'\i}a-Hern{\'a}ndez}, D.~A. and {Garc{\'\i}a P{\'e}rez}, A.~E. and {Ge}, Jian and {G{\'e}nova-Santos}, R. and {Gillespie}, Bruce A. and {Girardi}, L{\'e}o and {Gonz{\'a}lez Hern{\'a}ndez}, Jonay I. and {Grebel}, Eva K. and {Gunn}, James E. and {Guo}, Hong and {Haggard}, Daryl and {Hamilton}, Jean-Christophe and {Harris}, David W. and {Hawley}, Suzanne L. and {Hearty}, Frederick R. and {Ho}, Shirley and {Hogg}, David W. and {Holtzman}, Jon A. and {Honscheid}, Klaus and {Huehnerhoff}, J. and {Ivans}, Inese I. and {Ivezi{\'c}}, {\v{Z}}eljko and {Jacobson}, Heather R. and {Jiang}, Linhua and {Johansson}, Jonas and {Johnson}, Jennifer A. and {Kauffmann}, Guinevere and {Kirkby}, David and {Kirkpatrick}, Jessica A. and {Klaene}, Mark A. and {Knapp}, Gillian R. and {Kneib}, Jean-Paul and {Le Goff}, Jean-Marc and {Leauthaud}, Alexie and {Lee}, Khee-Gan and {Lee}, Young Sun and {Long}, Daniel C. and {Loomis}, Craig P. and {Lucatello}, Sara and {Lundgren}, Britt and {Lupton}, Robert H. and {Ma}, Bo and {Ma}, Zhibo and {MacDonald}, Nicholas and {Mack}, Claude E. and {Mahadevan}, Suvrath and {Maia}, Marcio A.~G. and {Majewski}, Steven R. and {Makler}, Martin and {Malanushenko}, Elena and {Malanushenko}, Viktor and {Manchado}, A. and {Mandelbaum}, Rachel and {Manera}, Marc and {Maraston}, Claudia and {Margala}, Daniel and {Martell}, Sarah L. and {McBride}, Cameron K. and {McGreer}, Ian D. and {McMahon}, Richard G. and {M{\'e}nard}, Brice and {Meszaros}, Sz. and {Miralda-Escud{\'e}}, Jordi and {Montero-Dorta}, Antonio D. and {Montesano}, Francesco and {Morrison}, Heather L. and {Muna}, Demitri and {Munn}, Jeffrey A. and {Murayama}, Hitoshi and {Myers}, Adam D. and {Neto}, A.~F. and {Nguyen}, Duy Cuong and {Nichol}, Robert C. and {Nidever}, David L. and {Noterdaeme}, Pasquier and {Nuza}, Sebasti{\'a}n E. and {Ogando}, Ricardo L.~C. and {Olmstead}, Matthew D. and {Oravetz}, Daniel J. and {Owen}, Russell and {Padmanabhan}, Nikhil and {Palanque-Delabrouille}, Nathalie and {Pan}, Kaike and {Parejko}, John K. and {Parihar}, Prachi and {P{\^a}ris}, Isabelle and {Pattarakijwanich}, Petchara and {Pepper}, Joshua and {Percival}, Will J. and {P{\'e}rez-Fournon}, Ismael and {P{\'e}rez-R{\`a}fols}, Ignasi and {Petitjean}, Patrick and {Pforr}, Janine and {Pieri}, Matthew M. and {Pinsonneault}, Marc H. and {Porto de Mello}, G.~F. and {Prada}, Francisco and {Price-Whelan}, Adrian M. and {Raddick}, M. Jordan and {Rebolo}, Rafael and {Rich}, James and {Richards}, Gordon T. and {Robin}, Annie C. and {Rocha-Pinto}, Helio J. and {Rockosi}, Constance M. and {Roe}, Natalie A. and {Ross}, Ashley J. and {Ross}, Nicholas P. and {Rossi}, Graziano and {Rubi{\~n}o-Martin}, J.~A. and {Samushia}, Lado and {Sanchez Almeida}, J. and {S{\'a}nchez}, Ariel G. and {Santiago}, Bas{\'\i}lio and {Sayres}, Conor and {Schlegel}, David J. and {Schlesinger}, Katharine J. and {Schmidt}, Sarah J. and {Schneider}, Donald P. and {Schultheis}, Mathias and {Schwope}, Axel D. and {Sc{\'o}ccola}, C.~G. and {Seljak}, Uros and {Sheldon}, Erin and {Shen}, Yue and {Shu}, Yiping and {Simmerer}, Jennifer and {Simmons}, Audrey E. and {Skibba}, Ramin A. and {Skrutskie}, M.~F. and {Slosar}, A. and {Sobreira}, Flavia and {Sobeck}, Jennifer S. and {Stassun}, Keivan G. and {Steele}, Oliver and {Steinmetz}, Matthias and {Strauss}, Michael A. and {Streblyanska}, Alina and {Suzuki}, Nao and {Swanson}, Molly E.~C. and {Tal}, Tomer and {Thakar}, Aniruddha R. and {Thomas}, Daniel and {Thompson}, Benjamin A. and {Tinker}, Jeremy L. and {Tojeiro}, Rita and {Tremonti}, Christy A. and {Vargas Maga{\~n}a}, M. and {Verde}, Licia and {Viel}, Matteo and {Vikas}, Shailendra K. and {Vogt}, Nicole P. and {Wake}, David A. and {Wang}, Ji and {Weaver}, Benjamin A. and {Weinberg}, David H. and {Weiner}, Benjamin J. and {West}, Andrew A. and {White}, Martin and {Wilson}, John C. and {Wisniewski}, John P. and {Wood-Vasey}, W.~M. and {Yanny}, Brian and {Y{\`e}che}, Christophe and {York}, Donald G. and {Zamora}, O. and {Zasowski}, Gail and {Zehavi}, Idit and {Zhao}, Gong-Bo and {Zheng}, Zheng and {Zhu}, Guangtun and {Zinn}, Joel C.},
        title = "{The Ninth Data Release of the Sloan Digital Sky Survey: First Spectroscopic Data from the SDSS-III Baryon Oscillation Spectroscopic Survey}",
      journal = {\apjs},
         year = 2012,
        month = dec,
       volume = {203},
       number = {2},
          eid = {21},
        pages = {21},
          doi = {10.1088/0067-0049/203/2/21},
archivePrefix = {arXiv},
       eprint = {1207.7137},
 primaryClass = {astro-ph.IM},
       adsurl = {https://ui.adsabs.harvard.edu/abs/2012ApJS..203...21A}
}

@ARTICLE{2011MNRAS.418.1587T,
       author = {{Taylor}, Edward N. and {Hopkins}, Andrew M. and {Baldry}, Ivan K. and {Brown}, Michael J.~I. and {Driver}, Simon P. and {Kelvin}, Lee S. and {Hill}, David T. and {Robotham}, Aaron S.~G. and {Bland-Hawthorn}, Joss and {Jones}, D.~H. and {Sharp}, R.~G. and {Thomas}, Daniel and {Liske}, Jochen and {Loveday}, Jon and {Norberg}, Peder and {Peacock}, J.~A. and {Bamford}, Steven P. and {Brough}, Sarah and {Colless}, Matthew and {Cameron}, Ewan and {Conselice}, Christopher J. and {Croom}, Scott M. and {Frenk}, C.~S. and {Gunawardhana}, Madusha and {Kuijken}, Konrad and {Nichol}, R.~C. and {Parkinson}, H.~R. and {Phillipps}, S. and {Pimbblet}, K.~A. and {Popescu}, C.~C. and {Prescott}, Matthew and {Sutherland}, W.~J. and {Tuffs}, R.~J. and {van Kampen}, Eelco and {Wijesinghe}, D.},
        title = "{Galaxy And Mass Assembly (GAMA): stellar mass estimates}",
      journal = {\mnras},
         year = 2011,
        month = dec,
       volume = {418},
       number = {3},
        pages = {1587-1620},
          doi = {10.1111/j.1365-2966.2011.19536.x},
archivePrefix = {arXiv},
       eprint = {1108.0635},
 primaryClass = {astro-ph.CO},
       adsurl = {https://ui.adsabs.harvard.edu/abs/2011MNRAS.418.1587T}
}

@ARTICLE{2011MNRAS.416.1654B,
       author = {{Bois}, Maxime and {Emsellem}, Eric and {Bournaud}, Fr{\'e}d{\'e}ric and {Alatalo}, Katherine and {Blitz}, Leo and {Bureau}, Martin and {Cappellari}, Michele and {Davies}, Roger L. and {Davis}, Timothy A. and {de Zeeuw}, P.~T. and {Duc}, Pierre-Alain and {Khochfar}, Sadegh and {Krajnovi{\'c}}, Davor and {Kuntschner}, Harald and {Lablanche}, Pierre-Yves and {McDermid}, Richard M. and {Morganti}, Raffaella and {Naab}, Thorsten and {Oosterloo}, Tom and {Sarzi}, Marc and {Scott}, Nicholas and {Serra}, Paolo and {Weijmans}, Anne-Marie and {Young}, Lisa M.},
        title = "{The ATLAS$^{3D}$ project - VI. Simulations of binary galaxy mergers and the link with fast rotators, slow rotators and kinematically distinct cores}",
      journal = {\mnras},
         year = 2011,
        month = sep,
       volume = {416},
       number = {3},
        pages = {1654-1679},
          doi = {10.1111/j.1365-2966.2011.19113.x},
archivePrefix = {arXiv},
       eprint = {1105.4076},
 primaryClass = {astro-ph.CO},
       adsurl = {https://ui.adsabs.harvard.edu/abs/2011MNRAS.416.1654B}
}

@ARTICLE{2011MNRAS.414..888E,
       author = {{Emsellem}, Eric and {Cappellari}, Michele and {Krajnovi{\'c}}, Davor and {Alatalo}, Katherine and {Blitz}, Leo and {Bois}, Maxime and {Bournaud}, Fr{\'e}d{\'e}ric and {Bureau}, Martin and {Davies}, Roger L. and {Davis}, Timothy A. and {de Zeeuw}, P.~T. and {Khochfar}, Sadegh and {Kuntschner}, Harald and {Lablanche}, Pierre-Yves and {McDermid}, Richard M. and {Morganti}, Raffaella and {Naab}, Thorsten and {Oosterloo}, Tom and {Sarzi}, Marc and {Scott}, Nicholas and {Serra}, Paolo and {van de Ven}, Glenn and {Weijmans}, Anne-Marie and {Young}, Lisa M.},
        title = "{The ATLAS$^{3D}$ project - III. A census of the stellar angular momentum within the effective radius of early-type galaxies: unveiling the distribution of fast and slow rotators}",
      journal = {\mnras},
         year = 2011,
        month = jun,
       volume = {414},
       number = {2},
        pages = {888-912},
          doi = {10.1111/j.1365-2966.2011.18496.x},
archivePrefix = {arXiv},
       eprint = {1102.4444},
 primaryClass = {astro-ph.CO},
       adsurl = {https://ui.adsabs.harvard.edu/abs/2011MNRAS.414..888E}
}

@ARTICLE{2011MNRAS.413..971D,
       author = {{Driver}, S.~P. and {Hill}, D.~T. and {Kelvin}, L.~S. and {Robotham}, A.~S.~G. and {Liske}, J. and {Norberg}, P. and {Baldry}, I.~K. and {Bamford}, S.~P. and {Hopkins}, A.~M. and {Loveday}, J. and {Peacock}, J.~A. and {Andrae}, E. and {Bland-Hawthorn}, J. and {Brough}, S. and {Brown}, M.~J.~I. and {Cameron}, E. and {Ching}, J.~H.~Y. and {Colless}, M. and {Conselice}, C.~J. and {Croom}, S.~M. and {Cross}, N.~J.~G. and {de Propris}, R. and {Dye}, S. and {Drinkwater}, M.~J. and {Ellis}, S. and {Graham}, Alister W. and {Grootes}, M.~W. and {Gunawardhana}, M. and {Jones}, D.~H. and {van Kampen}, E. and {Maraston}, C. and {Nichol}, R.~C. and {Parkinson}, H.~R. and {Phillipps}, S. and {Pimbblet}, K. and {Popescu}, C.~C. and {Prescott}, M. and {Roseboom}, I.~G. and {Sadler}, E.~M. and {Sansom}, A.~E. and {Sharp}, R.~G. and {Smith}, D.~J.~B. and {Taylor}, E. and {Thomas}, D. and {Tuffs}, R.~J. and {Wijesinghe}, D. and {Dunne}, L. and {Frenk}, C.~S. and {Jarvis}, M.~J. and {Madore}, B.~F. and {Meyer}, M.~J. and {Seibert}, M. and {Staveley-Smith}, L. and {Sutherland}, W.~J. and {Warren}, S.~J.},
        title = "{Galaxy and Mass Assembly (GAMA): survey diagnostics and core data release}",
      journal = {\mnras},
         year = 2011,
        month = may,
       volume = {413},
       number = {2},
        pages = {971-995},
          doi = {10.1111/j.1365-2966.2010.18188.x},
archivePrefix = {arXiv},
       eprint = {1009.0614},
 primaryClass = {astro-ph.CO},
       adsurl = {https://ui.adsabs.harvard.edu/abs/2011MNRAS.413..971D}
}

@ARTICLE{2011OExpr..19.2649B,
       author = {{Bland-Hawthorn}, Joss and {Bryant}, Julia and {Robertson}, Gordon and {Gillingham}, Peter and {O'Byrne}, John and {Cecil}, Gerald and {Haynes}, Roger and {Croom}, Scott and {Ellis}, Simon and {Maack}, Martin and {Skovgaard}, Peter and {Noordegraaf}, Danny},
        title = "{Hexabundles: imaging fiber arrays for low-light astronomical applications}",
      journal = {Optics Express},
         year = 2011,
        month = jan,
       volume = {19},
       number = {3},
        pages = {2649},
          doi = {10.1364/OE.19.002649},
       adsurl = {https://ui.adsabs.harvard.edu/abs/2011OExpr..19.2649B}
}

@ARTICLE{2010ApJ...723..818H,
       author = {{Hoffman}, Loren and {Cox}, Thomas J. and {Dutta}, Suvendra and {Hernquist}, Lars},
        title = "{Orbital Structure of Merger Remnants. I. Effect of Gas Fraction in Pure Disk Mergers}",
      journal = {\apj},
         year = 2010,
        month = nov,
       volume = {723},
       number = {1},
        pages = {818-844},
          doi = {10.1088/0004-637X/723/1/818},
archivePrefix = {arXiv},
       eprint = {1001.0799},
 primaryClass = {astro-ph.CO},
       adsurl = {https://ui.adsabs.harvard.edu/abs/2010ApJ...723..818H}
}

@ARTICLE{2010MNRAS.404..575L,
       author = {{Lotz}, Jennifer M. and {Jonsson}, Patrik and {Cox}, T.~J. and {Primack}, Joel R.},
        title = "{The effect of mass ratio on the morphology and time-scales of disc galaxy mergers}",
      journal = {\mnras},
         year = 2010,
        month = may,
       volume = {404},
       number = {2},
        pages = {575-589},
          doi = {10.1111/j.1365-2966.2010.16268.x},
archivePrefix = {arXiv},
       eprint = {0912.1590},
 primaryClass = {astro-ph.CO},
       adsurl = {https://ui.adsabs.harvard.edu/abs/2010MNRAS.404..575L}
}

@ARTICLE{2010MNRAS.404..590L,
       author = {{Lotz}, Jennifer M. and {Jonsson}, Patrik and {Cox}, T.~J. and {Primack}, Joel R.},
        title = "{The effect of gas fraction on the morphology and time-scales of disc galaxy mergers}",
      journal = {\mnras},
         year = 2010,
        month = may,
       volume = {404},
       number = {2},
        pages = {590-603},
          doi = {10.1111/j.1365-2966.2010.16269.x},
archivePrefix = {arXiv},
       eprint = {0912.1593},
 primaryClass = {astro-ph.CO},
       adsurl = {https://ui.adsabs.harvard.edu/abs/2010MNRAS.404..590L}
}

@ARTICLE{2009MNRAS.397.1202J,
       author = {{Jesseit}, Roland and {Cappellari}, Michele and {Naab}, Thorsten and {Emsellem}, Eric and {Burkert}, Andreas},
        title = "{Specific angular momentum of disc merger remnants and the {\ensuremath{\lambda}}$_{R}$-parameter}",
      journal = {\mnras},
         year = 2009,
        month = aug,
       volume = {397},
       number = {3},
        pages = {1202-1214},
          doi = {10.1111/j.1365-2966.2009.14984.x},
archivePrefix = {arXiv},
       eprint = {0810.0137},
 primaryClass = {astro-ph},
       adsurl = {https://ui.adsabs.harvard.edu/abs/2009MNRAS.397.1202J}
}

@ARTICLE{2008MNRAS.391.1137L,
       author = {{Lotz}, Jennifer M. and {Jonsson}, Patrik and {Cox}, T.~J. and {Primack}, Joel R.},
        title = "{Galaxy merger morphologies and time-scales from simulations of equal-mass gas-rich disc mergers}",
      journal = {\mnras},
         year = 2008,
        month = dec,
       volume = {391},
       number = {3},
        pages = {1137-1162},
          doi = {10.1111/j.1365-2966.2008.14004.x},
archivePrefix = {arXiv},
       eprint = {0805.1246},
 primaryClass = {astro-ph},
       adsurl = {https://ui.adsabs.harvard.edu/abs/2008MNRAS.391.1137L}
}

@ARTICLE{2008ApJ...683...94O,
       author = {{Oh}, Sang Hoon and {Kim}, Woong-Tae and {Lee}, Hyung Mok and {Kim}, Jongsoo},
        title = "{Physical Properties of Tidal Features in Interacting Disk Galaxies}",
      journal = {\apj},
         year = 2008,
        month = aug,
       volume = {683},
       number = {1},
        pages = {94-113},
          doi = {10.1086/588184},
archivePrefix = {arXiv},
       eprint = {0803.1893},
 primaryClass = {astro-ph},
       adsurl = {https://ui.adsabs.harvard.edu/abs/2008ApJ...683...94O}
}

@ARTICLE{2006MNRAS.372..839N,
       author = {{Naab}, Thorsten and {Jesseit}, Roland and {Burkert}, Andreas},
        title = "{The influence of gas on the structure of merger remnants}",
      journal = {\mnras},
         year = 2006,
        month = oct,
       volume = {372},
       number = {2},
        pages = {839-852},
          doi = {10.1111/j.1365-2966.2006.10902.x},
archivePrefix = {arXiv},
       eprint = {astro-ph/0605155},
 primaryClass = {astro-ph},
       adsurl = {https://ui.adsabs.harvard.edu/abs/2006MNRAS.372..839N}
}

@INPROCEEDINGS{2006SPIE.6269E..0GS,
       author = {{Sharp}, Robert and {Saunders}, Will and {Smith}, Greg and {Churilov}, Vladimir and {Correll}, David and {Dawson}, John and {Farrel}, Tony and {Frost}, Gabriella and {Haynes}, Roger and {Heald}, Ron and {Lankshear}, Allan and {Mayfield}, Don and {Waller}, Lew and {Whittard}, Dennis},
        title = "{Performance of AAOmega: the AAT multi-purpose fiber-fed spectrograph}",
    booktitle = {Ground-based and Airborne Instrumentation for Astronomy},
         year = 2006,
       editor = {{McLean}, Ian S. and {Iye}, Masanori},
       series = {Society of Photo-Optical Instrumentation Engineers (SPIE) Conference Series},
       volume = {6269},
        month = jun,
          eid = {62690G},
        pages = {62690G},
          doi = {10.1117/12.671022},
archivePrefix = {arXiv},
       eprint = {astro-ph/0606137},
 primaryClass = {astro-ph},
       adsurl = {https://ui.adsabs.harvard.edu/abs/2006SPIE.6269E..0GS}
}

@ARTICLE{2005AJ....130.2647V,
       author = {{van Dokkum}, Pieter G.},
        title = "{The Recent and Continuing Assembly of Field Elliptical Galaxies by Red Mergers}",
      journal = {\aj},
         year = 2005,
        month = dec,
       volume = {130},
       number = {6},
        pages = {2647-2665},
          doi = {10.1086/497593},
archivePrefix = {arXiv},
       eprint = {astro-ph/0506661},
 primaryClass = {astro-ph},
       adsurl = {https://ui.adsabs.harvard.edu/abs/2005AJ....130.2647V}
}

@ARTICLE{1996ApJ...471..115B,
       author = {{Barnes}, Joshua E. and {Hernquist}, Lars},
        title = "{Transformations of Galaxies. II. Gasdynamics in Merging Disk Galaxies}",
      journal = {\apj},
         year = 1996,
        month = nov,
       volume = {471},
        pages = {115},
          doi = {10.1086/177957},
       adsurl = {https://ui.adsabs.harvard.edu/abs/1996ApJ...471..115B}
}

@ARTICLE{1992AJ....103.1089B,
       author = {{Byrd}, Gene G. and {Howard}, Sethanne},
        title = "{Tidal Arms are Ubiquitous in Spiral Galaxies}",
      journal = {\aj},
         year = 1992,
        month = apr,
       volume = {103},
        pages = {1089},
          doi = {10.1086/116128},
       adsurl = {https://ui.adsabs.harvard.edu/abs/1992AJ....103.1089B}
}

@ARTICLE{1989ApJ...342....1H,
       author = {{Hernquist}, Lars and {Quinn}, P.~J.},
        title = "{Formation and Morphology of Shell Galaxies. II. Nonspherical Potentials}",
      journal = {\apj},
         year = 1989,
        month = jul,
       volume = {342},
        pages = {1},
          doi = {10.1086/167571},
       adsurl = {https://ui.adsabs.harvard.edu/abs/1989ApJ...342....1H}
}

@ARTICLE{1984ApJ...279..596Q,
       author = {{Quinn}, P.~J.},
        title = "{On the formation and dynamics of shells around elliptical galaxies.}",
      journal = {\apj},
         year = 1984,
        month = apr,
       volume = {279},
        pages = {596-609},
          doi = {10.1086/161924},
       adsurl = {https://ui.adsabs.harvard.edu/abs/1984ApJ...279..596Q}
}

@ARTICLE{1983ApJ...266..713O,
       author = {{Oke}, J.~B. and {Gunn}, J.~E.},
        title = "{Secondary standard stars for absolute spectrophotometry.}",
      journal = {\apj},
         year = 1983,
        month = mar,
       volume = {266},
        pages = {713-717},
          doi = {10.1086/160817},
       adsurl = {https://ui.adsabs.harvard.edu/abs/1983ApJ...266..713O}
}

@ARTICLE{1980ApJ...236..351D,
       author = {{Dressler}, A.},
        title = "{Galaxy morphology in rich clusters: implications for the formation and evolution of galaxies.}",
      journal = {\apj},
         year = 1980,
        month = mar,
       volume = {236},
        pages = {351-365},
          doi = {10.1086/157753},
       adsurl = {https://ui.adsabs.harvard.edu/abs/1980ApJ...236..351D}
}

@ARTICLE{1972ApJ...178..623T,
       author = {{Toomre}, Alar and {Toomre}, Juri},
        title = "{Galactic Bridges and Tails}",
      journal = {\apj},
         year = 1972,
        month = dec,
       volume = {178},
        pages = {623-666},
          doi = {10.1086/151823},
       adsurl = {https://ui.adsabs.harvard.edu/abs/1972ApJ...178..623T}
}

@ARTICLE{2018MNRAS.476.3137R,
       author = {{Robotham}, A.~S.~G. and {Davies}, L.~J.~M. and {Driver}, S.~P. and {Koushan}, S. and {Taranu}, D.~S. and {Casura}, S. and {Liske}, J.},
        title = "{ProFound: Source Extraction and Application to Modern Survey Data}",
      journal = {\mnras},
         year = 2018,
        month = may,
       volume = {476},
       number = {3},
        pages = {3137-3159},
          doi = {10.1093/mnras/sty440},
archivePrefix = {arXiv},
       eprint = {1802.00937},
 primaryClass = {astro-ph.IM},
       adsurl = {https://ui.adsabs.harvard.edu/abs/2018MNRAS.476.3137R}
}

@ARTICLE{2016ApJ...832...69O,
       author = {{Oh}, Sree and {Yi}, Sukyoung K. and {Cortese}, Luca and {van de Sande}, Jesse and {Mahajan}, Smriti and {Jeong}, Hyunjin and {Sheen}, Yun-Kyeong and {Allen}, James T. and {Bekki}, Kenji and {Bland-Hawthorn}, Joss and {Bloom}, Jessica V. and {Brough}, Sarah and {Bryant}, Julia J. and {Colless}, Matthew and {Croom}, Scott M. and {Fogarty}, L.~M.~R. and {Goodwin}, Michael and {Green}, Andy and {Konstantopoulos}, Iraklis S. and {Lawrence}, Jon and {L{\'o}pez-S{\'a}nchez}, {\'A}. R. and {Lorente}, Nuria P.~F. and {Medling}, Anne M. and {Owers}, Matt S. and {Richards}, Samuel and {Scott}, Nicholas and {Sharp}, Rob and {Sweet}, Sarah M.},
        title = "{The SAMI Galaxy Survey: Galaxy Interactions and Kinematic Anomalies in Abell 119}",
      journal = {\apj},
         year = 2016,
        month = nov,
       volume = {832},
       number = {1},
          eid = {69},
        pages = {69},
          doi = {10.3847/0004-637X/832/1/69},
archivePrefix = {arXiv},
       eprint = {1609.03595},
 primaryClass = {astro-ph.GA},
       adsurl = {https://ui.adsabs.harvard.edu/abs/2016ApJ...832...69O}
}

@ARTICLE{2023MNRAS.523.4381D,
       author = {{Desmons}, Alice and {Brough}, Sarah and {Mart{\'\i}nez-Lombilla}, Cristina and {De Propris}, Roberto and {Holwerda}, Benne and {L{\'o}pez-S{\'a}nchez}, {\'A}ngel R.},
        title = "{Galaxy and mass assembly (GAMA): comparing visually and spectroscopically identified galaxy merger samples}",
      journal = {\mnras},
         year = 2023,
        month = aug,
       volume = {523},
       number = {3},
        pages = {4381-4393},
          doi = {10.1093/mnras/stad1639},
archivePrefix = {arXiv},
       eprint = {2305.17894},
 primaryClass = {astro-ph.GA},
       adsurl = {https://ui.adsabs.harvard.edu/abs/2023MNRAS.523.4381D}
}

@ARTICLE{1996ApJ...462..563N,
       author = {{Navarro}, Julio F. and {Frenk}, Carlos S. and {White}, Simon D.~M.},
        title = "{The Structure of Cold Dark Matter Halos}",
      journal = {\apj},
         year = 1996,
        month = may,
       volume = {462},
        pages = {563},
          doi = {10.1086/177173},
archivePrefix = {arXiv},
       eprint = {astro-ph/9508025},
 primaryClass = {astro-ph},
       adsurl = {https://ui.adsabs.harvard.edu/abs/1996ApJ...462..563N}
}

@ARTICLE{2024ApJ...965..158Y,
       author = {{Yoon}, Yongmin and {Ko}, Jongwan and {Chung}, Haeun and {Byun}, Woowon and {Chun}, Kyungwon},
        title = "{Shell-type Tidal Features Are More Frequently Detected in Slowly Rotating Early-type Galaxies than Stream- and Tail-type Features}",
      journal = {\apj},
         year = 2024,
        month = apr,
       volume = {965},
       number = {2},
          eid = {158},
        pages = {158},
          doi = {10.3847/1538-4357/ad34ad},
archivePrefix = {arXiv},
       eprint = {2404.03459},
 primaryClass = {astro-ph.GA},
       adsurl = {https://ui.adsabs.harvard.edu/abs/2024ApJ...965..158Y}
}

@ARTICLE{2024MNRAS.530.4422K,
       author = {{Khalid}, A. and {Brough}, S. and {Martin}, G. and {Kimmig}, L.~C. and {Lagos}, C.~D.~P. and {Remus}, R. -S. and {Martinez-Lombilla}, C.},
        title = "{Characterizing tidal features around galaxies in cosmological simulations}",
      journal = {\mnras},
         year = 2024,
        month = jun,
       volume = {530},
       number = {4},
        pages = {4422-4445},
          doi = {10.1093/mnras/stae1064},
archivePrefix = {arXiv},
       eprint = {2404.12436},
 primaryClass = {astro-ph.GA},
       adsurl = {https://ui.adsabs.harvard.edu/abs/2024MNRAS.530.4422K}
}

@ARTICLE{2014MNRAS.441..274S,
       author = {{Scott}, Nicholas and {Davies}, Roger L. and {Houghton}, Ryan C.~W. and {Cappellari}, Michele and {Graham}, Alister W. and {Pimbblet}, Kevin A.},
        title = "{Distribution of slow and fast rotators in the Fornax cluster}",
      journal = {\mnras},
         year = 2014,
        month = jun,
       volume = {441},
       number = {1},
        pages = {274-288},
          doi = {10.1093/mnras/stu472},
archivePrefix = {arXiv},
       eprint = {1403.1705},
 primaryClass = {astro-ph.GA},
       adsurl = {https://ui.adsabs.harvard.edu/abs/2014MNRAS.441..274S}
}

@ARTICLE{1982ApJ...263..599S,
       author = {{Schwarzschild}, M.},
        title = "{Triaxial equilibrium models for elliptical galaxies with slow figure rotation}",
      journal = {\apj},
         year = 1982,
        month = dec,
       volume = {263},
        pages = {599-610},
          doi = {10.1086/160531},
       adsurl = {https://ui.adsabs.harvard.edu/abs/1982ApJ...263..599S}
}

@ARTICLE{2022A&A...667A..51T,
       author = {{Thater}, Sabine and {Jethwa}, Prashin and {Tahmasebzadeh}, Behzad and {Zhu}, Ling and {den Brok}, Mark and {Santucci}, Giulia and {Ding}, Yuchen and {Poci}, Adriano and {Lilley}, Edward and {Tim de Zeeuw}, P. and {Zocchi}, Alice and {Maindl}, Thomas I. and {Rigamonti}, Fabio and {Yang}, Meng and {Fahrion}, Katja and {van de Ven}, Glenn},
        title = "{Testing the robustness of DYNAMITE triaxial Schwarzschild modelling: The effects of correcting the orbit mirroring}",
      journal = {\aap},
         year = 2022,
        month = nov,
       volume = {667},
          eid = {A51},
        pages = {A51},
          doi = {10.1051/0004-6361/202243926},
archivePrefix = {arXiv},
       eprint = {2205.04165},
 primaryClass = {astro-ph.GA},
       adsurl = {https://ui.adsabs.harvard.edu/abs/2022A&A...667A..51T}
}

@article{ATLAS3DVII,
  title = {The {{ATLAS}}{{{\textsuperscript{3D}}}} Project - {{VII}}. {{A}} New Look at the Morphology of Nearby Galaxies: The Kinematic Morphology-Density Relation},
  author = {Cappellari, Michele and Emsellem, Eric and Krajnovi{\'c}, Davor and McDermid, Richard M. and Serra, Paolo and Alatalo, Katherine and Blitz, Leo and Bois, Maxime and Bournaud, Fr{\'e}d{\'e}ric and Bureau, M. and Davies, Roger L. and Davis, Timothy A. and {de Zeeuw}, P. T. and Khochfar, Sadegh and Kuntschner, Harald and Lablanche, Pierre-Yves and Morganti, Raffaella and Naab, Thorsten and Oosterloo, Tom and Sarzi, Marc and Scott, Nicholas and Weijmans, Anne-Marie and Young, Lisa M.},
  journal = {\mnras},
  year = {2011},
  month = sep,
  volume = {416},
  pages = {1680--1696},
  doi = {10.1111/j.1365-2966.2011.18600.x},
  adsurl = {https://ui.adsabs.harvard.edu/abs/2011MNRAS.416.1680C},
  archivePrefix = {arXiv},
  eprint = {1104.3545},
  eprinttype = {arxiv},
  number = {3},
  primaryClass = {astro-ph.CO}
}

@ARTICLE{2021MNRAS.508.2307V,
       author = {{van de Sande}, Jesse and {Croom}, Scott M. and {Bland-Hawthorn}, Joss and {Cortese}, Luca and {Scott}, Nicholas and {Lagos}, Claudia D.~P. and {D'Eugenio}, Francesco and {Bryant}, Julia J. and {Brough}, Sarah and {Catinella}, Barbara and {Foster}, Caroline and {Groves}, Brent and {Harborne}, Katherine E. and {L{\'o}pez-S{\'a}nchez}, {\'A}ngel R. and {McDermid}, Richard and {Medling}, Anne and {Owers}, Matt S. and {Richards}, Samuel N. and {Sweet}, Sarah M. and {Vaughan}, Sam P.},
        title = "{The SAMI galaxy survey: Mass and environment as independent drivers of galaxy dynamics}",
      journal = {\mnras},
         year = 2021,
        month = dec,
       volume = {508},
       number = {2},
        pages = {2307-2328},
          doi = {10.1093/mnras/stab2647},
archivePrefix = {arXiv},
       eprint = {2109.06189},
 primaryClass = {astro-ph.GA},
       adsurl = {https://ui.adsabs.harvard.edu/abs/2021MNRAS.508.2307V}
}

@ARTICLE{2018MNRAS.473.3000Z,
       author = {{Zhu}, Ling and {van den Bosch}, Remco and {van de Ven}, Glenn and {Lyubenova}, Mariya and {Falc{\'o}n-Barroso}, Jes{\'u}s and {Meidt}, Sharon E. and {Martig}, Marie and {Shen}, Juntai and {Li}, Zhao-Yu and {Yildirim}, Akin and {Walcher}, C. Jakob and {Sanchez}, Sebastian F.},
        title = "{Orbital decomposition of CALIFA spiral galaxies}",
      journal = {\mnras},
         year = 2018,
        month = jan,
       volume = {473},
       number = {3},
        pages = {3000-3018},
          doi = {10.1093/mnras/stx2409},
archivePrefix = {arXiv},
       eprint = {1709.06649},
 primaryClass = {astro-ph.GA},
       adsurl = {https://ui.adsabs.harvard.edu/abs/2018MNRAS.473.3000Z}
}

@ARTICLE{2020MNRAS.491.1690J,
       author = {{Jin}, Yunpeng and {Zhu}, Ling and {Long}, R.~J. and {Mao}, Shude and {Wang}, Lan and {van de Ven}, Glenn},
        title = "{SDSS-IV MaNGA: Internal mass distributions and orbital structures of early-type galaxies and their dependence on environment}",
      journal = {\mnras},
         year = 2020,
        month = jan,
       volume = {491},
       number = {2},
        pages = {1690-1708},
          doi = {10.1093/mnras/stz3072},
archivePrefix = {arXiv},
       eprint = {1911.00777},
 primaryClass = {astro-ph.GA},
       adsurl = {https://ui.adsabs.harvard.edu/abs/2020MNRAS.491.1690J}
}

@ARTICLE{2018NatAs...2..233Z,
       author = {{Zhu}, Ling and {van de Ven}, Glenn and {van den Bosch}, Remco and {Rix}, Hans-Walter and {Lyubenova}, Mariya and {Falc{\'o}n-Barroso}, Jes{\'u}s and {Martig}, Marie and {Mao}, Shude and {Xu}, Dandan and {Jin}, Yunpeng and {Obreja}, Aura and {Grand}, Robert J.~J. and {Dutton}, Aaron A. and {Macci{\`o}}, Andrea V. and {G{\'o}mez}, Facundo A. and {Walcher}, Jakob C. and {Garc{\'\i}a-Benito}, Rub{\'e}n and {Zibetti}, Stefano and {S{\'a}nchez}, Sebastian F.},
        title = "{The stellar orbit distribution in present-day galaxies inferred from the CALIFA survey}",
      journal = {Nature Astronomy},
         year = 2018,
        month = jan,
       volume = {2},
        pages = {233-238},
          doi = {10.1038/s41550-017-0348-1},
archivePrefix = {arXiv},
       eprint = {1711.06728},
 primaryClass = {astro-ph.GA},
       adsurl = {https://ui.adsabs.harvard.edu/abs/2018NatAs...2..233Z}
}

@ARTICLE{2013ExA....35...25D,
       author = {{de Jong}, Jelte T.~A. and {Verdoes Kleijn}, Gijs A. and {Kuijken}, Konrad H. and {Valentijn}, Edwin A.},
        title = "{The Kilo-Degree Survey}",
      journal = {Experimental Astronomy},
         year = 2013,
        month = jan,
       volume = {35},
       number = {1-2},
        pages = {25-44},
          doi = {10.1007/s10686-012-9306-1},
archivePrefix = {arXiv},
       eprint = {1206.1254},
 primaryClass = {astro-ph.CO},
       adsurl = {https://ui.adsabs.harvard.edu/abs/2013ExA....35...25D}
}

@ARTICLE{2011Msngr.146....8K,
       author = {{Kuijken}, K.},
        title = "{OmegaCAM: ESO's Newest Imager}",
      journal = {The Messenger},
         year = 2011,
        month = dec,
       volume = {146},
        pages = {8-11},
       adsurl = {https://ui.adsabs.harvard.edu/abs/2011Msngr.146....8K}
}

@ARTICLE{2019A&A...625A...2K,
       author = {{Kuijken}, K. and {Heymans}, C. and {Dvornik}, A. and {Hildebrandt}, H. and {de Jong}, J.~T.~A. and {Wright}, A.~H. and {Erben}, T. and {Bilicki}, M. and {Giblin}, B. and {Shan}, H. -Y. and {Getman}, F. and {Grado}, A. and {Hoekstra}, H. and {Miller}, L. and {Napolitano}, N. and {Paolilo}, M. and {Radovich}, M. and {Schneider}, P. and {Sutherland}, W. and {Tewes}, M. and {Tortora}, C. and {Valentijn}, E.~A. and {Verdoes Kleijn}, G.~A.},
        title = "{The fourth data release of the Kilo-Degree Survey: ugri imaging and nine-band optical-IR photometry over 1000 square degrees}",
      journal = {\aap},
         year = 2019,
        month = may,
       volume = {625},
          eid = {A2},
        pages = {A2},
          doi = {10.1051/0004-6361/201834918},
archivePrefix = {arXiv},
       eprint = {1902.11265},
 primaryClass = {astro-ph.GA},
       adsurl = {https://ui.adsabs.harvard.edu/abs/2019A&A...625A...2K}
}

@ARTICLE{1980MNRAS.193..189F,
       author = {{Fall}, S.~M. and {Efstathiou}, G.},
        title = "{Formation and rotation of disc galaxies with haloes.}",
      journal = {\mnras},
         year = 1980,
        month = oct,
       volume = {193},
        pages = {189-206},
          doi = {10.1093/mnras/193.2.189},
       adsurl = {https://ui.adsabs.harvard.edu/abs/1980MNRAS.193..189F}
}

@ARTICLE{1979MNRAS.189..831W,
       author = {{White}, S.~D.~M.},
        title = "{Further simulations of merging galaxies}",
      journal = {\mnras},
         year = 1979,
        month = dec,
       volume = {189},
        pages = {831-852},
          doi = {10.1093/mnras/189.4.831},
       adsurl = {https://ui.adsabs.harvard.edu/abs/1979MNRAS.189..831W}
}

@ARTICLE{2024MNRAS.529.3446C,
       author = {{Croom}, Scott M. and {van de Sande}, Jesse and {Vaughan}, Sam P. and {Rutherford}, Tomas H. and {Lagos}, Claudia del P. and {Barsanti}, Stefania and {Bland-Hawthorn}, Joss and {Brough}, Sarah and {Bryant}, Julia J. and {Colless}, Matthew and {Cortese}, Luca and {D'Eugenio}, Francesco and {Fraser-McKelvie}, Amelia and {Goodwin}, Michael and {Lorente}, Nuria P.~F. and {Richards}, Samuel N. and {Ristea}, Andrei and {Sweet}, Sarah M. and {Yi}, Sukyoung K. and {Zafar}, Tayyaba},
        title = "{The SAMI Galaxy Survey: galaxy spin is more strongly correlated with stellar population age than mass or environment}",
      journal = {\mnras},
         year = 2024,
        month = apr,
       volume = {529},
       number = {4},
        pages = {3446-3468},
          doi = {10.1093/mnras/stae458},
archivePrefix = {arXiv},
       eprint = {2402.06877},
 primaryClass = {astro-ph.GA},
       adsurl = {https://ui.adsabs.harvard.edu/abs/2024MNRAS.529.3446C}
}

@ARTICLE{1999ApJS..124..383C,
       author = {{Cretton}, N. and {de Zeeuw}, P. Tim and {van der Marel}, Roeland P. and {Rix}, Hans-Walter},
        title = "{Axisymmetric Three-Integral Models for Galaxies}",
      journal = {\apjs},
         year = 1999,
        month = oct,
       volume = {124},
       number = {2},
        pages = {383-401},
          doi = {10.1086/313264},
archivePrefix = {arXiv},
       eprint = {astro-ph/9902034},
 primaryClass = {astro-ph},
       adsurl = {https://ui.adsabs.harvard.edu/abs/1999ApJS..124..383C}
}

@ARTICLE{2003ApJ...583...92G,
       author = {{Gebhardt}, Karl and {Richstone}, Douglas and {Tremaine}, Scott and {Lauer}, Tod R. and {Bender}, Ralf and {Bower}, Gary and {Dressler}, Alan and {Faber}, S.~M. and {Filippenko}, Alexei V. and {Green}, Richard and {Grillmair}, Carl and {Ho}, Luis C. and {Kormendy}, John and {Magorrian}, John and {Pinkney}, Jason},
        title = "{Axisymmetric Dynamical Models of the Central Regions of Galaxies}",
      journal = {\apj},
         year = 2003,
        month = jan,
       volume = {583},
       number = {1},
        pages = {92-115},
          doi = {10.1086/345081},
archivePrefix = {arXiv},
       eprint = {astro-ph/0209483},
 primaryClass = {astro-ph},
       adsurl = {https://ui.adsabs.harvard.edu/abs/2003ApJ...583...92G}
}

@ARTICLE{2004ApJ...602...66V,
       author = {{Valluri}, Monica and {Merritt}, David and {Emsellem}, Eric},
        title = "{Difficulties with Recovering the Masses of Supermassive Black Holes from Stellar Kinematical Data}",
      journal = {\apj},
         year = 2004,
        month = feb,
       volume = {602},
       number = {1},
        pages = {66-92},
          doi = {10.1086/380896},
archivePrefix = {arXiv},
       eprint = {astro-ph/0210379},
 primaryClass = {astro-ph},
       adsurl = {https://ui.adsabs.harvard.edu/abs/2004ApJ...602...66V}
}

@ARTICLE{2015MNRAS.450.2842V,
       author = {{Vasiliev}, Eugene and {Athanassoula}, E.},
        title = "{Applying Schwarzschild's orbit superposition method to barred or non-barred disc galaxies}",
      journal = {\mnras},
         year = 2015,
        month = jul,
       volume = {450},
       number = {3},
        pages = {2842-2856},
          doi = {10.1093/mnras/stv805},
archivePrefix = {arXiv},
       eprint = {1505.03148},
 primaryClass = {astro-ph.GA},
       adsurl = {https://ui.adsabs.harvard.edu/abs/2015MNRAS.450.2842V}
}

@ARTICLE{2021MNRAS.500.1437N,
       author = {{Neureiter}, B. and {Thomas}, J. and {Saglia}, R. and {Bender}, R. and {Finozzi}, F. and {Krukau}, A. and {Naab}, T. and {Rantala}, A. and {Frigo}, M.},
        title = "{SMART: a new implementation of Schwarzschild's Orbit Superposition technique for triaxial galaxies and its application to an N-body merger simulation}",
      journal = {\mnras},
         year = 2021,
        month = jan,
       volume = {500},
       number = {1},
        pages = {1437-1465},
          doi = {10.1093/mnras/staa3014},
archivePrefix = {arXiv},
       eprint = {2009.08979},
 primaryClass = {astro-ph.GA},
       adsurl = {https://ui.adsabs.harvard.edu/abs/2021MNRAS.500.1437N}
}

@INCOLLECTION{1990dig..book..186B,
       author = {{Barnes}, J.},
        title = "{N-body studies of major mergers.}",
    booktitle = {Dynamics and Interactions of Galaxies},
         year = 1990,
       editor = {{Wielen}, Roland},
        pages = {186-195},
       adsurl = {https://ui.adsabs.harvard.edu/abs/1990dig..book..186B}
}

@ARTICLE{2005MNRAS.360.1185J,
       author = {{Jesseit}, R. and {Naab}, T. and {Burkert}, A.},
        title = "{Orbital structure of collisionless merger remnants: on the origin of photometric and kinematic properties of elliptical and S0 galaxies}",
      journal = {\mnras},
         year = 2005,
        month = jul,
       volume = {360},
       number = {4},
        pages = {1185-1200},
          doi = {10.1111/j.1365-2966.2005.09129.x},
archivePrefix = {arXiv},
       eprint = {astro-ph/0501418},
 primaryClass = {astro-ph},
       adsurl = {https://ui.adsabs.harvard.edu/abs/2005MNRAS.360.1185J}
}

@ARTICLE{2003ApJ...597..893N,
       author = {{Naab}, Thorsten and {Burkert}, Andreas},
        title = "{Statistical Properties of Collisionless Equal- and Unequal-Mass Merger Remnants of Disk Galaxies}",
      journal = {\apj},
         year = 2003,
        month = nov,
       volume = {597},
       number = {2},
        pages = {893-906},
          doi = {10.1086/378581},
archivePrefix = {arXiv},
       eprint = {astro-ph/0110179},
 primaryClass = {astro-ph},
       adsurl = {https://ui.adsabs.harvard.edu/abs/2003ApJ...597..893N}
}

@ARTICLE{2022ApJ...930..153S,
       author = {{Santucci}, Giulia and {Brough}, Sarah and {van de Sande}, Jesse and {McDermid}, Richard M. and {van de Ven}, Glenn and {Zhu}, Ling and {D'Eugenio}, Francesco and {Bland-Hawthorn}, Joss and {Barsanti}, Stefania and {Bryant}, Julia J. and {Croom}, Scott M. and {Davies}, Roger L. and {Green}, Andrew W. and {Lawrence}, Jon S. and {Lorente}, Nuria P.~F. and {Owers}, Matt S. and {Poci}, Adriano and {Richards}, Samuel N. and {Thater}, Sabine and {Yi}, Sukyoung},
        title = "{The SAMI Galaxy Survey: The Internal Orbital Structure and Mass Distribution of Passive Galaxies from Triaxial Orbit-superposition Schwarzschild Models}",
      journal = {\apj},
         year = 2022,
        month = may,
       volume = {930},
       number = {2},
          eid = {153},
        pages = {153},
          doi = {10.3847/1538-4357/ac5bd5},
archivePrefix = {arXiv},
       eprint = {2203.03648},
 primaryClass = {astro-ph.GA},
       adsurl = {https://ui.adsabs.harvard.edu/abs/2022ApJ...930..153S}
}

@ARTICLE{2009ApJS..182..543A,
       author = {{Abazajian}, Kevork N. and {Adelman-McCarthy}, Jennifer K. and {Ag{\"u}eros}, Marcel A. and {Allam}, Sahar S. and {Allende Prieto}, Carlos and {An}, Deokkeun and {Anderson}, Kurt S.~J. and {Anderson}, Scott F. and {Annis}, James and {Bahcall}, Neta A. and {Bailer-Jones}, C.~A.~L. and {Barentine}, J.~C. and {Bassett}, Bruce A. and {Becker}, Andrew C. and {Beers}, Timothy C. and {Bell}, Eric F. and {Belokurov}, Vasily and {Berlind}, Andreas A. and {Berman}, Eileen F. and {Bernardi}, Mariangela and {Bickerton}, Steven J. and {Bizyaev}, Dmitry and {Blakeslee}, John P. and {Blanton}, Michael R. and {Bochanski}, John J. and {Boroski}, William N. and {Brewington}, Howard J. and {Brinchmann}, Jarle and {Brinkmann}, J. and {Brunner}, Robert J. and {Budav{\'a}ri}, Tam{\'a}s and {Carey}, Larry N. and {Carliles}, Samuel and {Carr}, Michael A. and {Castander}, Francisco J. and {Cinabro}, David and {Connolly}, A.~J. and {Csabai}, Istv{\'a}n and {Cunha}, Carlos E. and {Czarapata}, Paul C. and {Davenport}, James R.~A. and {de Haas}, Ernst and {Dilday}, Ben and {Doi}, Mamoru and {Eisenstein}, Daniel J. and {Evans}, Michael L. and {Evans}, N.~W. and {Fan}, Xiaohui and {Friedman}, Scott D. and {Frieman}, Joshua A. and {Fukugita}, Masataka and {G{\"a}nsicke}, Boris T. and {Gates}, Evalyn and {Gillespie}, Bruce and {Gilmore}, G. and {Gonzalez}, Belinda and {Gonzalez}, Carlos F. and {Grebel}, Eva K. and {Gunn}, James E. and {Gy{\"o}ry}, Zsuzsanna and {Hall}, Patrick B. and {Harding}, Paul and {Harris}, Frederick H. and {Harvanek}, Michael and {Hawley}, Suzanne L. and {Hayes}, Jeffrey J.~E. and {Heckman}, Timothy M. and {Hendry}, John S. and {Hennessy}, Gregory S. and {Hindsley}, Robert B. and {Hoblitt}, J. and {Hogan}, Craig J. and {Hogg}, David W. and {Holtzman}, Jon A. and {Hyde}, Joseph B. and {Ichikawa}, Shin-ichi and {Ichikawa}, Takashi and {Im}, Myungshin and {Ivezi{\'c}}, {\v{Z}}eljko and {Jester}, Sebastian and {Jiang}, Linhua and {Johnson}, Jennifer A. and {Jorgensen}, Anders M. and {Juri{\'c}}, Mario and {Kent}, Stephen M. and {Kessler}, R. and {Kleinman}, S.~J. and {Knapp}, G.~R. and {Konishi}, Kohki and {Kron}, Richard G. and {Krzesinski}, Jurek and {Kuropatkin}, Nikolay and {Lampeitl}, Hubert and {Lebedeva}, Svetlana and {Lee}, Myung Gyoon and {Lee}, Young Sun and {French Leger}, R. and {L{\'e}pine}, S{\'e}bastien and {Li}, Nolan and {Lima}, Marcos and {Lin}, Huan and {Long}, Daniel C. and {Loomis}, Craig P. and {Loveday}, Jon and {Lupton}, Robert H. and {Magnier}, Eugene and {Malanushenko}, Olena and {Malanushenko}, Viktor and {Mandelbaum}, Rachel and {Margon}, Bruce and {Marriner}, John P. and {Mart{\'\i}nez-Delgado}, David and {Matsubara}, Takahiko and {McGehee}, Peregrine M. and {McKay}, Timothy A. and {Meiksin}, Avery and {Morrison}, Heather L. and {Mullally}, Fergal and {Munn}, Jeffrey A. and {Murphy}, Tara and {Nash}, Thomas and {Nebot}, Ada and {Neilsen}, Jr., Eric H. and {Newberg}, Heidi Jo and {Newman}, Peter R. and {Nichol}, Robert C. and {Nicinski}, Tom and {Nieto-Santisteban}, Maria and {Nitta}, Atsuko and {Okamura}, Sadanori and {Oravetz}, Daniel J. and {Ostriker}, Jeremiah P. and {Owen}, Russell and {Padmanabhan}, Nikhil and {Pan}, Kaike and {Park}, Changbom and {Pauls}, George and {Peoples}, Jr., John and {Percival}, Will J. and {Pier}, Jeffrey R. and {Pope}, Adrian C. and {Pourbaix}, Dimitri and {Price}, Paul A. and {Purger}, Norbert and {Quinn}, Thomas and {Raddick}, M. Jordan and {Re Fiorentin}, Paola and {Richards}, Gordon T. and {Richmond}, Michael W. and {Riess}, Adam G. and {Rix}, Hans-Walter and {Rockosi}, Constance M. and {Sako}, Masao and {Schlegel}, David J. and {Schneider}, Donald P. and {Scholz}, Ralf-Dieter and {Schreiber}, Matthias R. and {Schwope}, Axel D. and {Seljak}, Uro{\v{s}} and {Sesar}, Branimir and {Sheldon}, Erin and {Shimasaku}, Kazu and {Sibley}, Valena C. and {Simmons}, A.~E. and {Sivarani}, Thirupathi and {Allyn Smith}, J. and {Smith}, Martin C. and {Smol{\v{c}}i{\'c}}, Vernesa and {Snedden}, Stephanie A. and {Stebbins}, Albert and {Steinmetz}, Matthias and {Stoughton}, Chris and {Strauss}, Michael A. and {SubbaRao}, Mark and {Suto}, Yasushi and {Szalay}, Alexander S. and {Szapudi}, Istv{\'a}n and {Szkody}, Paula and {Tanaka}, Masayuki and {Tegmark}, Max and {Teodoro}, Luis F.~A. and {Thakar}, Aniruddha R. and {Tremonti}, Christy A. and {Tucker}, Douglas L. and {Uomoto}, Alan and {Vanden Berk}, Daniel E. and {Vandenberg}, Jan and {Vidrih}, S. and {Vogeley}, Michael S. and {Voges}, Wolfgang and {Vogt}, Nicole P. and {Wadadekar}, Yogesh and {Watters}, Shannon and {Weinberg}, David H. and {West}, Andrew A. and {White}, Simon D.~M. and {Wilhite}, Brian C. and {Wonders}, Alainna C. and {Yanny}, Brian and {Yocum}, D.~R.},
        title = "{The Seventh Data Release of the Sloan Digital Sky Survey}",
      journal = {\apjs},
         year = 2009,
        month = jun,
       volume = {182},
       number = {2},
        pages = {543-558},
          doi = {10.1088/0067-0049/182/2/543},
archivePrefix = {arXiv},
       eprint = {0812.0649},
 primaryClass = {astro-ph},
       adsurl = {https://ui.adsabs.harvard.edu/abs/2009ApJS..182..543A}
}

@ARTICLE{1989ApJ...343..617D,
       author = {{de Zeeuw}, Tim and {Franx}, Marijn},
        title = "{Kinematics of Gas in a Triaxial Galaxy}",
      journal = {\apj},
         year = 1989,
        month = aug,
       volume = {343},
        pages = {617},
          doi = {10.1086/167735},
       adsurl = {https://ui.adsabs.harvard.edu/abs/1989ApJ...343..617D}
}

@ARTICLE{2014MNRAS.441.3359D,
       author = {{Dutton}, Aaron A. and {Macci{\`o}}, Andrea V.},
        title = "{Cold dark matter haloes in the Planck era: evolution of structural parameters for Einasto and NFW profiles}",
      journal = {\mnras},
         year = 2014,
        month = jul,
       volume = {441},
       number = {4},
        pages = {3359-3374},
          doi = {10.1093/mnras/stu742},
archivePrefix = {arXiv},
       eprint = {1402.7073},
 primaryClass = {astro-ph.CO},
       adsurl = {https://ui.adsabs.harvard.edu/abs/2014MNRAS.441.3359D}
}

@ARTICLE{1911MNRAS..71..460P,
       author = {{Plummer}, H.~C.},
        title = "{On the problem of distribution in globular star clusters}",
      journal = {\mnras},
         year = 1911,
        month = mar,
       volume = {71},
        pages = {460-470},
          doi = {10.1093/mnras/71.5.460},
       adsurl = {https://ui.adsabs.harvard.edu/abs/1911MNRAS..71..460P}
}

@ARTICLE{2014ApJ...792...59Z,
       author = {{Zhu}, Ling and {Long}, R.~J. and {Mao}, Shude and {Peng}, Eric W. and {Liu}, Chengze and {Caldwell}, Nelson and {Li}, Biao and {Blakeslee}, John P. and {C{\^o}t{\'e}}, Patrick and {Cuillandre}, Jean-Charles and {Durrell}, Patrick and {Emsellem}, Eric and {Ferrarese}, Laura and {Gwyn}, Stephen and {Jord{\'a}n}, Andr{\'e}s and {Lan{\c{c}}on}, Ariane and {Mei}, Simona and {Mu{\~n}oz}, Roberto and {Puzia}, Thomas},
        title = "{The Next Generation Virgo Cluster Survey. V. Modeling the Dynamics of M87 with the Made-to-measure Method}",
      journal = {\apj},
         year = 2014,
        month = sep,
       volume = {792},
       number = {1},
          eid = {59},
        pages = {59},
          doi = {10.1088/0004-637X/792/1/59},
archivePrefix = {arXiv},
       eprint = {1407.2263},
 primaryClass = {astro-ph.GA},
       adsurl = {https://ui.adsabs.harvard.edu/abs/2014ApJ...792...59Z}
}

@ARTICLE{1985MNRAS.216..273D,
       author = {{de Zeeuw}, T.},
        title = "{Elliptical galaxies with separable potentials}",
      journal = {\mnras},
         year = 1985,
        month = sep,
       volume = {216},
        pages = {273-334},
          doi = {10.1093/mnras/216.2.273},
       adsurl = {https://ui.adsabs.harvard.edu/abs/1985MNRAS.216..273D}
}

@ARTICLE{2024MNRAS.533.1300D,
       author = {{Derkenne}, C. and {McDermid}, R.~M. and {Santucci}, G. and {Poci}, A. and {Thater}, S. and {Bellstedt}, S. and {Mendel}, J.~T. and {Foster}, C. and {Harborne}, K.~E. and {Lagos}, C.~D.~P. and {Wisnioski}, E. and {Croom}, S. and {Remus}, R. -S. and {Valenzuela}, L.~M. and {van de Sande}, J. and {Sweet}, S.~M. and {Ziegler}, B.},
        title = "{The MAGPI survey: evidence against the bulge-halo conspiracy}",
      journal = {\mnras},
         year = 2024,
        month = sep,
       volume = {533},
       number = {2},
        pages = {1300-1320},
          doi = {10.1093/mnras/stae1836},
archivePrefix = {arXiv},
       eprint = {2408.04834},
 primaryClass = {astro-ph.GA},
       adsurl = {https://ui.adsabs.harvard.edu/abs/2024MNRAS.533.1300D}
}

@BOOK{1974slsp.book.....L,
       author = {{Lawson}, Charles L. and {Hanson}, Richard J.},
        title = "{Solving least squares problems}",
         year = 1974,
       adsurl = {https://ui.adsabs.harvard.edu/abs/1974slsp.book.....L}
}

@ARTICLE{2015ApJ...813...82R,
       author = {{Reines}, Amy E. and {Volonteri}, Marta},
        title = "{Relations between Central Black Hole Mass and Total Galaxy Stellar Mass in the Local Universe}",
      journal = {\apj},
         year = 2015,
        month = nov,
       volume = {813},
       number = {2},
          eid = {82},
        pages = {82},
          doi = {10.1088/0004-637X/813/2/82},
archivePrefix = {arXiv},
       eprint = {1508.06274},
 primaryClass = {astro-ph.GA},
       adsurl = {https://ui.adsabs.harvard.edu/abs/2015ApJ...813...82R}
}

@ARTICLE{2020MNRAS.498..940M,
       author = {{Martel}, Hugo and {Richard}, Simon},
        title = "{Formation of counter-rotating stars during gas-rich disc-disc mergers}",
      journal = {\mnras},
         year = 2020,
        month = oct,
       volume = {498},
       number = {1},
        pages = {940-958},
          doi = {10.1093/mnras/staa2122},
       adsurl = {https://ui.adsabs.harvard.edu/abs/2020MNRAS.498..940M}
}

@ARTICLE{1969A&A.....3..455M,
       author = {{Moffat}, A.~F.~J.},
        title = "{A Theoretical Investigation of Focal Stellar Images in the Photographic Emulsion and Application to Photographic Photometry}",
      journal = {\aap},
         year = 1969,
        month = dec,
       volume = {3},
        pages = {455},
       adsurl = {https://ui.adsabs.harvard.edu/abs/1969A&A.....3..455M}
}

@ARTICLE{2022MNRAS.515.5335L,
       author = {{Li}, Jiaxuan and {Huang}, Song and {Leauthaud}, Alexie and {Moustakas}, John and {Danieli}, Shany and {Greene}, Jenny E. and {Abraham}, Roberto and {Ardila}, Felipe and {Kado-Fong}, Erin and {Lokhorst}, Deborah and {Lupton}, Robert and {Price}, Paul},
        title = "{Reaching for the Edge I: probing the outskirts of massive galaxies with HSC, DECaLS, SDSS, and Dragonfly}",
      journal = {\mnras},
         year = 2022,
        month = oct,
       volume = {515},
       number = {4},
        pages = {5335-5357},
          doi = {10.1093/mnras/stac2121},
archivePrefix = {arXiv},
       eprint = {2111.03557},
 primaryClass = {astro-ph.GA},
       adsurl = {https://ui.adsabs.harvard.edu/abs/2022MNRAS.515.5335L}
}




\appendix

\section{Example Galaxy Fits}
\label{sec:extra_gals}
\begin{figure*}
    \centering
    \includegraphics[width=\linewidth]{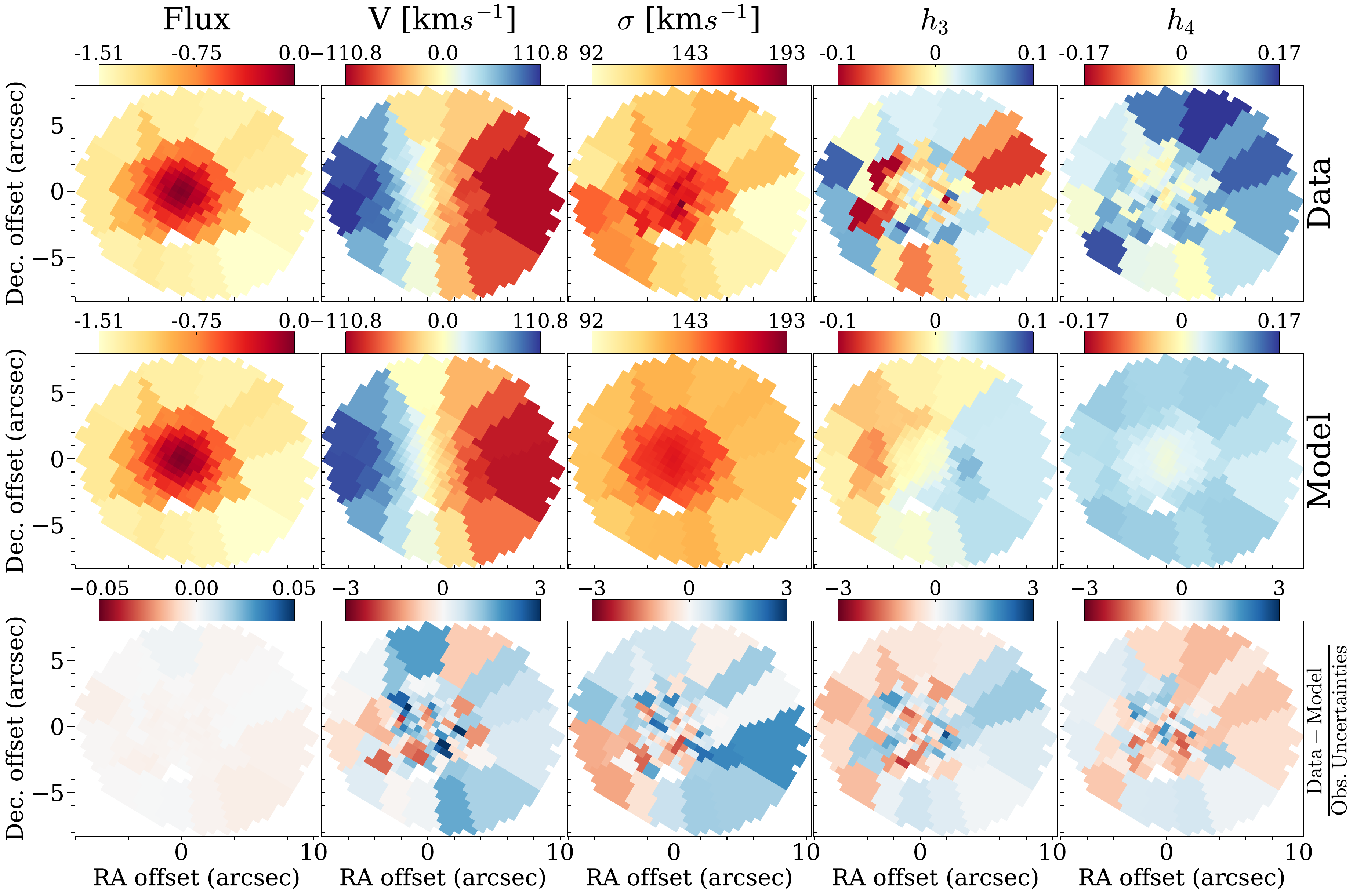}
    \caption[Schwarzschild Model for 511892]{The best-fitting Schwarzschild model for SAMI galaxy with CATID = 511892, and $\chi^2_{\text{red}}=1.13$. This galaxy displays a good fit at low spatial sampling, with a stellar mass of $\logm =10.86$, $\lre = 0.33$, $R_{\rm{max}}/R_{\rm{e}}=1.36$, and $N_{\rm{bins}} = 78$. Panels are arranged as in Figure \ref{fig:220394_velmaps}. The model reconstructs all velocity moments accurately, as the residual panels show low values and no structure.}
    \label{fig:511892_velmaps}
\end{figure*}

\begin{figure*}
    \centering
    \includegraphics[width=\linewidth]{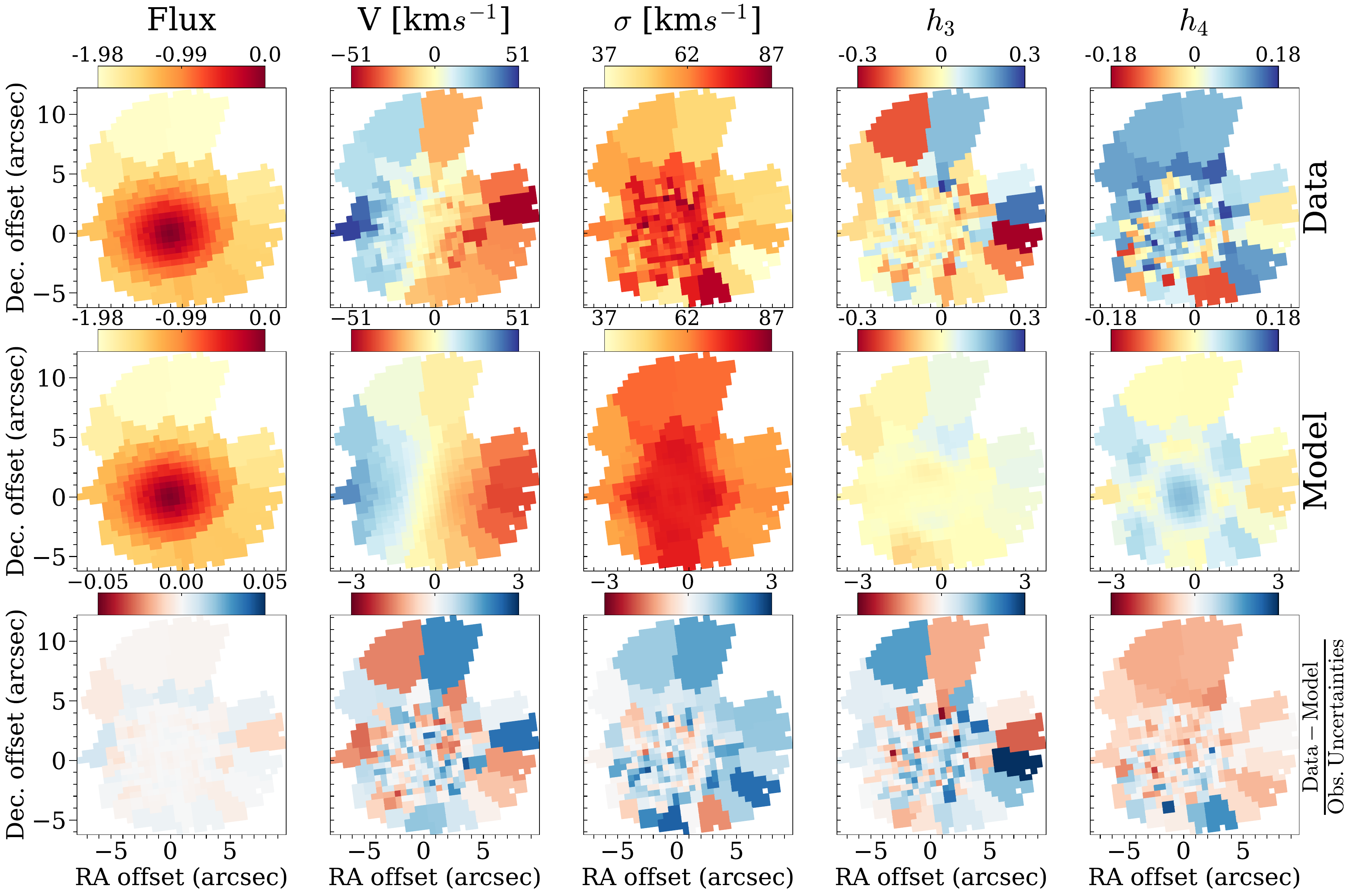}
    \caption[Schwarzschild Model for 79733]{The best-fitting Schwarzschild model for SAMI galaxy with CATID = 79733, and $\chi^2_{\text{red}}=0.86$. This galaxy displays a good fit with a triaxial shape, with a stellar mass of $\logm =10.23$, $\lre = 0.20$, $R_{\rm{max}}/R_{\rm{e}}=1.23$, and $N_{\rm{bins}} = 164$. Panels are arranged as in Figure \ref{fig:220394_velmaps}. The model reconstructs all velocity moments accurately, as the residual panels show low values and no structure.}
    \label{fig:79733_velmaps}
\end{figure*}

\begin{figure*}
    \centering
    \includegraphics[width=\linewidth]{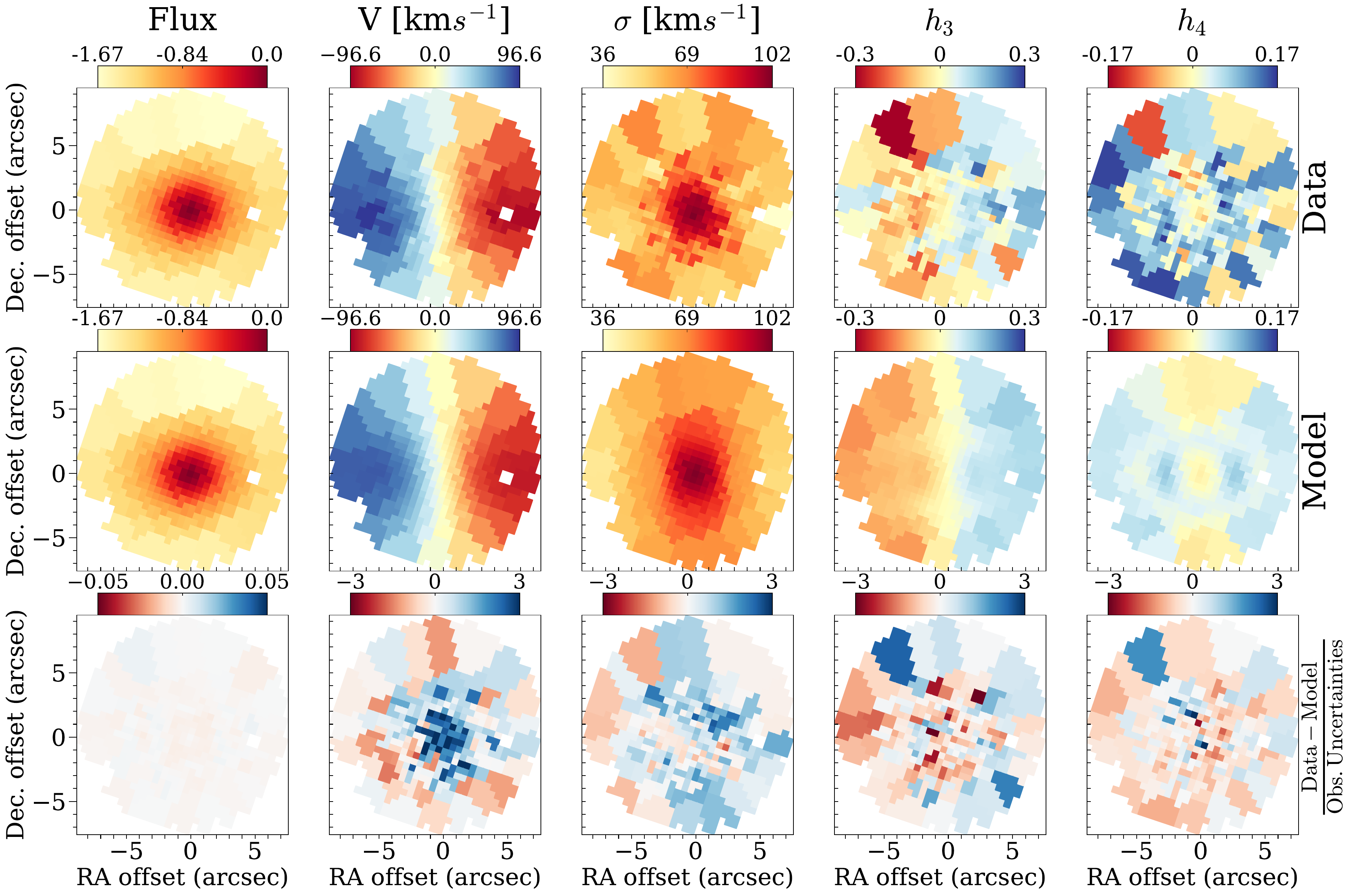}
    \caption[Schwarzschild Model for 289102]{The best-fitting Schwarzschild model for SAMI galaxy with CATID = 289102, and $\chi^2_{\text{red}}=1.07$. This galaxy displays a good fit with an oblate shape, with a stellar mass of $\logm =10.14$, $\lre = 0.46$, $R_{\rm{max}}/R_{\rm{e}}=1.56$, and $N_{\rm{bins}} = 120$. Panels are arranged as in Figure \ref{fig:220394_velmaps}. The model reconstructs all velocity moments accurately, as the residual panels show low values and no structure.}
    \label{fig:289102_velmaps}
\end{figure*}

\begin{figure*}
    \centering
    \includegraphics[width=\linewidth]{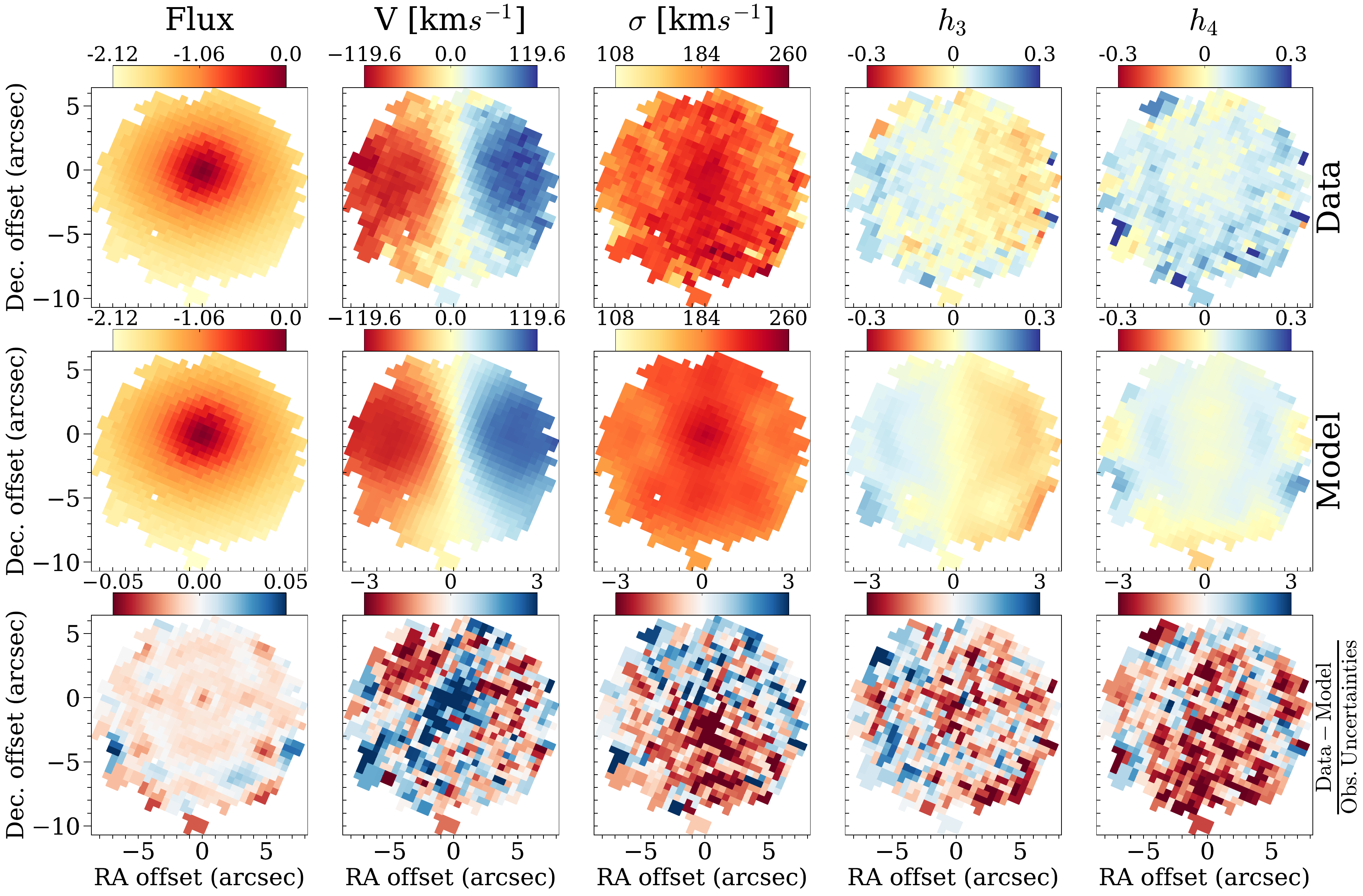}
    \caption[Schwarzschild Model for 137838]{The best-fitting Schwarzschild model for SAMI galaxy with CATID = 137838, and $\chi^2_{\text{red}}=5.34$. This galaxy has a stellar mass of $\logm =11.05$, $\lre = 0.29$, $R_{\rm{max}}/R_{\rm{e}}=1.23$, and $N_{\rm{bins}} = 459$. Panels are arranged as in Figure \ref{fig:220394_velmaps}. The model reconstructs the velocity maps with no residual structure, despite the higher $\chi^2_{\text{red}}$ value of 5.34.}
    \label{fig:137838_velmaps}
\end{figure*}

\begin{figure*}
    \centering
    \includegraphics[width=\linewidth]{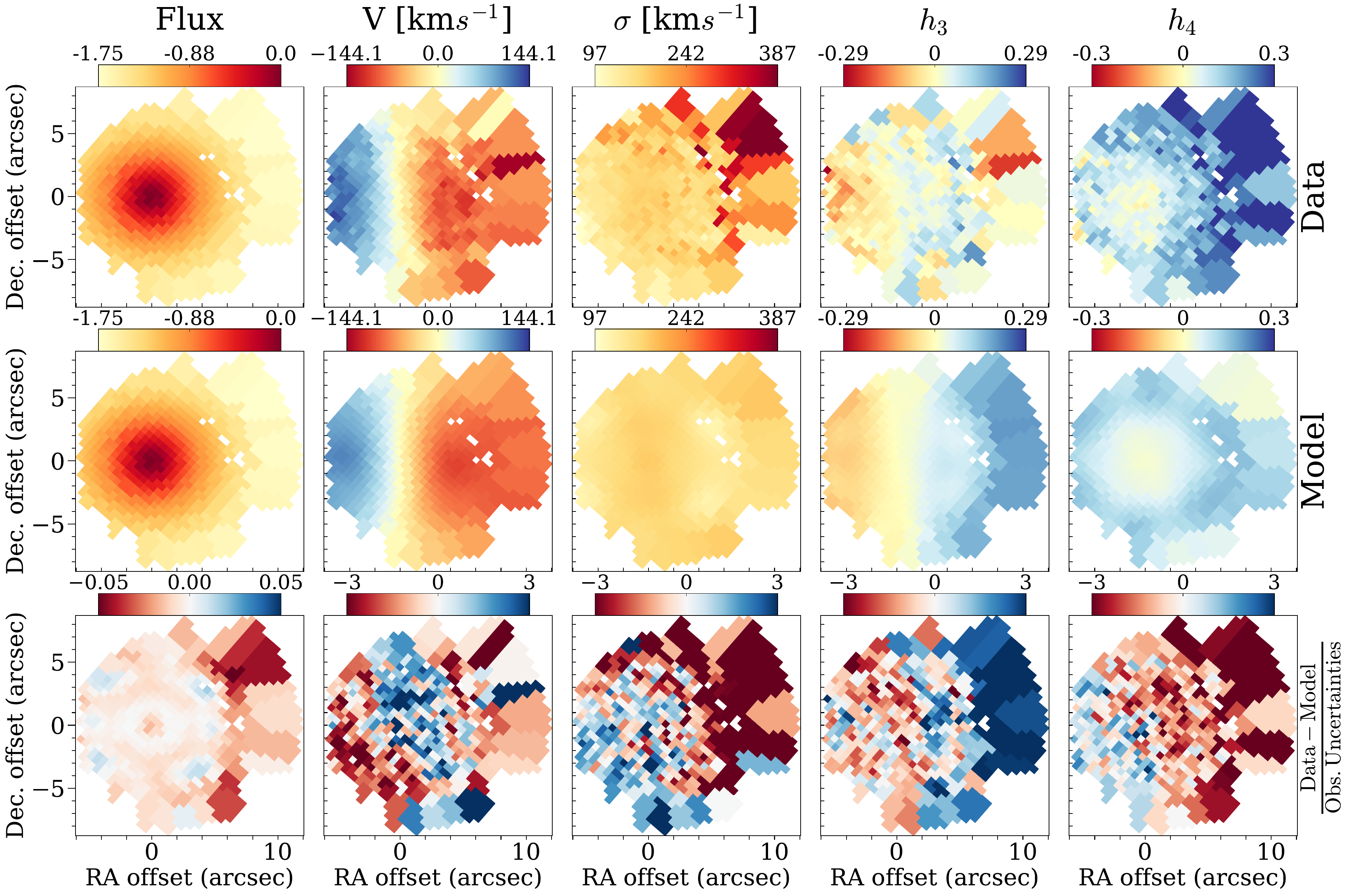}
    \caption[Schwarzschild Model for 137838]{The best-fitting Schwarzschild model for SAMI galaxy with CATID = 238922, and $\chi^2_{\text{red}}=8.19$. This galaxy has a stellar mass of $\logm =10.74$, $\lre = 0.36$, $R_{\rm{max}}/R_{\rm{e}}=1.47$, and $N_{\rm{bins}} = 208$. Panels are arranged as in Figure \ref{fig:220394_velmaps}. The model reconstructs the velocity maps with no residual structure, despite the higher $\chi^2_{\text{red}}$ value of 8.19.}
    \label{fig:238922_velmaps}
\end{figure*}

Here we present five further examples of Schwarzschild model fits to our sample, labelled by their CATID:
\begin{itemize}
    \item 511892: A galaxy close to the minimum number of spatial bins, 78 bins within $R_e$,  $\chi^2_{\text{red}}=1.13$. Shown in Figure \ref{fig:511892_velmaps}.
    \item 79733: A triaxial galaxy, 164 bins within $R_e$,  $\chi^2_{\text{red}}=0.86$. Shown in Figure \ref{fig:79733_velmaps}.
    \item 289102: An oblate galaxy, 120 bins within $R_e$,  $\chi^2_{\text{red}}=1.07$. Shown in Figure \ref{fig:289102_velmaps}.
    \item 137838: A galaxy with 459 bins within $R_e$, $\chi^2_{\text{red}}=5.34$. Shown in Figure \ref{fig:137838_velmaps}.
    \item 238922: A galaxy with 208 bins within $R_e$, $\chi^2_{\text{red}}=8.19$. Shown in Figure \ref{fig:238922_velmaps}.
\end{itemize}
\section{Example Parameter Spaces}
\label{sec:parameter_search}
\begin{figure}
    \centering
    \includegraphics[width=\linewidth]{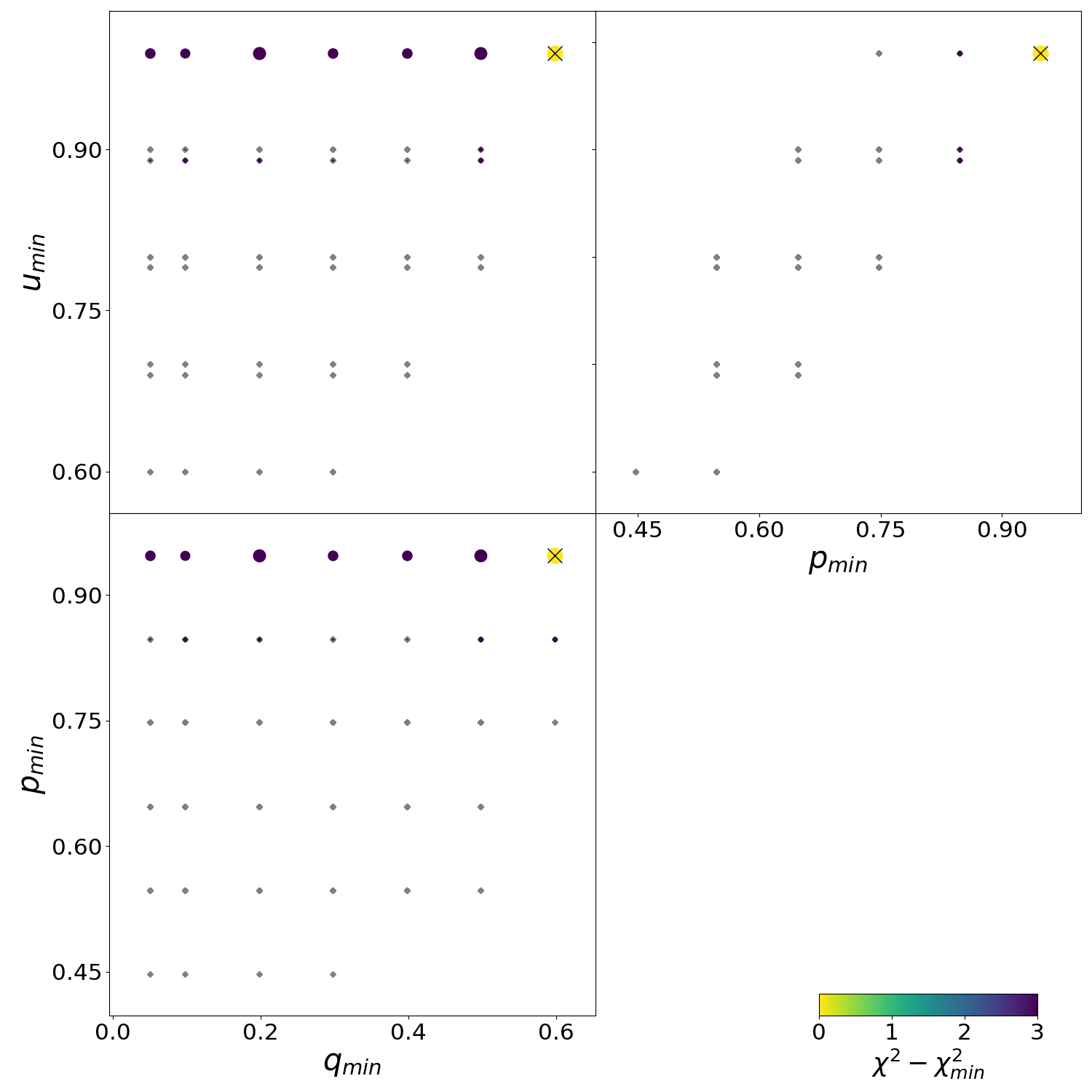}
    \caption{The parameter space searched over our coarse library of orbits for the galaxy with CATID 220394. Here we attempt to find initial values of the structural parameters $p_{\rm{min}},q_{\rm{min}}$ and $q_{\rm{min}}$. The colourbar represents the $\chi^2$ value of a point in our $p_{\rm{min}},q_{\rm{min}},q_{\rm{min}}$ grid. The point with an X represents the set of values taken as the best fit from this grid search.}
    \label{fig:220394_parm_coarse}
\end{figure}

\update{Here we present the parameter spaces searched by DYNAMITE in deriving the Schwarzschild model to the galaxy in our sample with CATID 220394, as described in Section \ref{sec:orbit_weighting}. We show the search over the coarse library of orbits to find initial values of the structural parameters $p_{\rm{min}},q_{\rm{min}}$ and $q_{\rm{min}}$ in Figure \ref{fig:220394_parm_coarse}. The colourbar represents the $\chi^2$ value of a point in our $p_{\rm{min}},q_{\rm{min}},q_{\rm{min}}$ grid, and the point with an X represents the set of values taken as the best fit from this grid search.}

\begin{figure}
    \centering
    \includegraphics[width=\linewidth]{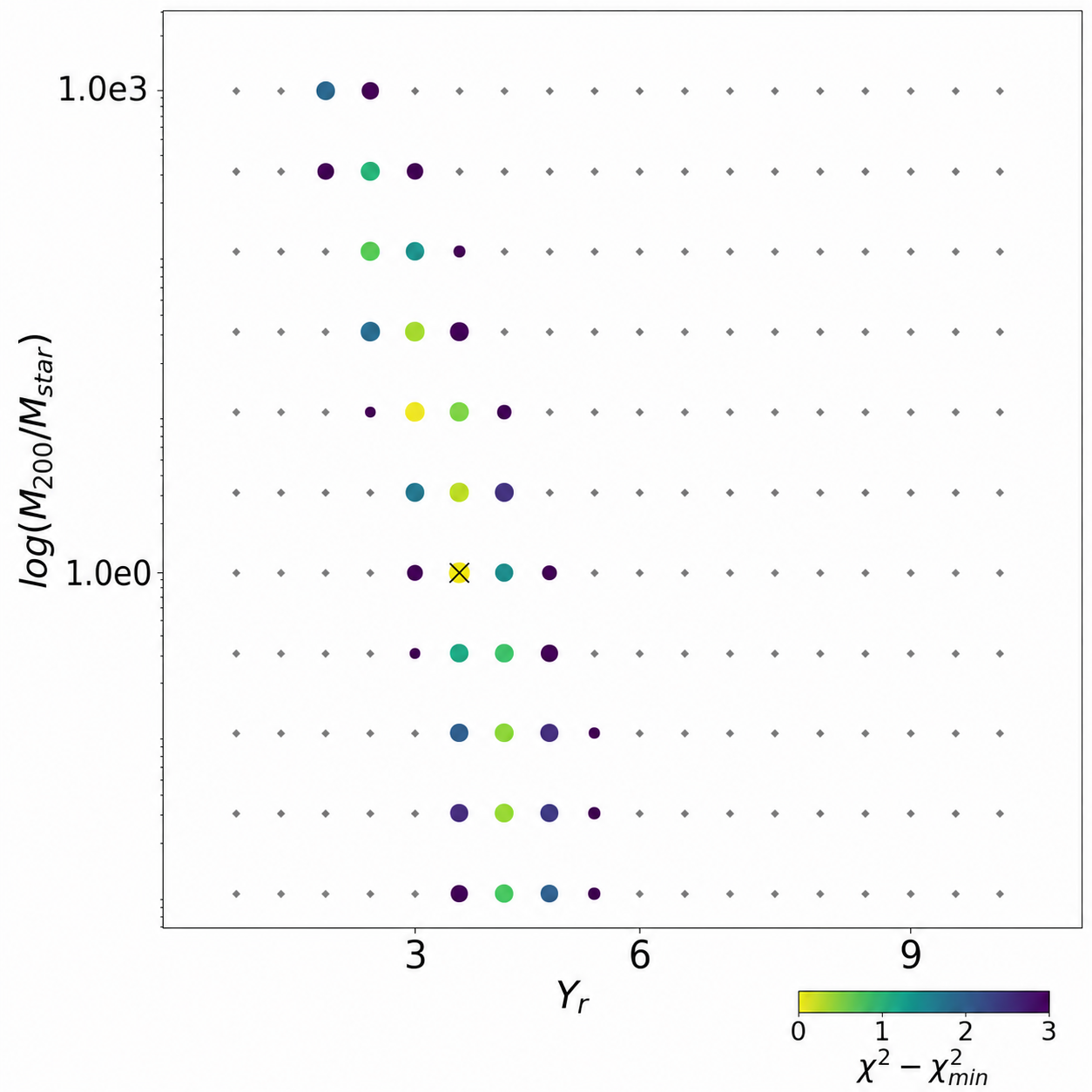}
    \caption{The parameter space searched over our second coarse library of orbits for the galaxy with CATID 220394. Here we attempt to find initial values of the parameters $\log f$ (or $\log(M_{200}/M_{\rm{star}})$) and $\Upsilon_*$. The colourbar represents the $\chi^2$ value of a point in our $\log f,\Upsilon_*$ grid. The point with an X represents the set of values taken as the best fit from this grid search.}
    \label{fig:220394_parm_middle}
\end{figure}

\update{In Figure \ref{fig:220394_parm_middle} we show the search over our coarse library of orbits to find initial values of $\log f$ and $\Upsilon_*$. The colourbar represents the $\chi^2$ value of a point in our $\log f,\Upsilon_*$ grid, and the point with an X represents the set of values taken as the best fit from this grid search.}

\begin{figure*}
    \centering
    \includegraphics[width=\linewidth]{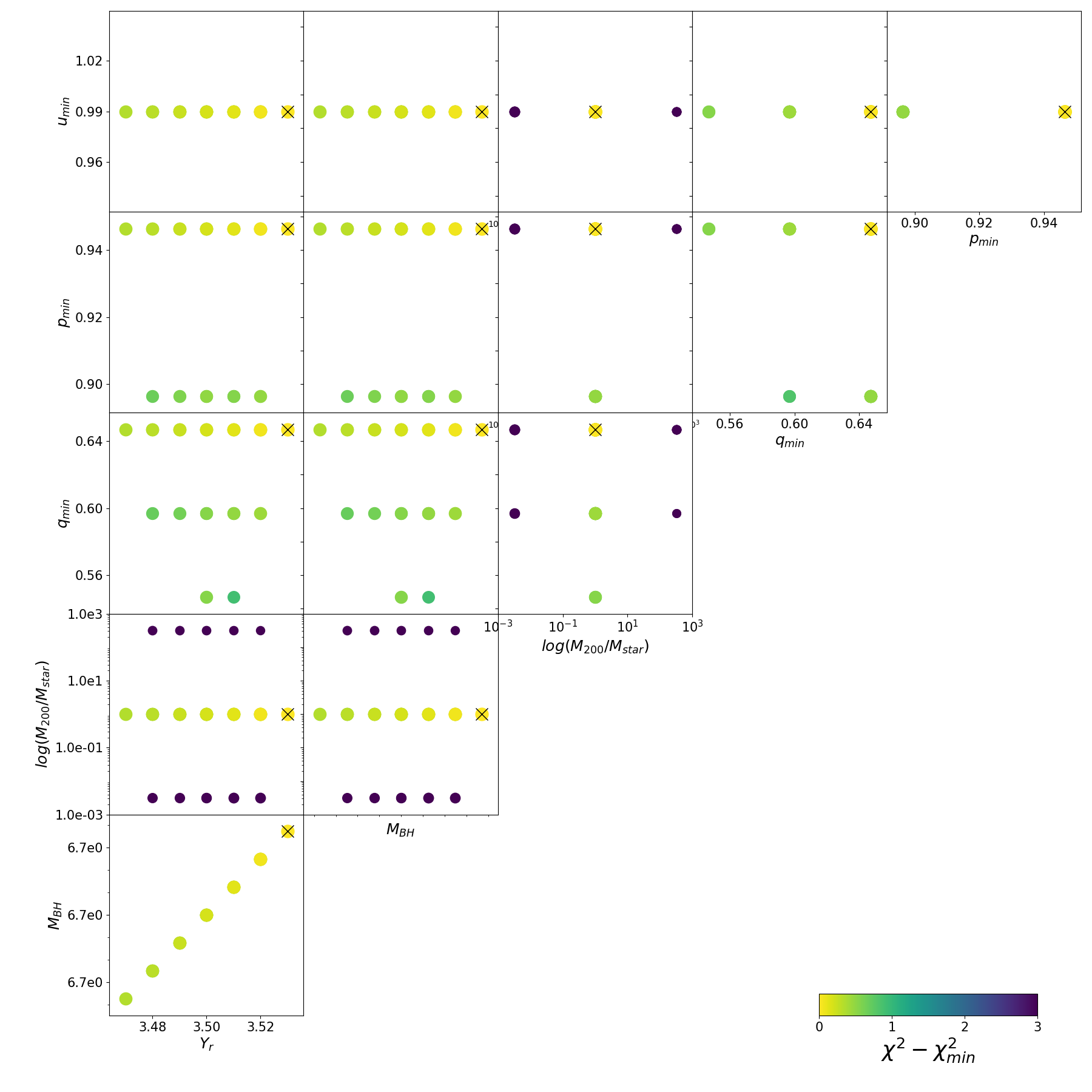}
    \caption{The parameter space searched over our fine library of orbits for the galaxy with CATID 220394. Here we attempt to find our best-fit values of the parameters $p_{\text{min}},q_{\text{min}},u_{\text{min}},\Upsilon_*,\log f,\text{ and }\log(\rm{M}_{BH})$. The colourbar represents the difference between the $\chi^2$ value of each grid point and the minimum $\chi^2$ value. The point with an X represents the set of values taken as the best fit from this grid search.}
    \label{fig:220394_parm_fine}
\end{figure*}

\update{In Figure \ref{fig:220394_parm_fine} we show the search over our fine library of orbits to find our best-fit values of the parameters $p_{\text{min}},q_{\text{min}},u_{\text{min}},\Upsilon_*,\log f,\text{ and }\log(\rm{M}_{BH})$. The colourbar represents the $\chi^2$ value of a point in our grid, and the point with an X represents the set of values taken as the best fit from this grid search. We find that $\rm{M}_{BH}$ is not constrained by our data in any of our galaxies, and thus the grid search never steps away from the initial value determined by the stellar mass scaling relation of \cite{2015ApJ...813...82R}.}

\begin{figure*}
    \centering
    \includegraphics[width=\linewidth]{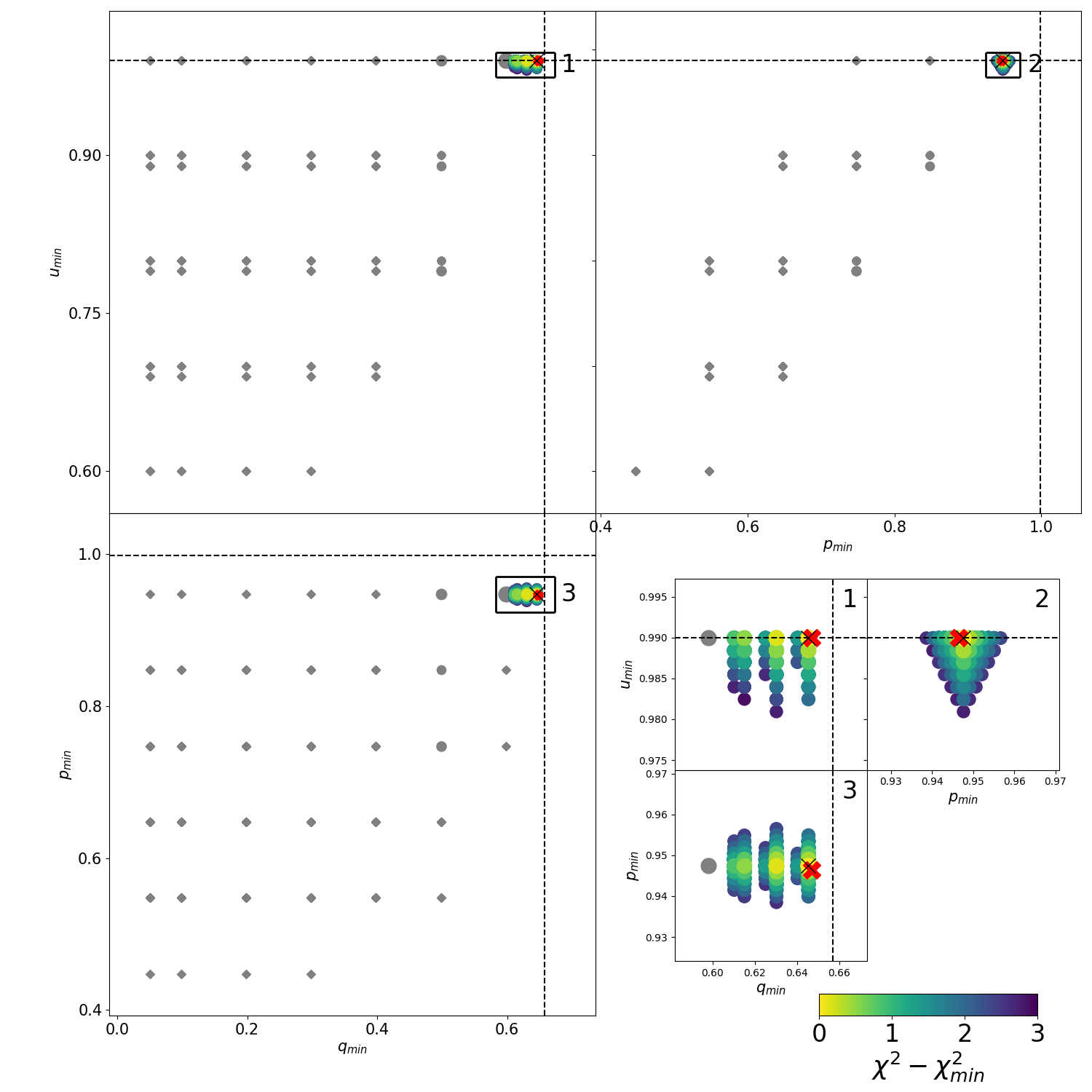}
    \caption{The parameter space searched over our fine library of orbits for the galaxy with CATID 220394, with a more fine grid than shown in Figure \ref{fig:220394_parm_fine}. Here we show the coarse parameter search over the $p$, $q$, and $u$ parameters, with the fine search overlaid. We additionally show a zoomed version of each panel in the bottom right of the Figure. The best fit parameter found is shown as a black X, with the best fit parameter from Figure \ref{fig:220394_parm_fine} shown as a red X. The maximum allowed values of each parameter are shown as dashed black lines. We find very similar best-fit values to Figure \ref{fig:220394_parm_fine}, demonstrating that our adopted sampling is sufficient to search the parameter space.}
    \label{fig:220394_convergence_pqu}
\end{figure*}

\begin{figure}
    \centering
    \includegraphics[width=\linewidth]{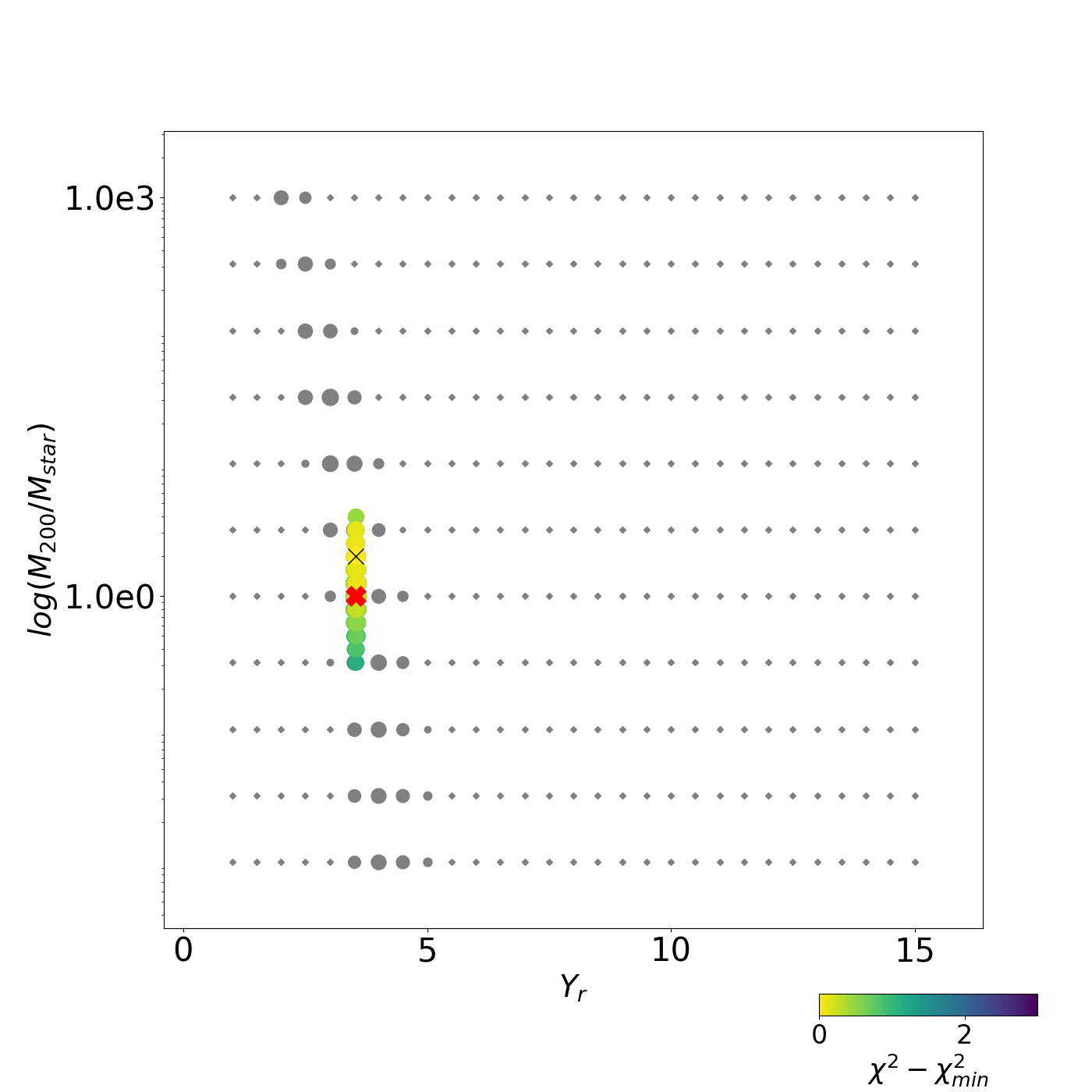}
    \caption{The parameter space searched over our fine library of orbits for the galaxy with CATID 220394, with a more fine grid than shown in Figure \ref{fig:220394_parm_fine}. Here we show the coarse parameter search over the $\log(M_{200}/M_{\rm{star}})$ and $\Upsilon_*$ parameters, with the fine search overlaid. The best fit parameter found is shown as a black X, with the best fit parameter from Figure \ref{fig:220394_parm_fine} shown as a red X. We find similar best-fit values to Figure \ref{fig:220394_parm_fine}. Although the $\log(M_{200}/M_{\rm{star}})$ value differs between the model fits, this only translates to a change in dark matter fraction within 1\re\ from 6\% to 7\%.}
    \label{fig:220394_convergence_dm_ml}
\end{figure}

\update{We note that the sampling of parameter space in \ref{fig:220394_parm_fine} may appear relatively coarse. This reflects the necessary trade off between resolution and computational expense, as the fitting of each model is computationally expensive and the full exploration of a fine grid parameter space for all galaxies in our sample would be prohibitive. To ensure that this choice does not bias our results, we performed a convergence test on the same galaxy shown in \ref{fig:220394_parm_fine} using a significantly finer sampling of the parameter space. We show this parameter search in Figures \ref{fig:220394_convergence_pqu} and \ref{fig:220394_convergence_dm_ml}. This finer search converges on very similar free parameters as compared to the search shown in Figure \ref{fig:220394_parm_fine}, demonstrating that our adopted sampling is sufficient to search the parameter space. Further, the orbit fractions derived are comparable within our adopted 10\% uncertainties: \hot\ = 0.84 vs 0.83, \warm\ = 0.09 vs 0.12, \cold\ = 0.04 vs 0.02, and \counter\ = 0.02 vs 0.03.}

\section{Previous Model Comparison}
\label{sec:model_comparison}
\begin{figure*}
    \centering
    \includegraphics[width=0.65\linewidth]{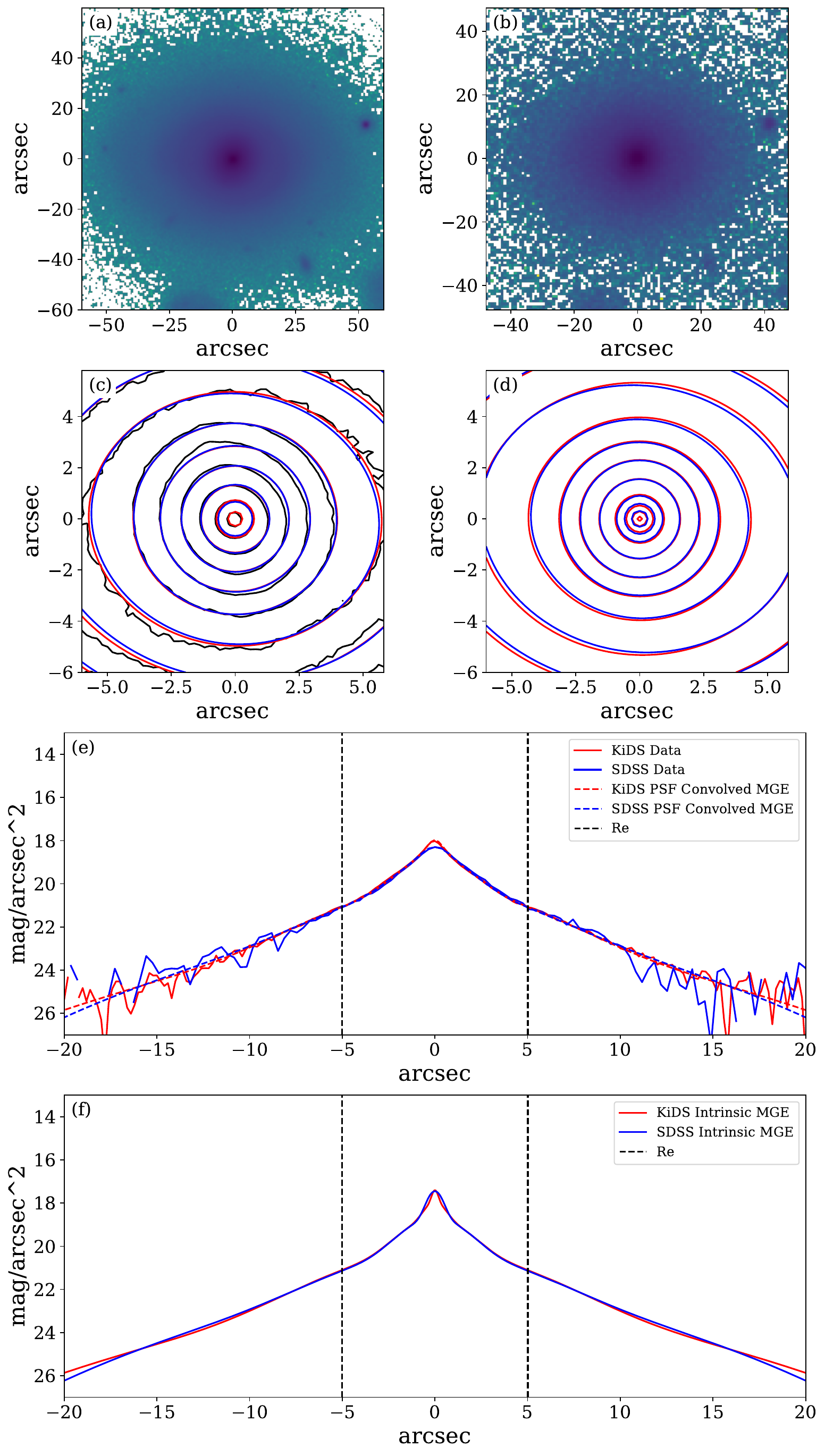}
    \caption{A comparison between the KiDS MGE used for models in this work, and the SDSS MGE used for the models by \protect\cite{2022ApJ...930..153S}, for the galaxy with CATID 220394. We show the r-band image for KiDS and for SDSS in panels (a) and (b) respectively. We show the fitted MGE convolved with the PSF for the central regions of KiDS and SDSS in panels (c) and (d) respectively. We show a one dimensional profile comparison in panel (e) where the photometric data for KiDS is shown in red (blue for SDSS) and the PSF convolved MGE for KiDS is shown in dashed red (dashed blue for SDSS). The effective radius is shown as a dashed black line. We show the intrinsic MGE, unconvolved with the PSF, in panel (f), with KiDS in red and SDSS in blue, and the effective radius in a dashed black line.}
    \label{fig:kids_sdss_mge_comparison_220394}
\end{figure*}

\begin{figure*}
    \centering
    \includegraphics[width=0.65\linewidth]{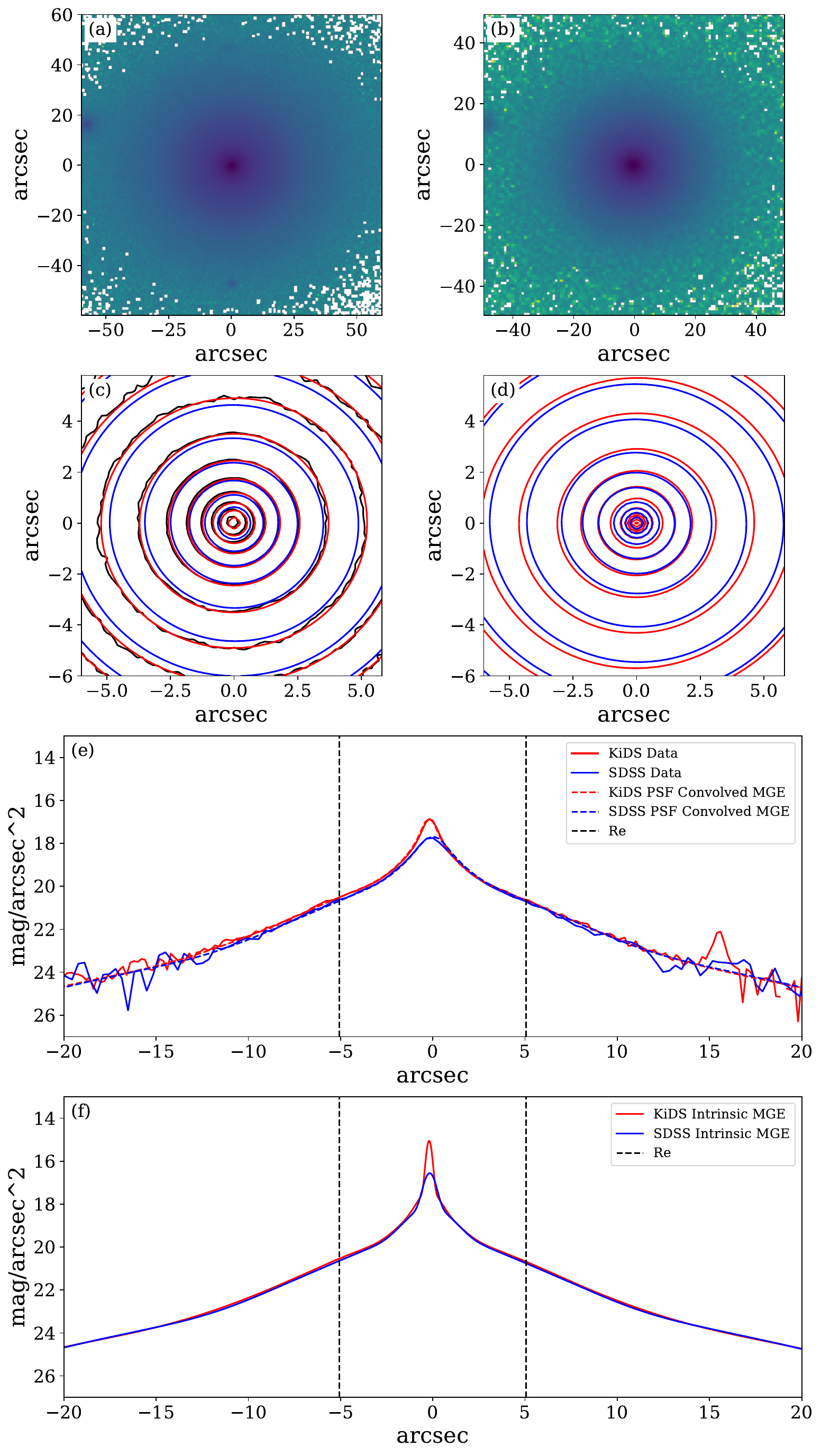}
    \caption{A comparison between the KiDS MGE used for models in this work, and the SDSS MGE used for the models by \protect\cite{2022ApJ...930..153S}, for the galaxy with CATID 39057. Data is displayed similarly to Figure \ref{fig:kids_sdss_mge_comparison_220394}.}
    \label{fig:kids_sdss_mge_comparison_39057}
\end{figure*}

\update{In this section, we present a comparison to the Schwarzschild models derived by \cite{2022ApJ...930..153S}. The primary difference between the models derived here and those derived by \cite{2022ApJ...930..153S} is the adopted photometric model, where we use KiDS r-band MGEs, and \cite{2022ApJ...930..153S} uses SDSS r-band MGEs. These MGE models are expected to be significantly different, particularly in the central regions, due to the difference in seeing (0.7" for KiDS, 1.4" for SDSS). We show this difference for two examples galaxies, 220394 and 39057, in Figures \ref{fig:kids_sdss_mge_comparison_220394} and \ref{fig:kids_sdss_mge_comparison_39057}. These figures clearly show the effect of seeing in the derived model, particularly in panel (f) of Figure \ref{fig:kids_sdss_mge_comparison_39057}, where the intrinsic KiDS MGE peaks significantly higher than SDSS.}

\begin{figure*}
    \centering
    \includegraphics[width=\linewidth]{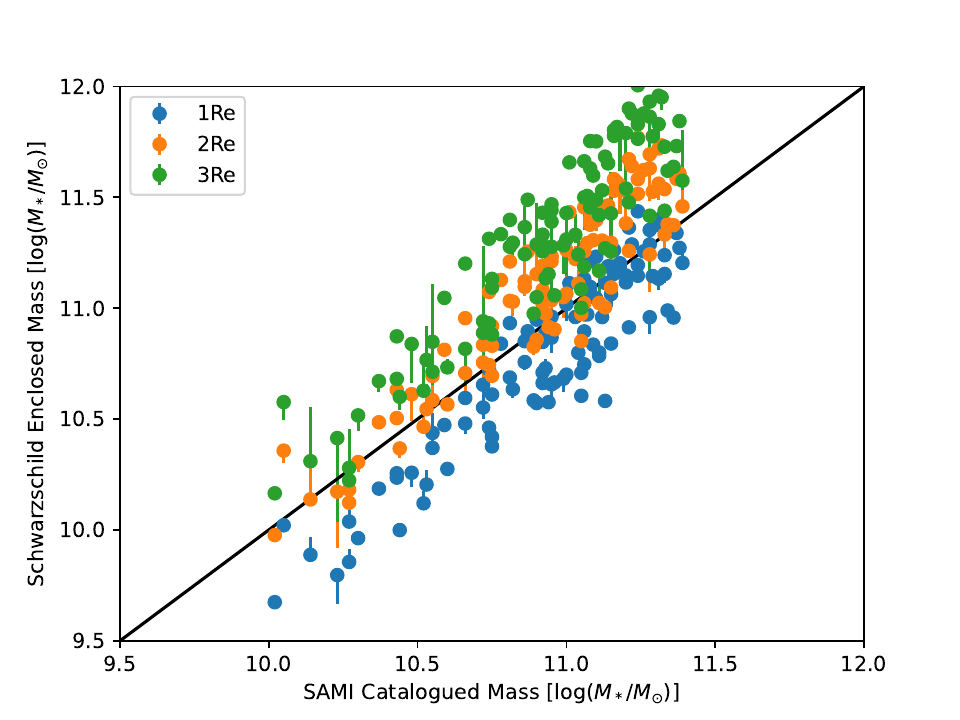}
    \caption{Enclosed dynamical masses within 1\re, 2\re, and 3\re\ for each galaxy in our sample, compared to the catalogued stellar mass. Uncertainties are taken as the $1\sigma$ uncertainties amongst tested models. The strong correlation suggests our derived enclosed masses, and thus models, are physically accurate.}
    \label{fig:enclosed_mass}
\end{figure*}

\update{To assess the sensitivity of our results to the adopted photometric model, we construct Schwarzschild models for the two galaxies 220394 and 39057 using the SDSS MGE model. We show the derived parameters in Table \ref{tab:kids_sdss_comparison}. We find consistent enclosed dynamical masses, but differences in orbit fraction between photometry for each galaxy. We further examine the reliability of our dynamical masses by comparing them with the stellar mass estimates from \cite{2015MNRAS.447.2857B} in Figure \ref{fig:enclosed_mass}. Here we present the enclosed dynamical masses within 1\re, 2\re, and 3\re\ for our sample. Uncertainties are taken as the $1\sigma$ distribution amongst tested models. Although stellar mass and dynamical mass are not equivalent, the total enclosed mass is tightly constrained.}

\update{These results suggest that while stellar kinematics are sufficient to constain dynamical mass for various photometric models, orbital fractions can vary significantly. We conclude that any observed differences in orbital fractions between this work and that of \cite{2022ApJ...930..153S} therefore reflect the sensitivity of the orbital decomposition to the assumed mass model, rather than indicating any inconsistency between the two modelling approaches.}

\begin{table}
	\centering
	\caption{A comparison of derived parameters for Schwarzschild models using different photometric models. For galaxies with CATIDs 220394 and 39057, we show parameters for models using a KiDS r-band MGE, and an SDSS r-band MGE. We show the orbit fractions for hot, warm, cold, and counter-rotating orbits within 1\re\ ($f_{\rm{Hot, e}}$, $f_{\rm{Warm, e}}$, $f_{\rm{Cold, e}}$, $f_{\rm{CR, e}}$, respectively). We also show the enclosed dynamical mass within 1\re, 2\re, and 3\re, in units of $\log(M/M_{\odot})$, as well as the reduced $\chi^2$.}
	\label{tab:kids_sdss_comparison}
	\begin{tabular}{ ccccc } 
		\hline
		\hline
		 & \multicolumn{2}{c}{220394} & \multicolumn{2}{c}{39057}\\
        & KiDS & SDSS & KiDS & SDSS\\
        \hline
        $f_{\rm{Hot, e}}$ &0.83 & 0.80& 0.81&0.72  \\
        $f_{\rm{Warm, e}}$ &0.12& 0.11& 0.15&0.20  \\
        $f_{\rm{Cold, e}}$ &0.02& 0.07& 0.01&0.06  \\
        $f_{\rm{CR, e}}$ &0.03& 0.01& 0.03&0.02  \\
        Enclosed Mass in 1\re & 10.00 & 9.98 & 10.00&10.00 \\
        Enclosed Mass in 2\re & 10.30 & 10.32 &10.37 &10.31 \\
        Enclosed Mass in 3\re & 10.47 & 10.53 &10.60 &10.47 \\
        $\chi^2_{\rm{red}}$ & 0.94 & 0.98 & 1.44 & 1.48 \\
		\hline
		\hline
	\end{tabular}
\end{table}

\bsp	
\label{lastpage}
\end{document}